\documentclass[twocolumn, tighten]{aastex62}

\usepackage{graphicx}	
\usepackage[fleqn]{amsmath}	
\usepackage{amssymb}	
\usepackage[caption=false]{subfig}
\usepackage{tablefootnote}
\usepackage{bm}

\received{February 4, 2019}
\revised{May 30, 2019}
\accepted{June 25, 2019}
\submitjournal{ApJ}

\shorttitle{NGC$\,$2244}
\shortauthors{Mu\v{z}i\'c et al.}

\begin{document}

\title{Looking deep into the Rosette Nebula's heart: The (sub)Stellar Content of the Massive Young Cluster NGC\,2244}

\correspondingauthor{Koraljka Mu\v{z}i\'c}
\email{kmuzic@sim.ul.pt}

\author[0000-0002-7989-2595]{Koraljka Mu\v{z}i\'c}
\affil{CENTRA/SIM, Faculdade de Ciencias de Universidade de Lisboa, Ed. C8, Campo Grande, P-1749-016 Lisboa, Portugal }

\author{Alexander Scholz}
\affiliation{SUPA, School of Physics \& Astronomy, St. Andrews University North Haugh, St Andrews, KY16 9SS, United Kingdom }

\author{Karla Pe\~{n}a Ram\'irez}
\affiliation{Centro de Astronom\'ia (CITEVA), Universidad de Antofagasta, Avenida Angamos 601, Antofagasta, Chile}

\author{Ray Jayawardhana}
\affiliation{Department of Astronomy, Cornell University, Ithaca, New York 14850, USA}
 
\author{Rainer Sch\"{o}del}
\affiliation{Instituto de Astrof\'isica de Andaluc\'ia (CSIC), Glorieta de la Astronom\'a s/n, 18008 Granada, Spain}

\author{Vincent C. Geers}
\affiliation{UK Astronomy Technology Centre, Royal Observatory Edinburgh, Blackford Hill, Edinburgh, EH9 3HJ, United Kingdom}
 
\author{Lucas A. Cieza}
\affiliation{N\'ucleo de Astronom\'ia, Facultad de Ingenier\'ia y Ciencias, Universidad Diego Portales, Av. Ejercito 441, Santiago, Chile}
 
\author{Amelia Bayo}
\affiliation{Instituto de F\'isica y Astronom\'ia, Universidad de Valpara\'iso, Av. Gran Breta\~{n}a 1111, Playa Ancha, Casilla 5030, Chile}

\begin{abstract}
As part of the ongoing effort to characterize the low-mass (sub)stellar population in a sample of massive young clusters, we have targeted the $\sim$2 Myr old
cluster NGC\,2244. The distance to NGC\,2244 from {\it Gaia DR2} parallaxes is 1.59 kpc, with errors of $1\%$ (statistical) and 11\% (systematic).
We used the Flamingos-2 near-infrared camera at the Gemini-South telescope for deep multi-band imaging of the central portion of the cluster ($\sim 2.4\,$pc$^2$).
We determined membership in a statistical manner, through a comparison of the cluster's color-magnitude diagram to that of a control field. Masses and extinctions of the candidate members are then calculated with the help of evolutionary models, leading to the first initial mass function (IMF) of the cluster extending into the substellar regime, with the 90\% completeness limit around $0.02$\,M$_{\sun}$.
The IMF is well represented by a broken power law ($dN/dM \propto M^{-\alpha}$), with
a break at $\sim$0.4\,M$_{\sun}$. 
The slope on the high-mass side ($0.4 - 7\,$M$_{\sun}$) is $\alpha=2.12\pm0.08$, close to the standard Salpeter's slope.
In the low-mass range ($0.02 - 0.4\,$M$_{\sun}$), we find a slope $\alpha=1.03\pm0.02$, which is on the high end of the typical
values obtained in nearby star-forming regions ($\alpha=0.5-1.0$), but still in agreement within the uncertainties.
Our results reveal no clear evidence for variations in the formation efficiency of brown dwarfs and very low-mass stars due to the presence of OB stars, or for a change in stellar densities. 
Our finding rules out photoevaporation and fragmentation of infalling filaments as substantial pathways for BD formation.

\end{abstract}



\section{Introduction}
\begin{figure*}
\centering
\resizebox{0.8\textwidth}{!}{\includegraphics{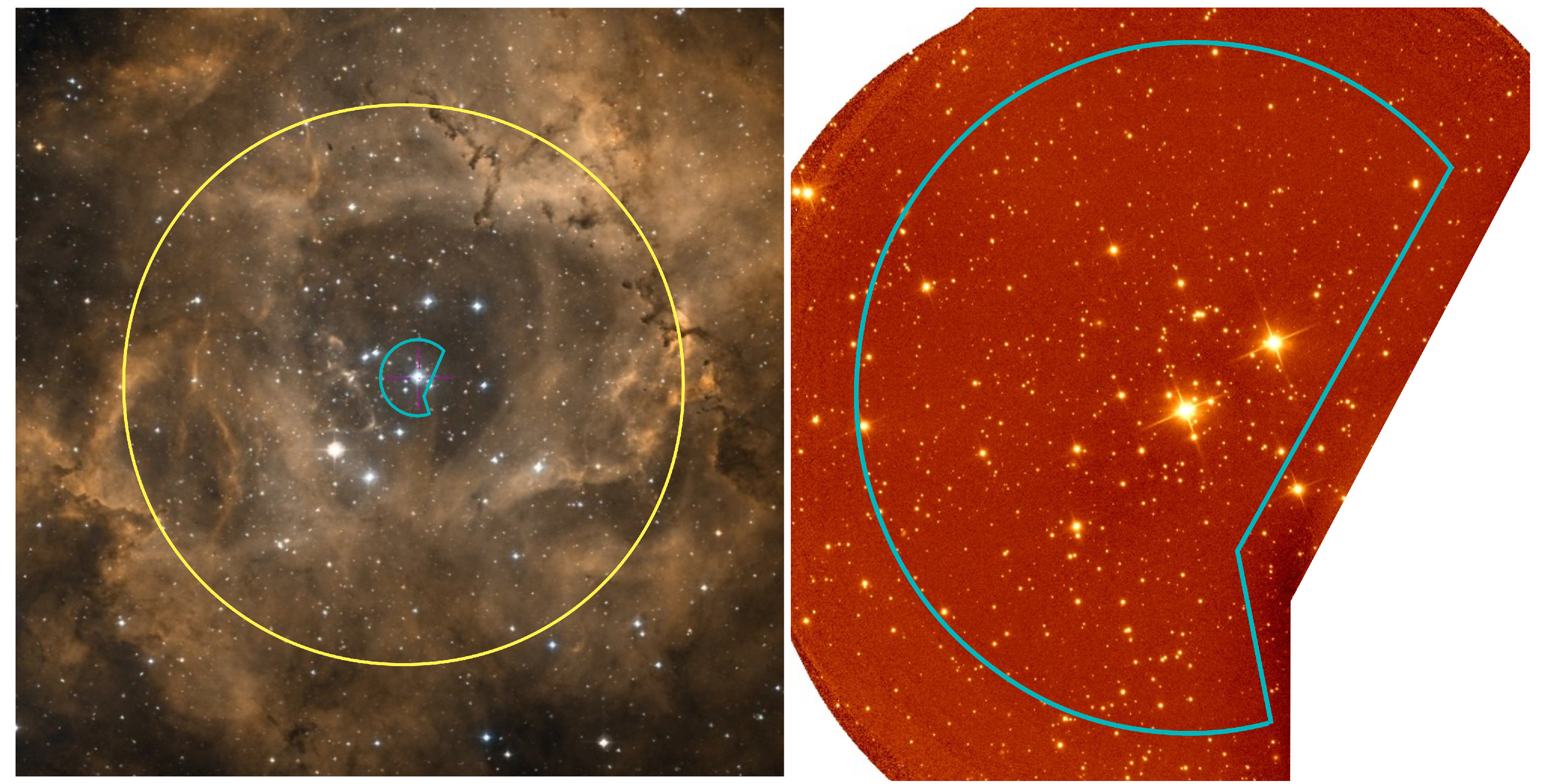}}
\caption{ 
{\bf Left:} $55'\times55'$ DSS image of the Rosette Nebula. The large circle has a $20'$ radius, and marks the extent of the NGC\,2244
cluster \citep{wang08}. The small cyan area is the one covered by our Gemini observations.
{\bf Right:} $Ks$-band image of the center of NGC\,2244 taken with Flamingos-2/Gemini-S. The marked area is the one analyzed in this paper, defined by the maximum overlap of the dithered frames ($\sim$11.1 arcmin$^2$). The instrument field of view is circular, and the linear cuts on the right-hand side of the image exclude the area vignetted by the telescope guide probe. The radius of the cyan (partial) circle is $2.4'$, or $1.1$\,pc at the distance of 1600\,pc. North is up, east to the left.
}
\label{fig_composite}
\end{figure*}

\subsection{Young star clusters and the IMF}
Star clusters are the primary locations of star formation, and the main
suppliers of stellar and substellar objects in the universe \citep{lada&lada03,portegieszwart10}. 
The youngest clusters 
are particularly important laboratories for understanding how star formation works. They give birth
to objects spanning at least four orders of magnitude in mass, from the most massive stars reaching several tens of solar masses, down
to the substellar objects below the deuterium-burning limit at $\sim\,$0.012\,M$_{\sun}$. Richly populated young clusters therefore
provide an ideal testbed in which to probe a variety of yet unsolved questions related to the formation of stars, brown dwarfs (BDs), and clusters. 

The distribution of stellar masses in young clusters, or the initial mass function (IMF), 
has been an important subject of many observational and theoretical efforts. On the high-mass side ($\gtrsim 1$M$_{\sun}$), it can be approximated by a power law form $dN/dM \propto M^{-\alpha}$, with $\alpha = 2.35$ \citep{salpeter55}. Below $\sim0.5$M$_{\sun}$, the mass distribution is significantly flatter (see, e.g., \citealt{luhman12} and references therein). 
Although the Salpeter's slope is often considered universal (see, e.g., \citealt{bastian10, offner14}), there are some notable examples of young clusters where the mass function appears to be flatter, with $\alpha$=1.3-1.8 \citep{stolte06,harayama08,bayo11,penaramirez12,habibi13,andersen16,muzic17}. While the flattening of the slope could be an effect of crowding for the clusters that are several kpc away (NGC 3603, Arches, Westerlund 1)\citep{ascenso09}, this is certainly not the case for the more nearby ones (Collinder~69, $\sigma\,$Ori, RCW 38). Employing Bayesian statistics, \citet{dib14} reports significant variations in IMF parameters of eight young Galactic clusters, including the slope 
of the IMF on the high-mass side. Recently, \citet{dib17} proposed a probabilistic formulation of the IMF, where the IMF is described by a tapered power law \citep{demarchi10}, and each parameter of this function is represented by a Gaussian probability distribution. The authors argue that the broad distribution of the IMF parameters might reflect the existence of equally broad distributions in the star forming conditions (e.g. \citealt{svoboda16}).

On the low-mass side, the observed IMFs in young clusters and star forming regions in general yield slopes $\alpha$ in the range 0.5 - 1,
for the masses below $\sim 1$\,M$_{\sun}$ and extending into the substellar regime (e.g., \citealt{luhman04c, luhman07, bayo11, penaramirez12, scholz12a, lodieu13, muzic15, muzic17}). The number ratio between low-mass stars and BDs is typically found in the range 2 - 5. 
The range of values reflects different sources of uncertainties that contribute to the derivation of the IMF slope and the star-to-BD ratio 
(uncertainties in distances, ages, use of different evolutionary models, reddening laws, membership issues, small sample sizes; see e.g. \citealt{scholz13}), rather than the variations between individual clusters. Although from the observations there is no convincing evidence for strong variations in the BD production efficiency in nearby star forming regions, theoretical expectations are somewhat different. Most of the current BD formation theories in fact predict
an increased production of substellar objects in dense environments, or close to very massive stars.
 The factors that facilitate BD formation are high gas densities \citep{bonnell08,padoan02,hennebelle09,jones18}, high stellar densities that favor BD formation through ejections \citep{bate12}, and the presence
of massive stars, where BDs can be formed through photoevaporation \citep{whitworth04}, or by fragmentation in their massive disks \citep{stamatellos11,vorobyov13}. 
Of the mentioned scenarios, a few give quantitative predictions of the impact of environment on the efficiency of BD production. In the scenario by \citet{bonnell08}, where new objects form by gravitational fragmentation of gas infalling into a
stellar cluster, BDs are preferentially formed in regions with high stellar density. An increase in object density by an order of magnitude would result in an increase in the BD fraction by a factor of about two. The predicted effect is more subtle in the hydrodynamical simulations of molecular cloud collapse by \citet{jones18}. An increase in gas density by a factor of 100 leads to BD frequencies larger by $\sim 35$\%.

To explore potential environmental influences on BD formation, we have initiated a photometric study 
in a sample of young, massive clusters, characterized by high stellar densities and/or a substantial population of massive stars. In this sense, the environments we aim to cover are significantly different from those in nearby star forming regions, with the exception of the Orion Nebula Cluster (ONC), the only site of massive star formation within 500\,pc from the Sun. The latest results from the ONC show a bimodal form of an IMF, with a secondary peak in the substellar regime \citep{drass16}. The authors 
interpret the substellar peak as possibly formed by BDs ejected from multiple systems or circumstellar disks at an early stage of star formation. The spectroscopic confirmation of this result, however, is still pending. 
In the first paper resulting from study of massive clusters, we studied the core of RCW\,38, a young (1 Myr), embedded cluster located at the distance of 1.7 kpc \citep{muzic17}. RCW\,38 is characterized both by high stellar densities (core stellar surface density $\Sigma\sim$2500\,pc$^{-2}$, i.e. twice as dense as the ONC and more than ten times denser than NGC\,1333; \citealt{muzic17}), and a substantial population of massive OB stars (60 OB candidates in total, and a dozen in our small field of view of $0.5\times0.5$\,pc$^2$; \citealt{wolk08, winston11}). Given these characteristics, and the absence of more extreme environments that are close enough to allow the detection of BD candidates, this cluster was thought to be a good starting point in which to look for possible environmental differences in the substellar regime. Unlike what has been reported for the ONC, we found that RCW\,38's IMF is consistent with those in nearby star forming regions ($\alpha\,\sim0.7$ for the range 0.02 - 0.5\,M$_{\sun}$). 

\subsection{NGC\,2244}
In this second paper of the series, we present the observation of the young cluster NGC\,2244, located at the center of Rosette Nebula, one of the most prominent features of the Mon OB2 Cloud. The cluster contains more than 70 massive OB stars, which are presumed to be responsible for the evacuation of the central part of the nebula \citep{romanzunigalada08}. Estimates of the distance to NGC\,2244 vary from 1.4 to 1.7 kpc \citep{ogura&ishida81, perez87,park02, hensberge00,lombardi11, martins12, bell13, kharchenko13}, whereas most of the authors agree on $\sim$\,2\,Myr for the age of the cluster \citep{perez87,park02, hensberge00,bell13}. 

The stellar content of NGC\,2244 has been extensively studied from X-ray to mid-infrared (MIR) wavelengths. The most sensitive X-ray data to date are the $Chandra$ observations published in \citet{wang08}. The authors identify more than 900 X-ray sources in a $\sim\,17'\times17'$ field, of which $77$\% have optical or near-infrared (NIR) counterparts and are considered young cluster members. The X-ray-selected population by \citet{wang08} is nearly complete down to 0.5\,M$_{\sun}$. The deepest published NIR data on NGC\,2244 are those taken with Flamingos at the 2.1-m telescope at the Kitt Peak National Observatory, as part of the imaging survey of the Rosette complex (complete down to $J\sim 18.5$\,mag; \citealt{romanzuniga08}). The analysis of several clusters in the complex reveals NGC\,2244 to be the largest one.
That paper, however, deals with the entire Rosette complex, and does not present further specific analysis of the NGC\,2244 cluster population. The early attempts to derive the IMF of the cluster reported a shallow slope $\alpha\sim1.7 - 1.8$ \citep{massey95,park02} for stars with $m>3\,$M$_{\sun}$, but at the same time \citet{park02} cautioned about the incompleteness  at the intermediate and low masses in their sample. More recently, \citet{wang08} derive the slope $\alpha\sim2.1$ for the mass range 0.5 - 30\,M$_{\sun}$, close to the Salpeter slope. 

The
Rosette Nebula has also been studied within the MYStIX project \citep[e.g.][]{kuhn14, kuhn15}
who identify 255 probable members in the regions Rosette D and E, overlapping with the position of the NGC\,2244 cluster.
They estimate the peak surface density of $\Sigma \approx 300\,$stars\,pc$^{-2}$ in the region Rosette D, centered close to the star HD46150, located also in the center of our observed field. In the circular area with a radius of 0.5\,pc around the peak position, 
the average surface density is $\sim 100\,$stars\,pc$^{-2}$, assuming a distance of 1.4 kpc. 
This value is similar to those found in nearby star forming regions. 

This paper is structured as follows. Section~\ref{Obs&DR} contains the details of the observations and data reduction.
The data analysis, including the point-spread function (PSF) fitting, photometric calibration, and completeness analysis, is presented in Section~\ref{analysis}. 
In Section~\ref{sec_gaia_dist}, we derive the distance to the cluster.
In Section~\ref{results}, we discuss the cluster membership, derive the masses and their distribution, estimate BD frequencies
for NGC\,2244, and discuss the stellar densities in various clusters and star forming regions.
The results are discussed in Section~\ref{sec_discussion}. Summary and conclusions are given in Section~\ref{summary}.

\section{Observations and Data Reduction}
\label{Obs&DR}

\begin{deluxetable*}{lccccccc}
\tablecaption{Summary of the observations.}
\tablehead{
\colhead{Object} &
\colhead{$\alpha$ (J2000)} &
\colhead{$\delta$ (J2000)} &
\colhead{date} &
\colhead{filter} &
\colhead{exposure time} &
\colhead{IQ\tablenotemark{a}} &
\colhead{airmass}
}
\tablecolumns{8}
\startdata 
NGC\,2244 & 06:31:55.0 & +04:56:34 & 2015-12-29 & $J$ & $83\times20\,$s & 0\farcs50 & 1.23--1.34\\
NGC\,2244 & 06:31:55.0 & +04:56:34 & 2015-11-28 & $H$ & $128\times13\,$s & 0\farcs45 & 1.22--1.26\\
NGC\,2244 & 06:31:55.0 & +04:56:34 & 2015-11-28 & $Ks$ & $22\times15\,$s & 0\farcs40 & 1.27--1.29\\
control field & 06:31:25.0 & +03:37:57 & 2015-12-23 2015-12-29 & $J$ & $(35+57)\times20\,$s & 0\farcs60 & 1.28--1.34 1.33--1.50\\
control field & 06:31:25.0 & +03:37:57 & 2015-11-28 & $H$ & $128\times13\,$s & 0\farcs45 & 1.22--1.35 \\
control field & 06:31:25.0 & +03:37:57 & 2015-11-28 & $Ks$ & $22\times15\,$s & 0\farcs45 & 1.21 
\enddata
\tablenotetext{a}{Image quality (stellar FWHM measured in the reduced images)}
\label{tab:obs}
\end{deluxetable*}

Observations of NGC~2244 and another field outside of the cluster area (the control field) were performed using the Flamingos-2 (F2) imager at the Gemini-South telescope \citep{eikenberry04}, with
the program number GS-2015B-Q-46. The $F2$ camera provides a circular field of view (FoV) with a radius of $\sim 3.2'$, and a pixel scale of 0.18 arcsec per pixel,
meaning that the instrument becomes critically sampled in very good seeing conditions.
For the guiding, only the Peripheral Wavefront Sensor (PWS2) was available at the time of our observations. PWS2 vignetted part of the
FoV in our observations, and these parts of the detector were masked during the data reduction.
The data in $J-$, $H-$, and $Ks$-bands were obtained in queue mode in November and December 2015. 
All the observations were taken at 70-percentile conditions for image quality, and 50-percentile cloud cover (i.e. $J$-band image quality better than 0\farcs6 70\% of the time and photometric sky at least 50\% of the time)\footnote{\url{http://www.gemini.edu/sciops/telescopes-and-sites/observing-condition-constraints}}.
The summary of the observations is given in Table~\ref{tab:obs}.

A control field was observed to estimate the degree of contamination by field stars.
A suitable control field should be far enough from the cluster not to contain any of its stars, yet close enough to trace the same background population. We selected a field located away from the main molecular cloud emission of the Rosette complex (see Fig.~2 of \citealt{romanzuniga08}).
The field is located $\sim 1.3\,$deg from the center of NGC\,2244, along the line parallel to the Galactic plane. 

Standard near-infrared data reduction techniques were applied using our home-brewed {\sc IDL} routines, including dark subtraction, sky subtraction, flat-fielding, bad-pixel correction,
and mosaic construction by a simple shift-and-add. Due to dithering and vignetting by the PWS2, the constructed mosaic varies in depth across the field. To avoid issues with different photometric completeness limits, we constrain the analysis only to the part of the mosaic constructed from the maximum number of frames in all three bands, spanning the area of $\sim$11.1 arcmin$^2$.
The  $Ks$-band image of the observed region of NGC\,2244 is shown in Fig.~\ref{fig_composite}.

\section{Data analysis}
\label{analysis}

\subsection{Photometry}
\label{phot}

\begin{figure*}
\centering
\resizebox{17cm}{!}{\includegraphics{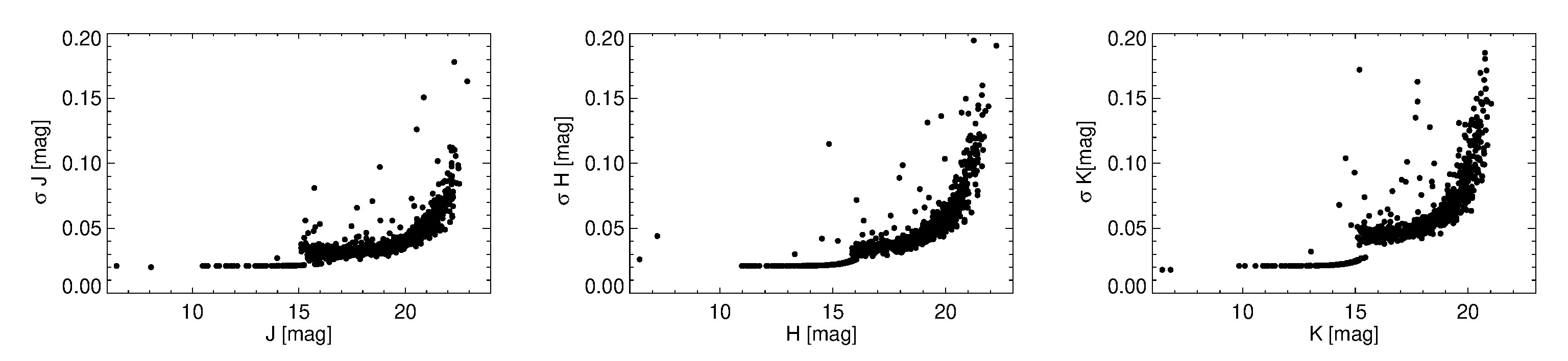}}
\caption{
Photometric uncertainties as a function of magnitude in $J$ (left), $H$ (middle) and $K$ (right), for the cluster dataset. Photometry of the bright sources whose photon counts exceeded the linearity limit of the detector was replaced with the photometry from UKIDSS or 2MASS, causing the discontinuity.}
\label{fig_photcal}
\end{figure*}
 
The photometry was performed using the {\sc DaophotII/Allstar} PSF-fitting algorithm  \citep{stetson87}, with a Gaussian PSF that varies quadratically with position in the frame ($VARIABLE$ $PSF$ parameter set to 2). We excluded all the sources with the {\sc Daophot} sharpness parameter $|sh|>0.7$, which helps to get rid of galaxies, clumps, knots, and spurious detections caused by ghosts in our images.
The photometric zero-points and color terms were calculated from the comparison with the data from the
United Kingdom InfraRed Telescope (UKIRT) Infrared Deep Sky Survey (UKIDSS; \citealt{lawrence07}) Galactic Plane Survey (GPS; \citealt{lucas08}) Data Release 10. This catalog was chosen because of its photometric depth, which allows a significant overlap with the F2 source lists.
The photometric uncertainties provided by the UKIDSS pipeline only include random photon noise error, and are therefore underestimated. We use a relation provided by \citet{hodgkin09} to calculate more realistic photometric errors that also include the systematic
calibration error component. The stars with counts surpassing the linear regime of the $F2$ detector were excluded from the comparison. The transition to the non-linear regime occurs at $J\sim 15$, $H\sim 16$, and $K\sim 15\,$mag. 
To avoid potential issues due different spatial resolutions of the $F2$ and UKIDSS data, we have discarded those sources with a neighbor at distances $<1.0''$ and brightness contrast $\Delta H<2$\,mag. Furthermore, we have rejected objects with uncertainties larger than 0.1 mag in the UKIDSS data, or in the $F2$ instrumental magnitudes.
Finally, the photometric zero-points and color terms were calculated using the following
equations:
\begin{equation}
\begin{aligned}
 J= J_{instr} + ZP_1 + c_1*(J-K)\\
 H = H_{instr} + ZP_2 + c_2*(J-K)\\
 K = Ks_{instr} + ZP_3 + c_3*(J-K)\\
\end{aligned}
\end{equation}

For the cluster field, the zero-points and the color terms are ZP$_1$=$25.07\pm0.02$, ZP$_2=25.25\pm0.02$, ZP$_3=24.68\pm0.03$, c$_1=0.06\pm0.02$, c$_2=0.04\pm0.02$, and c$_3=-0.08\pm0.03$.
For the control field we obtain ZP$_1$=$25.00\pm0.04$, ZP$_2=25.28\pm0.04$, ZP$_3=24.64\pm0.05$, c$_1=0.05\pm0.04$, c$_2=-0.01\pm0.04$, and c$_3=-0.06\pm0.05$.
The photometric uncertainties were calculated by combining the uncertainties of the zero-points,
color terms, and the measurement uncertainties supplied by {\sc Daophot}, and are shown in Fig.~\ref{fig_photcal} for the cluster field. The median uncertainties of the $F2$ photometry are 0.04\,mag in $J-$ and $H-$, and 0.05\,mag in the $K-$band.
The photometry of the non-linear stars was replaced by the UKIDSS measurements, with the exception of the two brightest stars in the field which are saturated even in UKIDSS. Their photometry was taken from 2MASS \citep{2mass}, and transformed to the UKIDSS photometric system using the relations given in \citet{hodgkin09}. 
Finally, we removed all the sources identified as galaxies in the UKIDSS catalog, verifying visually that they are not resolved into multiple sources in $F2$ images (8 galaxies removed). 
The final $JHK$ catalog contains 793 objects (Table~\ref{tab:catalog}).

\begin{deluxetable*}{lcccccccc}
\tablecaption{Near-infrared photometry of sources in the NGC 2244 field covered by Flamingos-2 observations.}
\tablehead{
\colhead{ID} &
\colhead{$\alpha$ (J2000)} &
\colhead{$\delta$ (J2000)} &
\colhead{J} &
\colhead{$\sigma_J$} &
\colhead{H} &
\colhead{$\sigma_H$} &
\colhead{K} &
\colhead{$\sigma_K$} 
}
\tablecolumns{9}
\startdata 
  0 & 06:31:48.95 & 04:58:20.1 & 19.480 & 0.034 & 18.706 & 0.039 & 18.461 & 0.051 \\
  1 & 06:31:50.08 & 04:57:56.5 & 16.255 & 0.027 & 15.885 & 0.026 & 15.757 & 0.041 \\
  2 & 06:31:50.58 & 04:57:29.3 & 17.388 & 0.037 & 16.740 & 0.040 & 16.150 & 0.052 \\
  3 & 06:31:50.64 & 04:58:35.8 & 15.598 & 0.032 & 14.933 & 0.022 & 14.601 & 0.023 \\
  4 & 06:31:50.74 & 04:57:50.8 & 17.653 & 0.031 & 17.027 & 0.033 & 16.781 & 0.045 \\
  5 & 06:31:50.85 & 04:57:38.0 & 17.673 & 0.040 & 16.912 & 0.042 & 16.288 & 0.055 \\
  6 & 06:31:50.86 & 04:57:20.0 & 17.274 & 0.034 & 16.717 & 0.036 & 16.263 & 0.047 \\
  7 & 06:31:51.18 & 04:58:22.0 & 20.515 & 0.040 & 19.800 & 0.050 & 19.264 & 0.058 \\
  8 & 06:31:51.24 & 04:57:35.4 & 18.535 & 0.036 & 17.763 & 0.038 & 17.436 & 0.050 \\
  9 & 06:31:51.25 & 04:58:16.4 & 16.429 & 0.027 & 15.993 & 0.027 & 15.821 & 0.041 
\enddata
\tablecomments{This table is published in its entirety in the machine-readable format.
      A portion is shown here for guidance regarding its form and content.}
\label{tab:catalog}
\end{deluxetable*}



\begin{figure}
\centering
\resizebox{8cm}{!}{\includegraphics{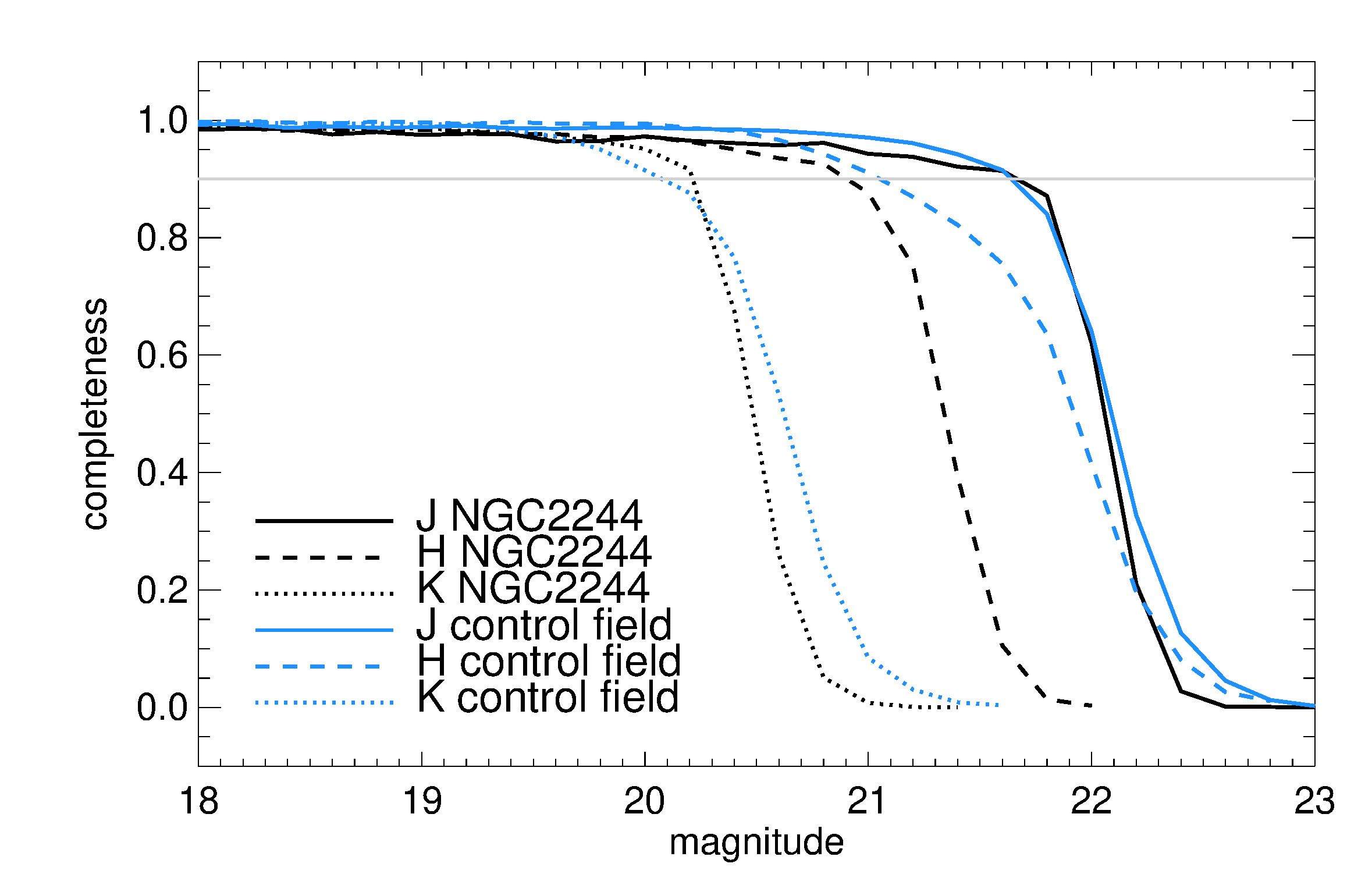}}
\caption{Completeness of the $F2$ photometry calculated with an artificial star experiment for NGC\,2244 (black) and the control field (blue). Different line styles represent different photometric bands: full, dashed, and dotted lines for $J$-, $H-$, and $K-$bands, respectively.}
\label{fig_compl}
\end{figure}

The completeness of the photometry was assessed with an artificial star test. 
Artificial stars were added to the original image using the {\sc Daophot}'s function $ADDSTAR$.
The resulting image was then used as the input for a routine identical to the one used to obtain the photometry for the cluster 
and the control field (Section~\ref{phot}), using the same PSF solution.
We inserted only 50 stars at a time to avoid crowding, 
and repeated the procedure 40 times at each magnitude (in steps of 0.2 mag) to improve statistics. 
The ratio between the number of recovered and inserted artificial stars is plotted against magnitude is shown in Fig.~\ref{fig_compl}, for both the cluster (black lines) and the control field (blue lines).
The 90\% completeness limits are $J=21.7\,$mag, $H=20.9\,$mag and $K=20.2\,$mag for NGC\,2244, and $J=21.7\,$mag, $H=21.1\,$mag and $K=20.1\,$mag for the control field.

\subsection{Distance to NGC\,2244 from $Gaia$ DR2}
\label{sec_gaia_dist}

\begin{figure*}
\centering
\resizebox{15cm}{!}{\includegraphics{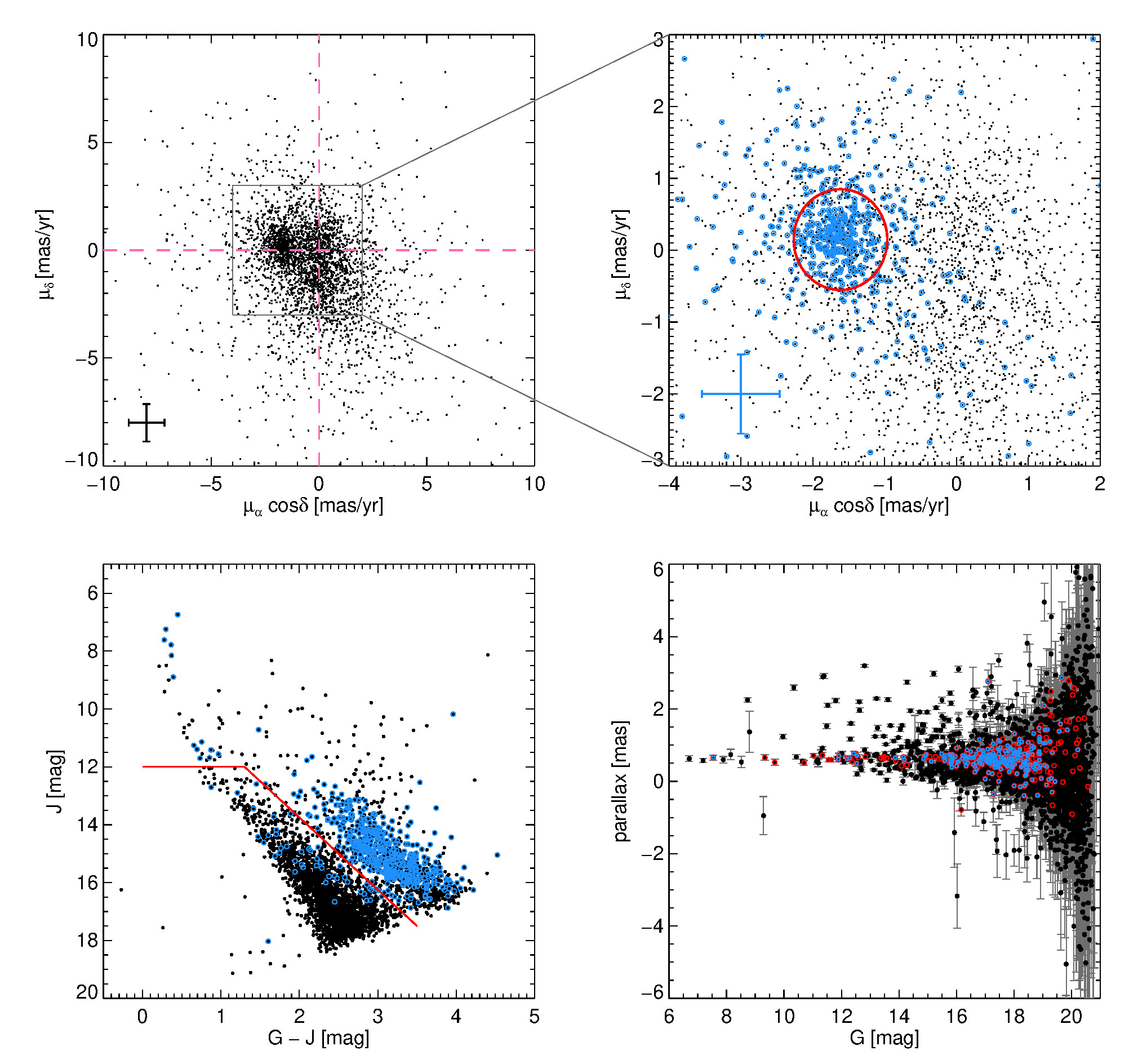}}
\caption{
Astrometric and photometric data for the sources in the NGC\,2244 field, located within a circle with radius $10'$ centered on the star HD46150. 
{\bf Top left:} $Gaia$ DR2 proper motions. The error bars representing the mean uncertainty of the sources plotted with the corresponding color. 
{\bf Top right:} enlarged version of the top left panel. Blue open circles in this and other panels mark the members identified by \citet{meng17}. The red ellipse marks our selection criterion for the sources to be included in the parallax calculation.
{\bf Bottom left:} $Gaia$ DR2/UKIDSS CMD of the same set of sources.
{\bf Bottom right:} $Gaia$ DR2 parallaxes. 
Sources selected as candidate members for the calculation of cluster distance are located within the red proper motion ellipse and to the right of and above the red lines in the CMD. These sources are marked with open red circles.
}
\label{fig_gaia}
\end{figure*}

The distance to NGC\,2244 was previously estimated to be between 1.4 kpc to 1.7 kpc \citep{ogura&ishida81, perez87,park02, hensberge00,lombardi11, martins12, bell13, kharchenko13,dias14}. The recently published $Gaia$ Data Release 2 (DR2, \citealt{GaiaDR2}) provides precise positions, proper motions, parallaxes, and $G-$band magnitudes for more than 1.3 billion sources. We queried the $Gaia$ DR2 catalog within $10'$ radius from the brightest star in the field (HD 46150; $06:31:55.52$, $+04:56:34.3$). 
This area contains the region observed with F2, but covers a much larger portion of the cluster.
The resulting list contains 3498 objects. We then cross-matched the $Gaia$ DR2 result with the catalog of 718 clusters members from \citet{meng17}, constructed
based on X-ray detection, infrared excess, proper motion, optical color selection, and a cut-off in the
clustrocentric distance. The total number of matched sources with a valid five-point astrometric solution is 528. 

The proper motions of the $Gaia$ sources in the field are shown in the top panels of Fig.~\ref{fig_gaia}. There is a distinct overabundance of sources close to the position $\mu_{\alpha}$, $\mu_{\delta}\approx (-1.5,0.2)\,$mas\,yr$^{-1}$, corresponding to the location of the cluster members. An enlarged version of this plot is shown in the upper right panel, with the blue open circles marking the matched members from \citet{meng17}. The mean proper motion of the matched members are 
$\mu_{\alpha} = -1.62 \pm 0.65\,$mas\,yr$^{-1}$ and
$\mu_{\delta} = 0.15 \pm 0.70\,$mas\,yr$^{-1}$. 

The color-magnitude diagram (CMD) of the same set of sources constructed by matching the $Gaia$ DR2 to Uthe KIDSS survey shows the probable member sequence clearly standing out to the right of the bulk of the field stars. We can therefore use the proper motions and the position in the CMD as criteria to assign cluster membership. To consider a source a member in this exercise, we require it to be located inside the red ellipse in the proper motion space, to the right of or above the dividing line in the CMD, and to be classified as stars in UKIDSS (class=-1). The semimajor and semiminor axes of the red ellipse are equal to $1\sigma$ widths of the proper motion distributions in R.A. and Dec. The transition from the pre-main sequence to the main sequence at the age of NGC\,2244 occurs roughly at $J=12$ (see Figs.~\ref{fig_iso} and \ref{fig_cmdccd}), motivating the horizontal red line. After applying the cuts mentioned above, we keep 345 out of the initial 3498 sources. These 345 sources are shown with red open circles in the parallax plot in the lower right panel of Fig.~\ref{fig_gaia}. 242 of the 345 selected sources belong also to the members list of \citet{meng17}. We note that our criteria may cause us to miss some members (mainly due to the relatively narrow allowed proper motion space, especially considering individual proper motion uncertainties), but the idea here is to minimize the number of interlopers that could bias our distance analysis, rather than to do a sophisticated membership analysis at this point.

We estimate the distance to the cluster through a maximum likelihood procedure (see e.g. \citealt{cantat18}), maximizing the following quantity:

\begin{equation}
\mathcal{L}\propto \prod_{i}P(\varpi_{i}|d,\sigma_{\varpi_i})=\prod_{i}\frac{1}{\sqrt{2\pi \sigma_{\varpi_i}^2}}exp\Big(-\frac{(\varpi_i-\frac{1}{d})^2}{2\sigma_{\varpi_i^2}}\Big),
 \end{equation}
 
 where $P(\varpi_{i}|d,\sigma_{\varpi_i})$ is the 
 probability of measuring a value $\varpi_i$ for the parallax of star $i$, if its true distance is $d$, and its measurement uncertainty is $\sigma_{\varpi_i}$. Here we assume that there are no correlations between individual parallax measurements, and that the likelihood for the cluster distance can be represented as the product of the individual likelihoods of all its members. We neglect any cluster spread, i.e. all the stars are assumed to be located at the same distance, which should have a negligible effect on our distance determination, given the relatively large distance to the cluster. In this way we obtain a distance to NGC\,2244 of $d=1.59\pm0.02\,$kpc. If we restrict ourselves only to the members identified by \citet{meng17} that also pass our cuts (242 objects), we obtain a very similar value of $d=1.57\pm0.02\,$kpc.
 Our distance estimate is in agreement with the most recent value $1.55^{+0.1}_{-0.09}$\,kpc from \citet{kuhn19}, derived also from the $Gaia$ DR2 data, and assuming the systematic parallax uncertainty of 0.04\,mas.
 
It has been shown that $Gaia$ DR2 parallaxes suffer systematic uncertainties, with a global zero-point offset of -0.029 mas \citep{lindegren18, arenou18,luri18}. However, applying this correction to small areas is not advisable, since the local variations can be significantly different 
\citep{arenou18}. Moreover, a low number of observed QSOs prevents a calculation of the local bias close to the galactic plane. 
\citet{arenou18} report the median systematic offset in parallax of $\sim$-0.065\,mas for the cluster samples of \citet{kharchenko13} and \citet{dias14}, with a dependence on cluster distance and stellar colors. Judging from 
the right-hand side panels of Figure~16 of \citet{arenou18}, an appropriate offset value for NGC\,2244 is $\sim$-0.08\,mas, i.e. the Gaia distance derived here might be overestimated by $\sim$180\,pc, or 11\%. 
Considering this potential systematics, it might seem that the distance constrained by $Gaia$ does not provide a significant improvement with respect to previous measurement; however, we have to keep in mind that this is only the second data release, and that by the end of the mission, the systematics are expected to be at a microarcsecond level even at distances of a few kiloparsecs.
To understand the effect this uncertainty might have on the results, we will report them for two values of the distance, 1.6 and 1.4 kpc.

As a side note, in the bottom left panel of Fig.~\ref{fig_gaia}, there is a group of relatively bright sources not recognized as members by \citet{meng17}, located to the right and above the red lines (roughly $J<13$ and $G-J>1$). Only about $10$\% of these sources pass all our membership cuts. The parallaxes of those that do not, signal that they are mostly located behind the cluster, and their position in absolute color-magnitude diagrams is consistent with a giant branch. About a quarter of these objects have been observed by the LAMOST survey\footnote{\url{http://dr5.lamost.org/}}, and have determinations of surface gravity (log$g$) and effective temperature (T$_{\mathrm{eff}}$). About $75$\% of those have T$_{\mathrm{eff}}$=4500-5000\,K and log$g<$3, i.e. they are background giants.

\section{Identification of cluster members and the IMF}
\label{results}
\subsection{Outline}
Since this is a long and relatively complicated section, here we present a short summary of the methodology, giving readers a chance to skip some sections that might be too technical or detailed.
\begin{itemize}
\item
{\it The goal of this section} is to derive the IMF and the star-to-BD number ratio for NGC\,2244 (Sections~\ref{sec_imf} and \ref{sec_ratio}). The results are presented for three different isochrones and two distances (1400\,pc and 1600\,pc), in order to estimate the effect that different assumptions have on the results.
\item
{\it Membership determination} (Section~\ref{membership}). 
The first step in the analysis is to separate cluster stars from the contaminants sharing the same line of sight.
For the sources brighter than J$\sim 18$, this can be done using the color, proper motion, and parallax cuts on the $Gaia$ DR2 and $Pan-STARRS$ data (Section~\ref{membership_gaiaps1}). However, our NIR catalog extends more than 3 magnitudes deeper, and in this regime we employ a statistical determination of membership, comparing the CMD of the population observed in direction of the cluster to that of the control field (Section~\ref{sec_cleaning}). 
In this procedure, there are several parameters that are varied (cell sizes and positions, difference in extinction between the cluster and the control field), and each combination of these parameters can give a slightly different solution. We keep all the possible solutions (324 member lists) and use them later to evaluate the uncertainties that this approach introduces to the IMF and star-to-BD ratio. 

\item
{\it Isochrones} (Appendix~\ref{append_isochrones}). Stellar parameters are derived in comparison to theoretical models. The basic products of the stellar models are the bolometric luminosity and effective temperature, which are converted into magnitudes and colors, by applying bolometric corrections (BC) and $T_{\mathrm{eff}}$-color relations. The conversion can be done by using the atmosphere models, or empirical relations. Here we decide to test both approaches, and we derive our results using three different isochrones.
 
\item
{\it Derivation of stellar parameters} (Section~\ref{sec_parameters}). We compare three different methods to derive effective temperature, mass, and extinctions. The first method is based on fitting the spectral energy distribution (SED) over a large wavelength range (optical to mid-infrared, where available; Section~\ref{sec_sedfit}). The other two methods explore the derivation of parameters from the three NIR colors only ($JHK$), in one case assuming that the observed colors are photospheric (Section~\ref{sec_masses_nir_noexcess}), and in the other case assuming that some intrinsic excess at the NIR wavelengths is present (Section~\ref{sec_masses_nir_excess}). We find that the results are consistent with each other, validating the use of only NIR magnitudes and colors in mass determination at the faint end, where this is our only option (Section~\ref{sec_method_comparison}).
\end{itemize}

\subsection{Cluster membership}
\label{membership}

\begin{figure*}
\centering
\resizebox{15cm}{!}{\includegraphics{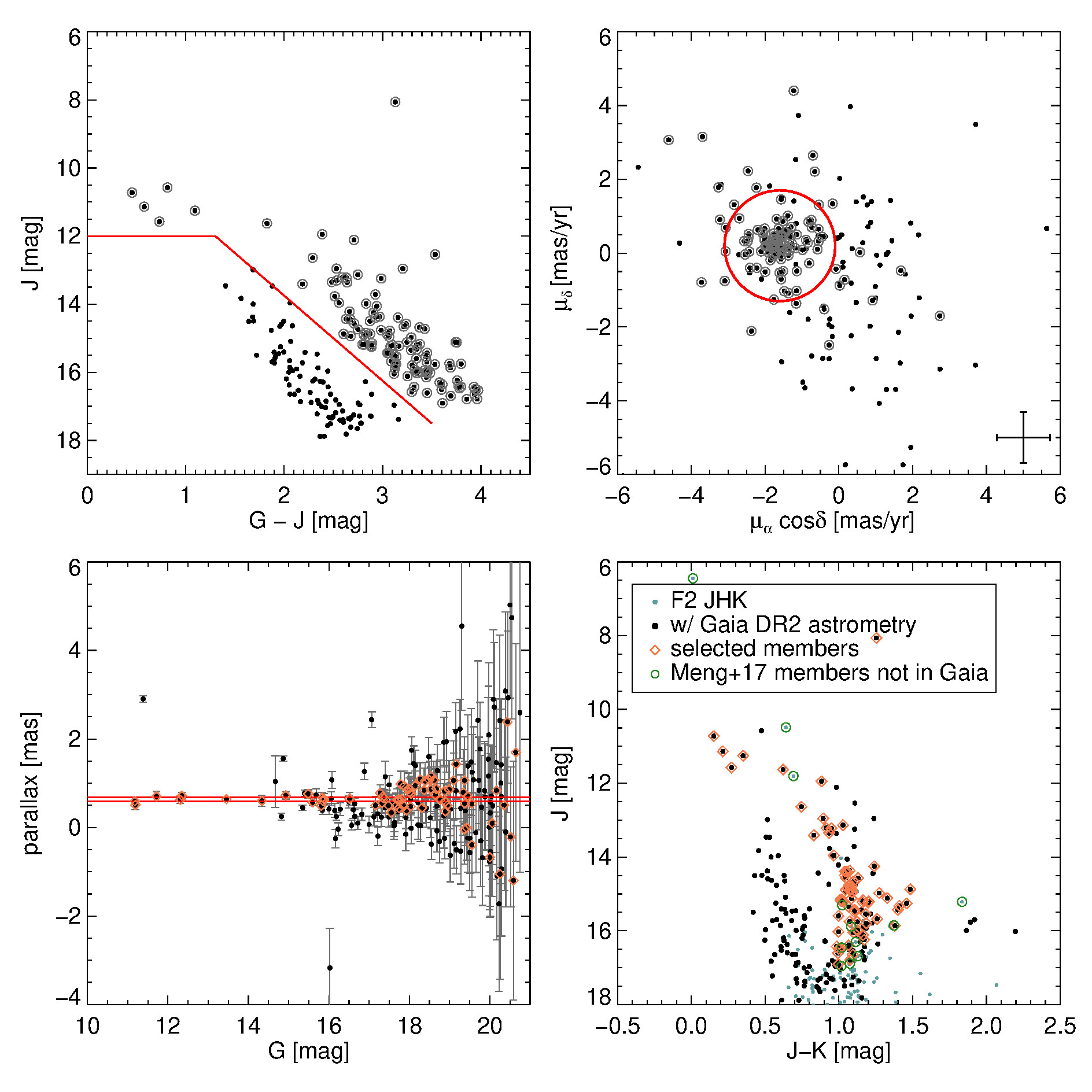}}
\caption{Selection of candidate members in the $F2$ $JHK$ catalog using the $Gaia$ DR2 data. 
{\bf Top left:} $J$, $G$ - $J$ CMD for the $JHK$ sources with a five-point astrometric solution from $Gaia$ DR2. 
{\bf Top right:} Proper motion for the same set of objects.
{\bf Bottom left:} Parallax as a function $G$ magnitude.
{\bf Bottom right:} $J$, $J$-$K$ CMD for the objects in the $F2$ field-of-view. The small green dots mark the sources found in the $F2$ $JHK$ catalog and the black dots those with a valid $Gaia$ five-point astrometric solution.
We select objects located above the red lines in the $J, (J-G)$ diagram (open circles in the top panels), having 1$\sigma$ proper motion errors
intersecting the circular area outlined by the red line in the top right panel, and 1$\sigma$ errors in parallax intersecting one of the red lines in the 
bottom left panel. Selected candidate members are shown as orange diamonds. Green open circles mark the members from \citet{meng17} not found in the $Gaia$ catalog.
}
\label{fig_gaia_mem}
\end{figure*}

\begin{figure*}
\centering
\resizebox{17cm}{!}{\includegraphics{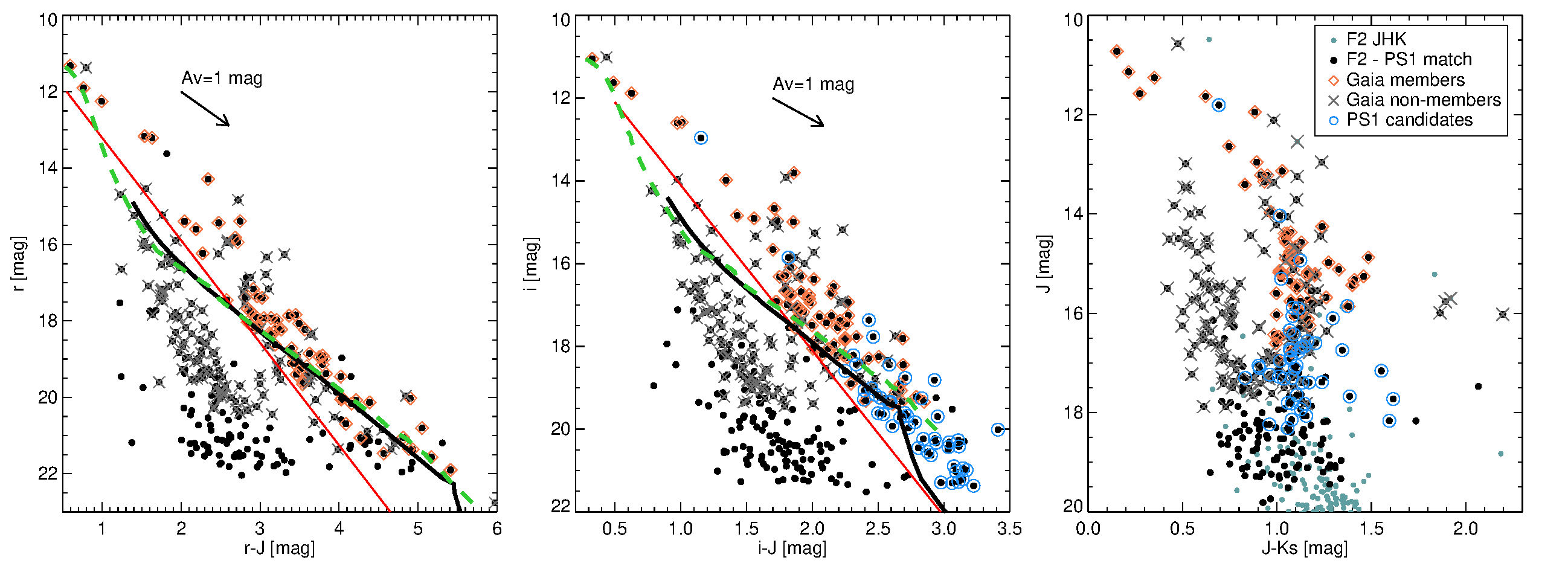}}
\caption{
Selection of candidate members using the $PS1$ and $F2$ data. Black dots mark all the sources from the $F2$ catalog having a $PS1$ counterpart with valid $r$ or $i$ photometry. Orange diamonds in all panels mark the candidate members selected with help of $Gaia$ DR2, and grey crosses those rejected in the procedure. 
Left and middle panels: $r$, $r$-$J$ and $i$, $i$-$J$ CMDs for the $PS1/F2$ match. The 2 Myr evolutionary models shifted to the distance of 1.6 kpc are shown with the green dashed (PARSEC) and black solid (BT-Settl) lines. 
We select as candidate members all the remaining sources located to the right of the red lines in both CMDs (blue open circles).
Right panel: $J$, $J$-$K$ CMD. The small green dots mark the sources found in the $F2$ $JHK$ catalog, black dots those with a optical counterpart from $PS1$.
}
\label{fig_ps1_mem}
\end{figure*}
\subsubsection{Gaia DR2 and Pan-STARRS1}
\label{membership_gaiaps1}
For the brighter portion of sources in our field of view, data from $Gaia$ DR2 and Pan-STARRS1 ($PS1$; \citealt{PS1}) are available.
199 sources from our $JHK$ catalog have a counterpart with a five-parameter astrometric solution in $Gaia$ DR2 (down to $J\approx17.5-18$) within the matching radius of 0\farcs5. The average distance between the matched entries in the two catalogs is 0\farcs03, justifying our choice of the relatively small matching radius.
We consider a source to be member candidate if it satisfies the following criteria, as depicted in Fig.~\ref{fig_gaia_mem}: 
(1) it is located above and to the right of the red lines in the $J$, $(G-J)$ diagram (open grey circles in top panels);
(2) its 1 $\sigma$ proper motion errors intersect the circular area with the radius of 1.5 mas\,yr$^{-1}$, centered at the average value of the proper motions derived in section~\ref{sec_gaia_dist} (red circle in the top right panel);
(3) its 1 $\sigma$ errors in parallax intersect one of the red lines plotted in bottom left panel (marking 1.4 and 1.7 kpc distance, see Section~\ref{sec_gaia_dist}). 
In total 72/199 of sources pass our cuts and are considered member candidates. They are marked with orange diamonds in the bottom right panel of Fig.~\ref{fig_gaia_mem}.

$PS1$ contains neither proper motions nor parallaxes, but its optical photometry is deeper than $Gaia$, and can further help in selection of candidate members. As shown in the left and middle panels of Fig.~\ref{fig_ps1_mem}, the sources are separated in two bands when combining optical with NIR photometry
\footnote{$PS1$ AB photometry was converted to Vega system using the filter zero-points available from \url{http://svo2.cab.inta-csic.es/theory/fps3/}}. Orange diamonds mark the member candidates selected by $Gaia$, and grey crosses those rejected in the same procedure. Following the sequence of the $Gaia$ member candidates, we draw a dividing lines in $r, r-J$ and $i, i-J$ CMDs (solid red lines). We select the sources not classified by $Gaia$ and located to the right hand side of the red lines in both diagrams as member candidates (blue open circles). 47 out of 362 $F2-PS1$ matches (matching radius 0\farcs5) with valid $r$ or $i$ magnitudes are selected in this way. Judging from the $Gaia$ color versus astrometric selection, $\sim 15$ of these sources might still be interlopers. The selected sources are identified in Table~\ref{tab:sed}.

The selection using $Gaia$ and $PS1$ is nearly complete down to $J\sim 18.5$. About 6\% of the $F2$ sources with $J<18.5$ (21/337) cannot be classified as members or non-members in this way, because no counterpart with valid $r$ or $i$ magnitude is found in $PS1$, or no astrometry is given in $Gaia$. Most of these sources are located very close to the two brightest stars in our FoV, which might have affected their measurements in $Gaia$ or $PS1$. 

We matched the sources selected in this section with the member list of \citet{meng17}, who select as member candidates all sources detected in both X-rays and $Spitzer/IRAC$, or those that show excess at wavelengths $8-24\,\mu m$. The depth of their survey seems to be fairly complete down to the 2MASS completeness limit at $J=15.8\,$mag, and we restrict the comparison down to that value. Of the 61 $Gaia/PS1$ selected candidates brighter than $J=15.8$, 47 are found in the member list of Meng et al. Of the 14 sources that are not in their list, 10 are not in the X-ray/NIR combined catalog of \citet{wang08}, which was used for selection in the former work. The remaining 4 objects were observed by $Spitzer$ \citep{balog07}, but they do not show an excess at these wavelengths, which is possibly the reason why they were rejected by \citet{meng17}.
On the other hand, of the 59 objects rejected by $Gaia/PS1$, 20 are found in the member list of Meng et al. Of these 20, six were rejected based on either colors or proper motions, and the remainder based on parallax. It might be worth noting that the relative error of the parallax for these sources is relatively large (median $\sim0.3$), but only for 4 of them the lower and upper bounds of the distance confidence interval from $Gaia$ DR2 \citep{bailerjones18} agree with the distance to NGC\,2244.

\subsubsection{Flamingos-2}
\label{sec_cleaning}

\begin{figure*}
\centering
\resizebox{15cm}{!}{\includegraphics{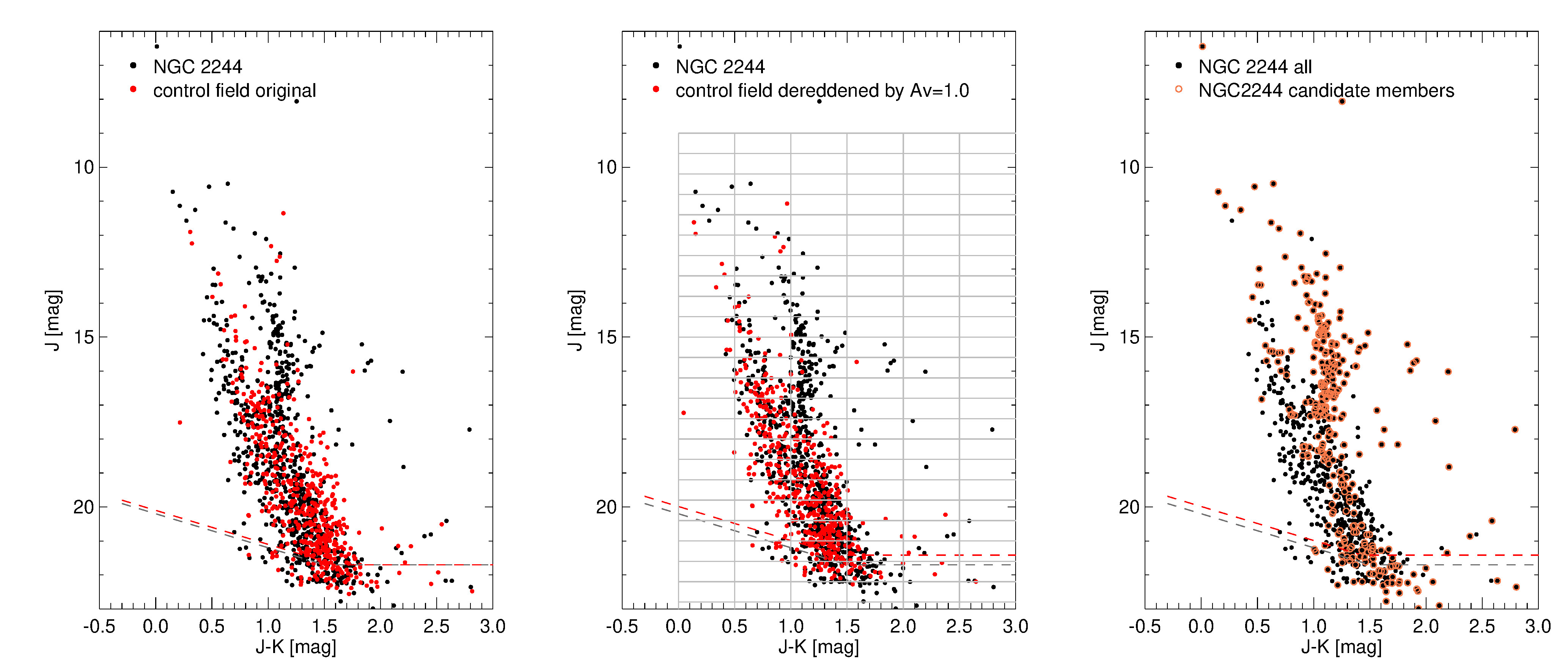}}
\caption{
A demonstration of the statistical membership determination procedure. Here we show only one of total 324 realizations, which were obtained by varying several parameters, e.g. the difference in extinction between the cluster and the control field, cell size, etc. (see Section~\ref{sec_cleaning} for details).
{\bf Left panel:} CMD towards the cluster (black) and the control field (red). Grey and red dashed lines represent the $90$\% completeness limits for the cluster and the control field, respectively. 
{\bf Middle panel:} Here the control field has been dereddened by A$_V$=1, to approximately match the probable field sequence at $J-K<0.8$ and $J<17$. The completeness limit for the control field has been modified accordingly. The grid cell size is $0.5\times0.6$\,mag.
{\bf Right panel:} We overplot (orange circles) the candidate members that ``survived" the statistical CMD cleaning in this particular step.
}
\label{fig_cleaning}
\end{figure*}

Our $F2$ $JHK$ photometry extends more than 3 magnitudes in $J$ deeper from the objects detected in $PS1$. In the magnitude range covered by $Gaia$/$PS1$ the probable member sequence separates nicely from the background even in the NIR CMD, but below $J\sim17$ the cluster members occupy the same space in the CMD as the foreground and background field stars. To estimate the relative contribution of the contaminants to the stellar sample observed in the direction of the cluster, we
compare the cluster CMD with that of the control field (see Fig.~\ref{fig_cleaning}). The method is similar to the one we applied to the cluster RCW~38 \citep{muzic17}, but includes a few modifications. 
The main steps of the method are as follows.
\begin{enumerate}
    \item 
    The CMD is subdivided into grid cells with a step size ($\Delta col$, $\Delta mag$) in the color and magnitude axes, respectively.  
    \item
    For each CMD cell, we calculate the expected number densities of stars in the cluster field, and the control field. The number density of field stars has in principle to be scaled to account for different on-sky areas covered by the images, although in our case this number is close to one.
    \item
     A number of objects corresponding to the expected field object population is then randomly removed from corresponding cells of the cluster CMD.
\end{enumerate}

To calculate the number densities of stars in each cell, we use the method outlined in \citet{bonatto07}. Each star is represented by a Gaussian probability distribution of magnitude and color, with the widths determined by the uncertainties. The expected number density of stars in each cell is calculated as the sum of each individual star's probability of being in a corresponding cell. This way we take into account the fact that stars with large uncertainties have a non-negligible probability of populating more than one CMD cell.  

The choice of the cell size is an important parameter to take into account, and results from a compromise between having a sufficient number of stars in each cell and preserving the morphology of different CMD evolutionary sequences. We have tested the cell size of $\Delta col$ = (0.4, 0.5, 0.6) mag, and $\Delta mag$ = (0.4, 0.6, 0.8) mag. Furthermore, for each combination of the color and magnitude cell sizes, we repeat the procedure with the grid shifted by $\pm$ 1/3 of the cell width in each dimension. This results in 81 different configurations.

The control field population seems to have, on average, a slightly higher extinction than the cluster field. This can be appreciated by looking at the upper sequence in the CMD, where there is a slight shift between the left-hand side sequence of the cluster and the control field. We estimate that the difference is somewhere in the range 0.5 - 1 mag. The control field sequence is therefore dereddened at the beginning of the procedure by $\Delta A_V$=(0.5, 1.0)\,mag.
Finally, we applied the procedure to both the ($J$, $J-H$) and ($J$, $J-K$) CMDs. In total, this procedure results in $81\times2\times2=324$ different member lists, which can then be used to calculate statistical errors of the IMF and the star-to-BD number ratio, that are introduced by the decontamination procedure.

We estimate that $61\pm2$\% of the sources observed toward the cluster are actually field contaminants. The number of cluster stars surviving the procedure is $310\pm18$ (the uncertainty is the standard deviation of the 324 outcomes of the above procedure).

\begin{figure*}
\centering
\resizebox{17cm}{!}{\includegraphics{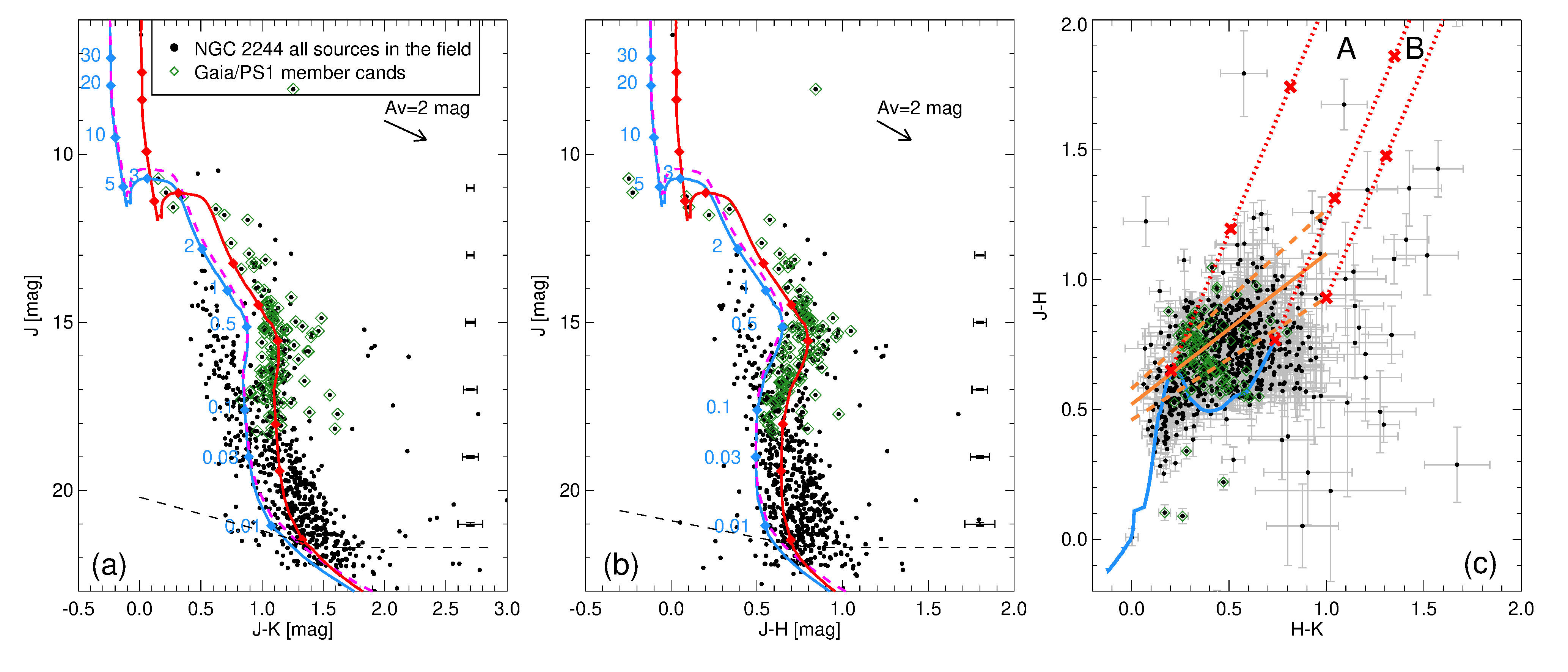}}
\caption{ 
$J$, $J-K$ (a) and $J$, $J-H$ (b) color-magnitude diagrams, and color-color diagram (c) of all the sources detected towards NGC\,2244 (black dots), with the green diamonds marking
the candidate members from the $Gaia$ and $PS1$ selection.
The blue solid line shows the 2 Myr isochrone ($isochrone$\,2) shifted to the distance of 1600\,pc, while the red solid line represents the same isochrone reddened to the average extinction of the known cluster members (A$_V$=1.5\,mag). The dashed magenta isochrone is the unreddened isochrone at the distance of 1400\,pc. 
The solid and dashed orange lines represent the locus of T-Tauri stars and the corresponding uncertainties \citep{meyer97}. The dotted red lines show the reddening vectors \citep{cardelli89}, and the red crosses along the reddening lines mark A$_V=0$, 5 and 10\,mag. The black dashed lines mark the 90\% completeness limit. See text for explanation of zones A and B in panel (c).}
\label{fig_cmdccd}
\end{figure*}

Our procedure successfully recovers the member sequence at $J\lesssim 17\,$mag, which appears separated from the field objects located to the left (see Fig.~\ref{fig_cmdccd}).
Green diamonds in this plot show the candidate members from the $Gaia$ and $PS1$ selection.

\subsection{Derivation of stellar parameters}  
\label{sec_parameters}

In this section we explore different approaches to derive stellar effective temperatures, masses, and extinction (A$_V$) of candidate members of NGC\,2244 from multi-band photometry. 
For the candidate members selected with the help of $Gaia$ and $Pan-STARRS1$, we can take advantage of the existence of photometry over a large wavelength range, and derive these parameters by fitting their SEDs to the synthetic photometry obtained from atmosphere models. 
This, however, only works for the bright part of our catalog. For the objects with $J\gtrsim18.5$ where we only have $JHK$ photometry available, the two-parameter SED fitting becomes degenerate. Instead, to simultaneously derive the extinction and the masses of the cluster members, we compare the $JHK$ photometry to model isochrones. The first approach here is to assume that the observed colors come from the reddened stellar photosphere. However, as argued by
\citet{cieza05}, classical T-Tauri stars (CTTSs) can present an excess already in the $J$-band, and therefore a simple dereddening of the photometry to the model isochrones can overestimate the extinction, and consequently also the stellar luminosity and mass. About 70\% of the stellar members of NGC\,2244 host dusty disks or envelopes \citep{meng17}, which can contribute some excess flux to their near-infrared colors. 
For this reason we also develop a method which takes 
into account this possible intrinsic excess when deriving the extinction toward individual stars in the cluster.

Throughout this section, we assume an age of 2 Myr (when employing the isochrones; \citealt{perez87,park02, hensberge00,bell13}), and adopt the extinction law from \citet{cardelli89} with the standard value of the total-to-selective extinction ratio R$_V=3.1$ \citep{fernandes12}.
For a detailed description of the isochrones used in Sections~\ref{sec_masses_nir_noexcess} and \ref{sec_masses_nir_excess}, see Appendix~\ref{append_isochrones}. 
We derive the parameters assuming the cluster distance of 1.6 kpc derived from the $Gaia$ data, and use this for the comparison of the different methods. At the end, we also derive the masses for the distance of 1.4 kpc, which should be a lower limit on the distance judged from the $Gaia$ systematics.

\subsubsection{Optical to near-infrared SED fitting}
\label{sec_sedfit}

\begin{figure}
\centering
\resizebox{0.5\textwidth}{!}{\includegraphics{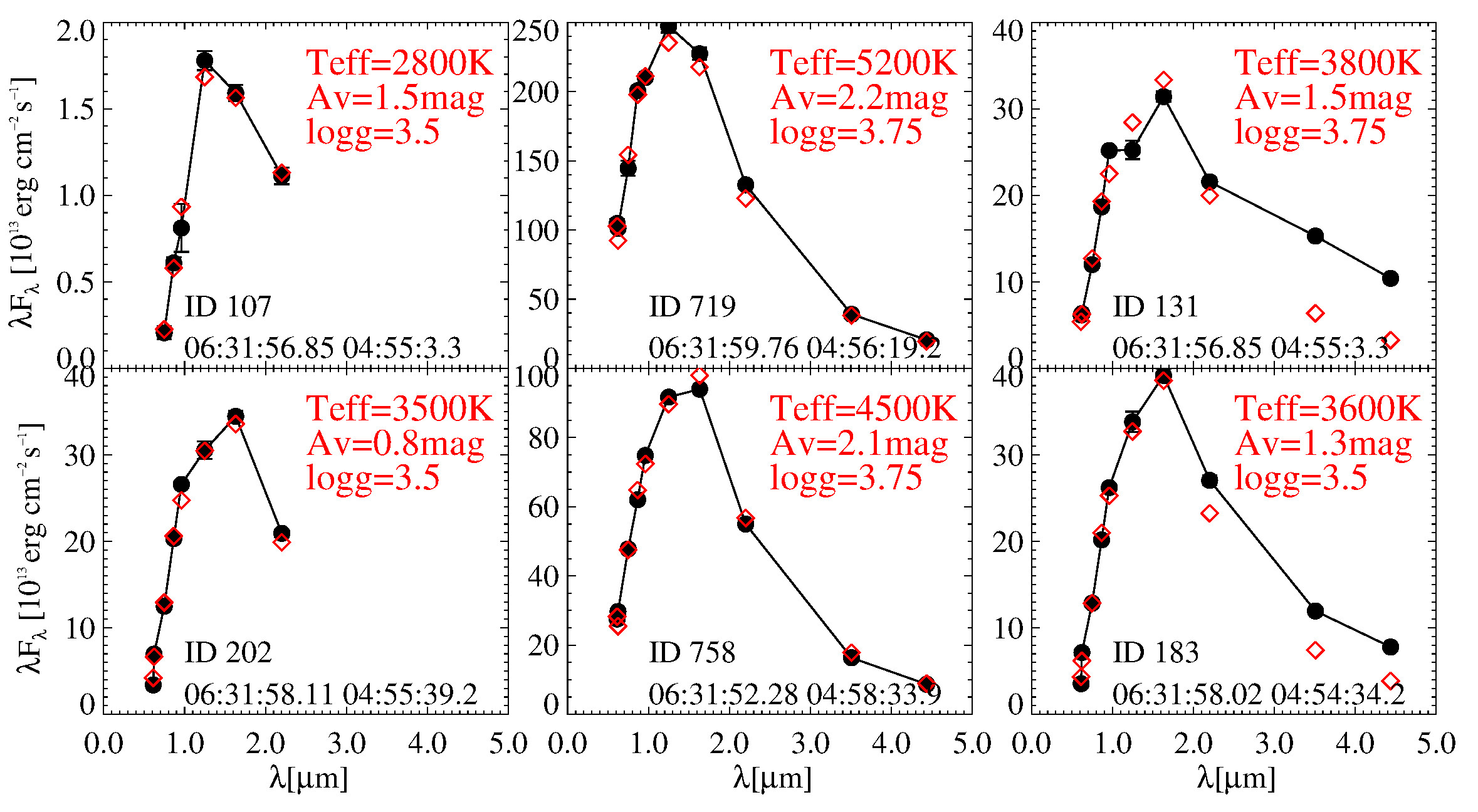}}
\caption{A subset of candidate member SEDs, using the $Gaia$, $Pan-STARRS1$, $F2$ $JHK$ photometry, and $Spitzer/IRAC$ photometry where available (black dots). The best-fit model are shown in red. For the sources showing infrared excess (rightmost panels), the $Spitzer$ photometry was not used in the fit. }
\label{fig_seds}
\end{figure}

We construct the SEDs of the $Gaia/PS1$ candidate members using the available optical and NIR photometry mentioned in the previous sections.
We also use the $Spitzer/IRAC$ photometry from \citet{balog07} where available, but only for the sources not showing excess at mid-infrared wavelengths (15 sources). Whether a source exhibits an excess was determined with the help of VOSA \citep{bayo08}, which detects mid-infrared excess based on the slope of the infrared SED. The validity of this selection was also visually inspected.
We retrieved the BT-Settl atmosphere models \citep{baraffe15} for log$g$ values of between 3.5 and 4.5, and T$_{\mathrm{eff}}$ between 2000 K and 25000 K\footnote{retrieved from \begin{sloppypar}\url{http://svo2.cab.inta-csic.es/svo/theory/newov2/syph.php}\end{sloppypar}}
and interpolate the models to the intermediate log$g$ values of 3.75 and 4.25 (young stars with an age of up to a few million years and masses below 10\,M$_{\sun}$ should have log$g$ between 3.5 and 4.25).
The metallicity is set to the solar value. 
The extinction was varied between 0 and 10 in steps of 0.1\,mag.
We find the model best matching the data by minimizing the $\chi^2$ parameter calculated as

\begin{equation}
	\chi^2=\sum\limits_{i=1}^N\frac{(F_{i\_obs}-D\times F_{i\_model})^2}{\sigma_{i\_obs}^2},
\end{equation}

where $F_{obs}$ and $F_{model}$ are observed and model fluxes, respectively, and $\sigma_{obs}$ are observed flux uncertainties. $D$ is the dilution factor, $D=(R/d)^2$, with $R$ being the stellar radius and $d$ the cluster distance (fixed to 1600 pc). The value of $R$ is fixed for each T$_{\mathrm{eff}}$ and log$g$ combination and was extracted from the models. This is motivated by the relation $R=\sqrt{G\times m/g}$, where $m$ stands for the object's mass, which is directly related to T$_{\mathrm{eff}}$ at a particular value of log$g$.
{We visually inspected all the fits, and rejected two of them due to a clear discrepancy between the observed and the best-fit model fluxes.
In Fig.~\ref{fig_seds} we show a subset of the object SEDs, together with the best-fit model synthetic photometry (red). In the leftmost panels we show two objects with optical to near-infrared photometry, in the middle panels those with no excess in $Spitzer$ photometry, and in the rightmost panels those showing the excess (the excess points were not used to fit the SED). 
The best-fit parameters are given in Table~\ref{tab:sed}.

Finally, the masses were obtained by interpolating the T$_{\mathrm{eff}}$ - mass relation from the 2 Myr models to the best-fit T$_{\mathrm{eff}}$ values.

\begin{deluxetable*}{lcccccl}
\tablecaption{Member candidates from Gaia DR2 and PS1, along with the best-fit parameters from the SED.}
\tablehead{
\colhead{ID} &
\colhead{$\alpha$ (J2000)} &
\colhead{$\delta$ (J2000)} &
\colhead{T$_{\mathrm{eff}}$ [K]} &
\colhead{A$_V$ [mag]} &
\colhead{log(g [cm\,s$^{-2}$])} &
\colhead{selection}
}
\tablecolumns{7}
\startdata 
  2 & 06:31:50.58 & 04:57:29.3 & 3100 & 1.9 & 3.75 & PS1 \\      
  3 & 06:31:50.64 & 04:58:35.8 & 6000 & 4.5 & 4.25 & Gaia DR2 \\
  5 & 06:31:50.85 & 04:57:38.0 & 3100 & 2.9 & 3.75 & PS1      \\
  6 & 06:31:50.86 & 04:57:20.0 & 2900 & 0.2 & 3.50 & PS1     \\ 
 18 & 06:31:52.28 & 04:58:33.9 & 3000 & 1.2 & 3.50 & PS1      
\enddata
\tablecomments{This table is published in its entirety in the machine-readable format.
      A portion is shown here for guidance regarding its form and content.}
\label{tab:sed}
\end{deluxetable*}

\subsubsection{From NIR Colors, without Excess}
\label{sec_masses_nir_noexcess}

In this section we derive the stellar parameters by using only the $JHK$ photometry, assuming that the colors are coming directly from the stellar photosphere, i.e. the objects present no excess due to dusty disks in the NIR.
To simultaneously derive T$_{\mathrm{eff}}$, masses, and A$_V$ of the cluster members, we simply deredden the photometry in the $J,(J-H)$ diagram to the isochrones (see Fig~\ref{fig_cmdccd}). 
We also check the source position in the color-color diagram (CCD; right panel in Fig.~\ref{fig_cmdccd}), where the solid blue line represents evolutionary models and the dotted red lines are the reddening vectors. If a source
is not located in a strip defined by the two reddening lines encompassing the region compatible with evolutionary models (two dotted red lines to the left, region A), then the parameters cannot be derived.

To estimate the effect that the photometric uncertainties have on this determination of extinction and mass, we apply the Monte Carlo method:
for each source we create a set of 1000 magnitudes in each band, assuming a normal distribution with a standard deviation equal
to the respective photometric uncertainty. For each of the 1000 realizations we then derive the mass, T$_{\mathrm{eff}}$ and A$_V$ by dereddening the photometry to the isochrone.
The final A$_V$ for each source is calculated as an average, and its uncertainty as the standard deviation of all realizations. The resulting mass distributions, however, in most cases will not be well represented by a normal distribution because the magnitude 
is not a linear function of mass. The resulting mass distributions are typically skewed toward higher masses, meaning that by taking a mean of all the realizations as the final mass, we might in fact be overestimating it. For this reason, we save the resulting mass and T$_{\mathrm{eff}}$ distributions for each source. These distributions are later used to estimate the confidence interval for mass and T$_{\mathrm{eff}}$ when comparing different derivation approaches, and to draw masses in the Monte Carlo simulation used to determine the IMF.

\subsubsection{From NIR colors, with excess}
\label{sec_masses_nir_excess}

In this section we derive the stellar parameters by using only the $JHK$ photometry, but this time allowing that some of the objects might have an intrinsic NIR excess.
While this method makes various assumptions concerning the disk properties, it also might be more realistic, given the large disk fraction in the cluster.
Fig.~\ref{fig_cmdccd} shows the CMD and the CCD used to derive the source extinction and masses. 
To estimate the effect that the photometric uncertainties have on our determination of extinction and mass, we apply the same Monte Carlo method as in the previous section, only using a different method to derive individual T$_{\mathrm{eff}}$, mass and A$_V$, consisting of the following steps.

\begin{enumerate}
\item 
We first check the source's position in the CCD.  
In Fig.~\ref{fig_cmdccd}~(c), the solid blue line represents evolutionary models, the solid and dashed orange lines
the locus of T-Tauri stars and the corresponding uncertainties \citep{meyer97}. The dotted red lines are the reddening vectors, encompassing the regions where the colors are consistent with reddened evolutionary models or CTTSs (region A), and CTTSs only (region B).
\item 
Each star is represented by a distribution of colors determined by the value of photometry and the corresponding uncertainty (sampled 1000 times), which will typically fall into different regions of the CCD. 
If a point falls into the region A, but below the CTTS locus, the extinction and corresponding mass are derived by dereddening the photometry to the isochrone in the $J, (J-H)$ CMD ($case~1$). 
In region B, the extinction is derived by dereddening the colors to the CTTS locus ($case~2$; \citealt{meyer97}).  
If the object falls to the left of the region A, or to the right of the region B, the extinction
and mass cannot be derived.

\item
In $case~2$, in addition to the interstellar extinction, we also correct for an excess due to the 
circumstellar disk or envelope, chosen randomly in the interval r$_J=0 - 0.7\,$mag for the $J$-band and r$_H=0-1.1\,$mag for the H-band \citep{cieza05}\footnote{As previously discussed in \citet{muzic17}, we assume that the excess values are valid in the BD regime as well.}, with a condition r$_H>$r$_J$ \citep{meyer97}.
If a point falls into the region A, but above the CTTS line (named A1 in the following), we have two options: deredden the photometry to the isochrone (as in $case~1$), or to the CTTS locus (as in $case~2$).
To decide, we first look at in what fraction of 1000 realizations a star's photometry ends up above the CTTS line. If this fraction is less than  
70\%, we apply the $case~2$ procedure to all the points that fall into the region A1. Otherwise, we randomly pick a number of points (from A1) to which the $case~2$ procedure would be applied, so that the total number of times we allow for an infrared excess amounts to 70\%. In the remainder of the cases we apply the same procedure as in $case~1$.
The mentioned percentage is chosen to match the 
disk frequency of the cluster \citep{meng17}. 

\item
The derived extinction is then used to convert the $J$-band photometry to the absolute $J$-band magnitude, which directly corresponds to a certain T$_{\mathrm{eff}}$ and mass as given by the models.

\end{enumerate}


\subsubsection{Comparison of the three methods}
\label{sec_method_comparison}
\begin{figure*}
\centering
\resizebox{\textwidth}{!}{\includegraphics{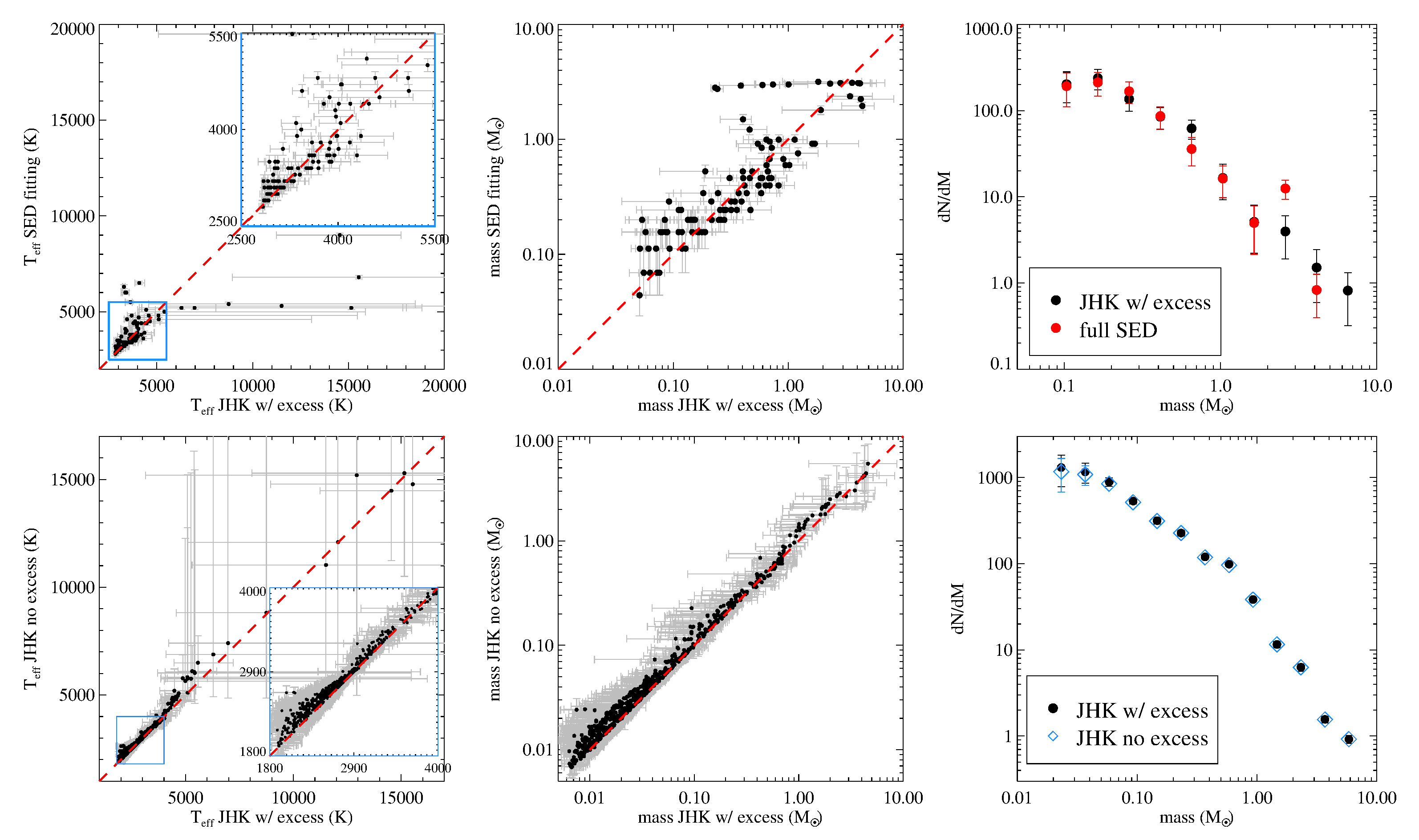}}
\caption{Comparison of effective temperatures (left), masses (middle), and the resulting IMF (right) derived by the three methods (see Section~\ref{sec_parameters} for details). The dashed red lines show a linear relation with a slope of 1 (not a fit). The uncertainties of some points in the right bottom panel are omitted as they are smaller than the size of the symbols. Note that the top panels include only the sources marked as candidate members by the $PS1-Gaia$ selection, while the bottom panels show the derived parameters for all the sources from the $JHK$ catalog for which the mass and mass/T$_{\mathrm{eff}}$ could be derived, assuming the age and the distance of the cluster. 
}
\label{fig_tmcompare}
\end{figure*}

\begin{figure}
\centering
\resizebox{0.45\textwidth}{!}{\includegraphics{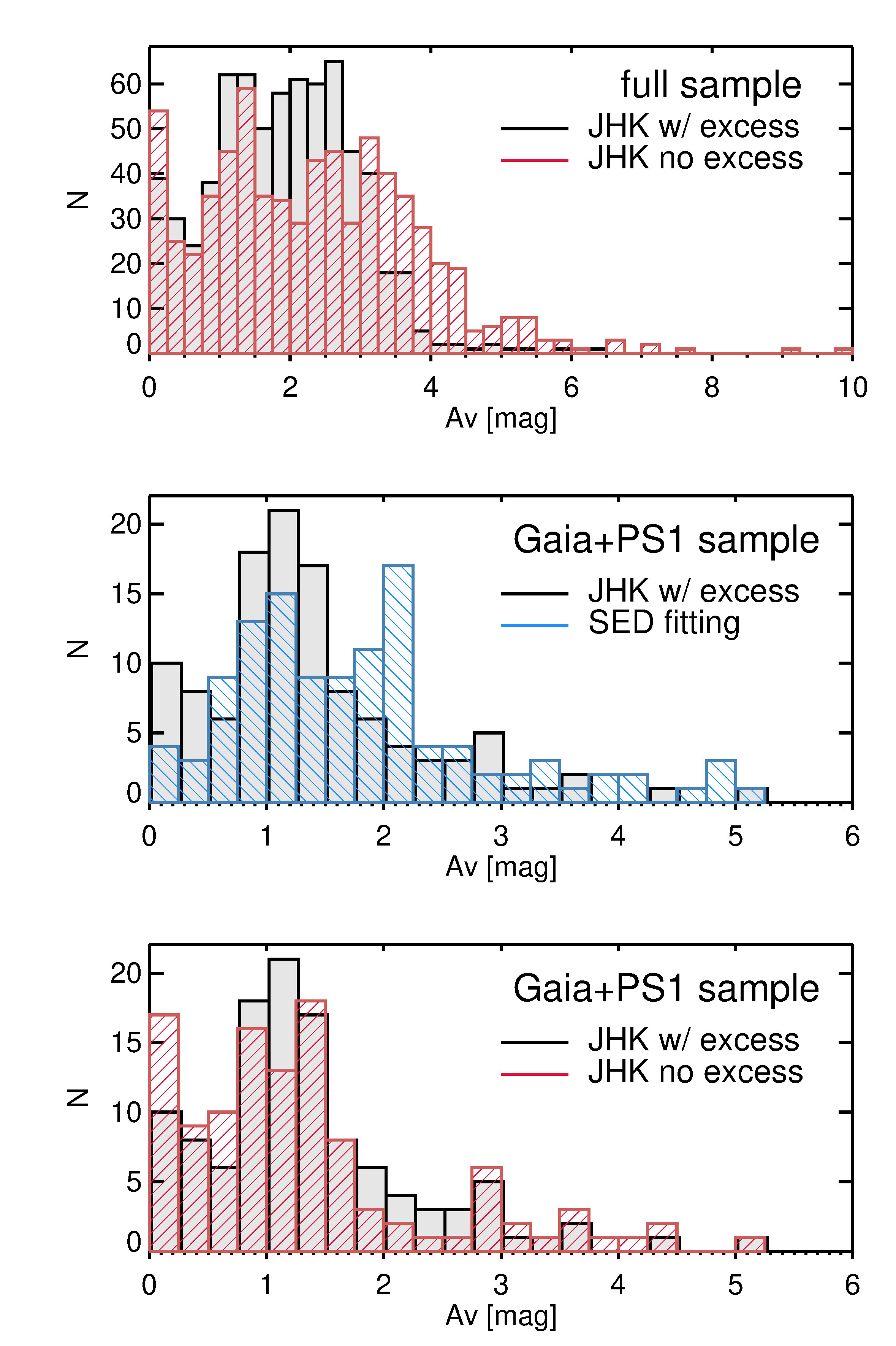}}
\caption{Histograms of extinction values derived in Section~\ref{sec_parameters}. {\bf Top:} A$_V$ derived by \textit{methods 2} and {\it 3} on the full $JHK$ sample, assuming that all the objects belong to the cluster. {\bf Middle:} A$_V$ derived by \textit{methods 1} and {\it 3}, for the $Gaia/PS1$ candidate member sample. {\bf Bottom:} A$_V$ derived by \textit{methods 2} and {\it 3}, for the $Gaia/PS1$ candidate member sample. 
}
\label{fig_av}
\end{figure}

In Fig.~\ref{fig_tmcompare}, we compare the three mass/T$_{\mathrm{eff}}$ methods described in Sections~\ref{sec_sedfit} (SED fitting; $method~1$ for the rest of the section), \ref{sec_masses_nir_noexcess} ($JHK$ without excess; $method~2$), and \ref{sec_masses_nir_excess} ($JHK$ with excess; $method~3$). 
The top panels compare the results of the $method~3$ with those of $method~1$, and the bottom panels $method~3$ with $method~2$.
In the left and middle panels, we show the comparison of T$_{\mathrm{eff}}$ and masses, where in case of $methods~2$ and $3$ we show only the results using the $isochrone~1$.
The T$_{\mathrm{eff}}$ error bars for the SED fitting are set to 100 K, reflecting the spacing of the fitting grid.   
For the methods based on $JHK$ photometry only ($methods~2$ and $3$), we have a distribution of T$_{\mathrm{eff}}$ and masses for each object, which do not necessarily follow the normal distribution, and in most cases are not symmetric. The T$_{\mathrm{eff}}$ and mass for the plot were calculated as the weighted average, with the weights provided by the distribution function. The error bars represent 95\% confidence limits. We note that the large T$_{\mathrm{eff}}$ error bars for the massive stars in $methods~2$ and $3$ arise from the T$_{\mathrm{eff}}$/A$_V$ degeneracy at the main sequence transition region of the isochrones. Basically, there is more than one combination of T$_{\mathrm{eff}}$ and A$_V$ that can be used to deredden star's photometry onto the isochrone, and the span of T$_{\mathrm{eff}}$ value in this region is large ($\sim 14000$\,K).

The right hand panels of Fig.~\ref{fig_tmcompare} show the IMFs derived using the masses obtained by different methods.
The IMFs are derived in the same way as explained in the following section. For the upper plot the input object list contains the $Gaia$-$PS1$ candidate members, while in the lower plot the IMF is averaged over 324 lists resulting from the control-field-based membership analysis (Section~\ref{sec_cleaning}).

\begin{itemize}
\item \underline{$Method~3$ versus $method~1$} (top panels of Fig.~\ref{fig_tmcompare})\\

The values of T$_{\mathrm{eff}}$ obtained by the two methods are in reasonable agreement below $\sim 6000\,$K, where most of the objects are located. 
Above this temperature there is a set of objects with significantly higher T$_{\mathrm{eff}}$ from the $JHK$ method (albeit with very large error bars). These are the objects that occupy the space in the CMD around the pre-main sequence transition (masses roughly in the 2-6\,M$_{\sun}$ range, see Fig.~\ref{fig_cmdccd}). Their T$_{\mathrm{eff}}$ and mass distribution tends to be bimodal, resulting in the larger error bars, and the values typically higher than those obtained by the SED fitting.  
There is also a handful of sources just above the smaller blue box in the left panel of Fig.~\ref{fig_tmcompare}. SED fitting for these sources prefers higher T$_{\mathrm{eff}}$ and A$_V$ than the $JHK$ method. There is no obvious explanation for this discrepancy, but we note that most of them (4/5) are either at the faint end in $Gaia$ and have very large parallax errors, or were selected in $PS1$ (color selection only). They might simply be contaminants.

For the objects located inside the blue box of the top left plot of Fig.~\ref{fig_tmcompare}, we obtain the average T$_{\mathrm{eff}}$ difference ($method~3$ - $method~1$) and its standard deviation $\Delta$T$_{\mathrm{eff}}$$=-80\pm310$\,K.
The resulting IMFs are very similar. The difference in the bin at around 3\,M$_{\sun}$ comes from the five objects discussed above, where SED fitting yields higher masses. Removing those objects, the red point becomes consistent with the black one.

To conclude, the overall good agreement between the parameters obtained by the two methods, and in particular by the similar IMFs, is reassuring, and validates the methods based on $JHK$ colors only. This is important, because it represents the only option to derive T$_{\mathrm{eff}}$ and mass for the fainter sources.

\item 
\underline{$Method~3$ versus $method~2$} (bottom panels of Fig.~\ref{fig_tmcompare})\\

As expected, the method ignoring the NIR excess predicts systematically higher T$_{\mathrm{eff}}$ and masses than the method allowing for the infrared excess. 
The average T$_{\mathrm{eff}}$ difference ($method~3$ - $method~2$) and its standard deviation are $\Delta$T$_{\mathrm{eff}}$$=-80\pm150$\,K.

The results agree well within the errors, and the impact on the resulting IMF is minimal, and only at the lowest-mass bins. Clearly, ignoring the NIR excesses in the determination of mass does not seems to be a crucial point for studying the IMF, but anyway for the remainder of the paper we decide to use the masses derived from $method~3$.

\end{itemize}

In Fig.~\ref{fig_av} we show histograms of the extinction values derived by the three methods. The top panel is comparing the A$_V$ values from \textit{methods 2} and \textit{3}, for the objects from the full $JHK$ sample. Note that the A$_V$ derivation relies on dereddening the photometry to the 2 Myr isochrone shifted to the distance of the cluster. The middle and bottom panels show the histograms of the three methods, but only for the objects in the $Gaia/PS1$ sample of candidate members. The average extinction of the cluster from the SED fitting ($method~1$) is $1.7\pm1.1\,$mag, from $JHK$ photometry without excess ($method~2$) $1.3\pm1.1\,$mag, and from $JHK$ photometry with excess ($method~3$) $1.4\pm0.9\,$mag. The listed uncertainties are standard deviations of the three distributions. The extinctions derived by the three methods are in general consistent with each other, and we see no large intracluster spatially variable spread of extinctions. 

\subsection{Initial Mass Function}
\label{sec_imf}

\begin{figure*}
\gridline{\fig{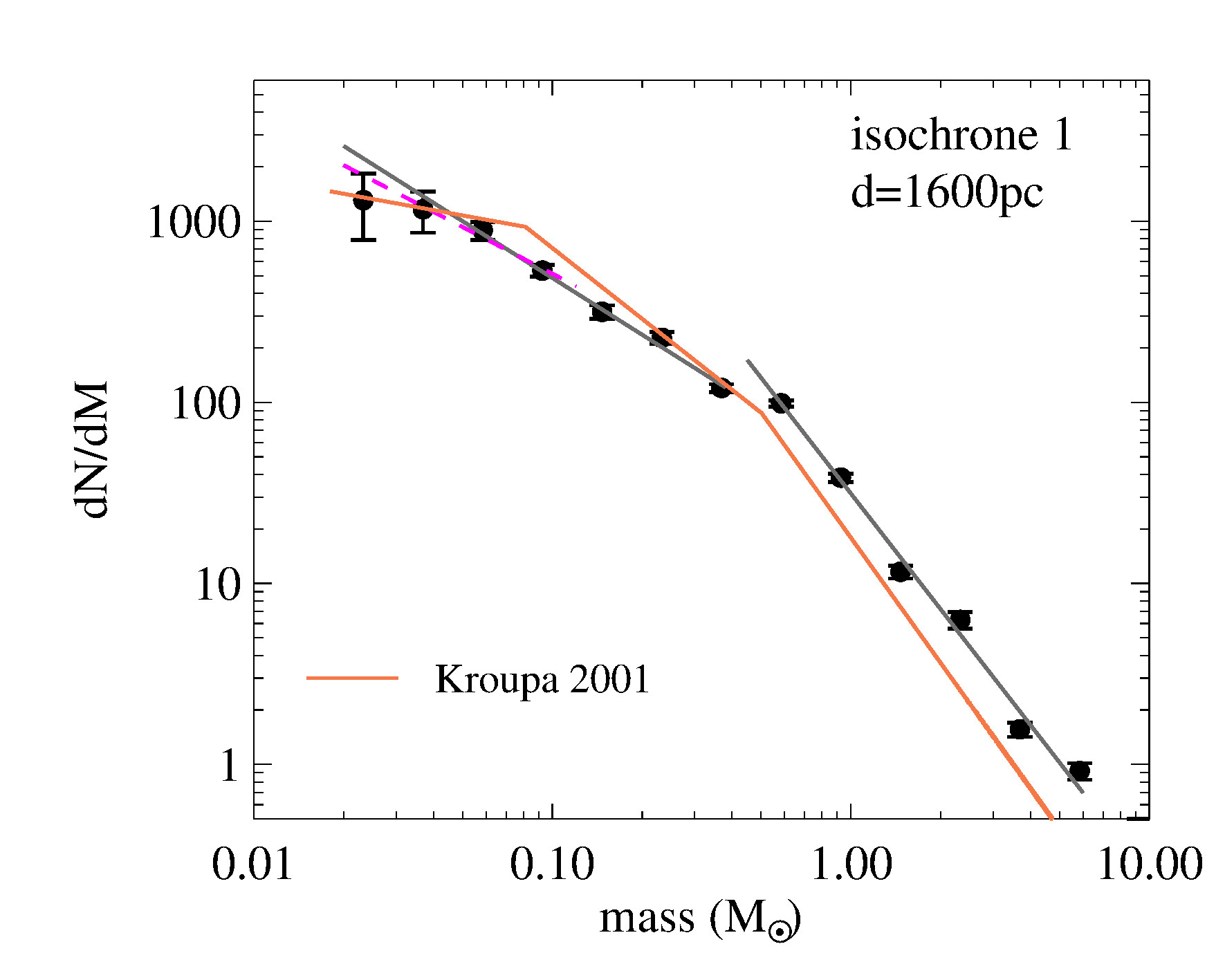}{0.33\textwidth}{}
          \fig{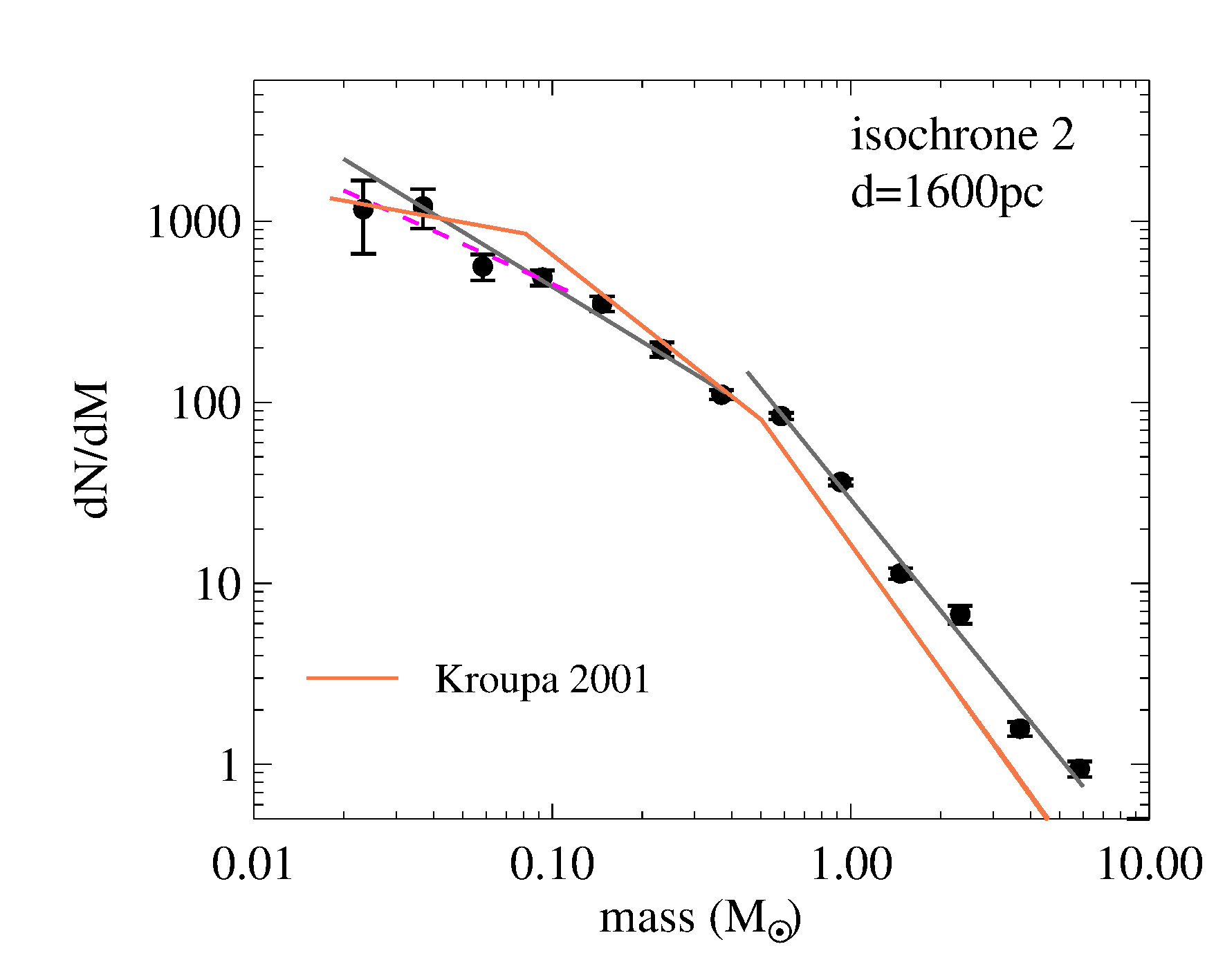}{0.33\textwidth}{}
          \fig{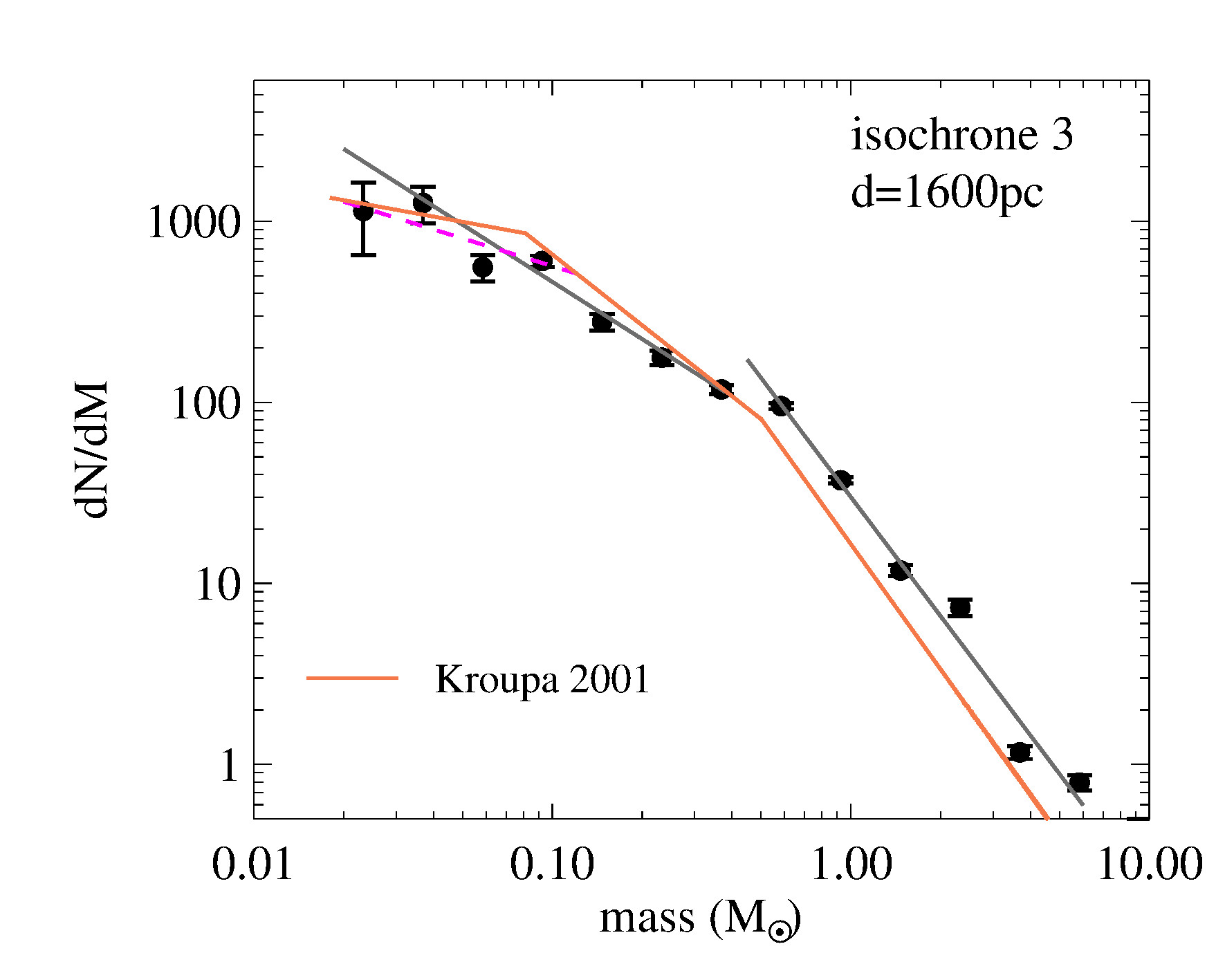}{0.33\textwidth}{}}
\gridline{\fig{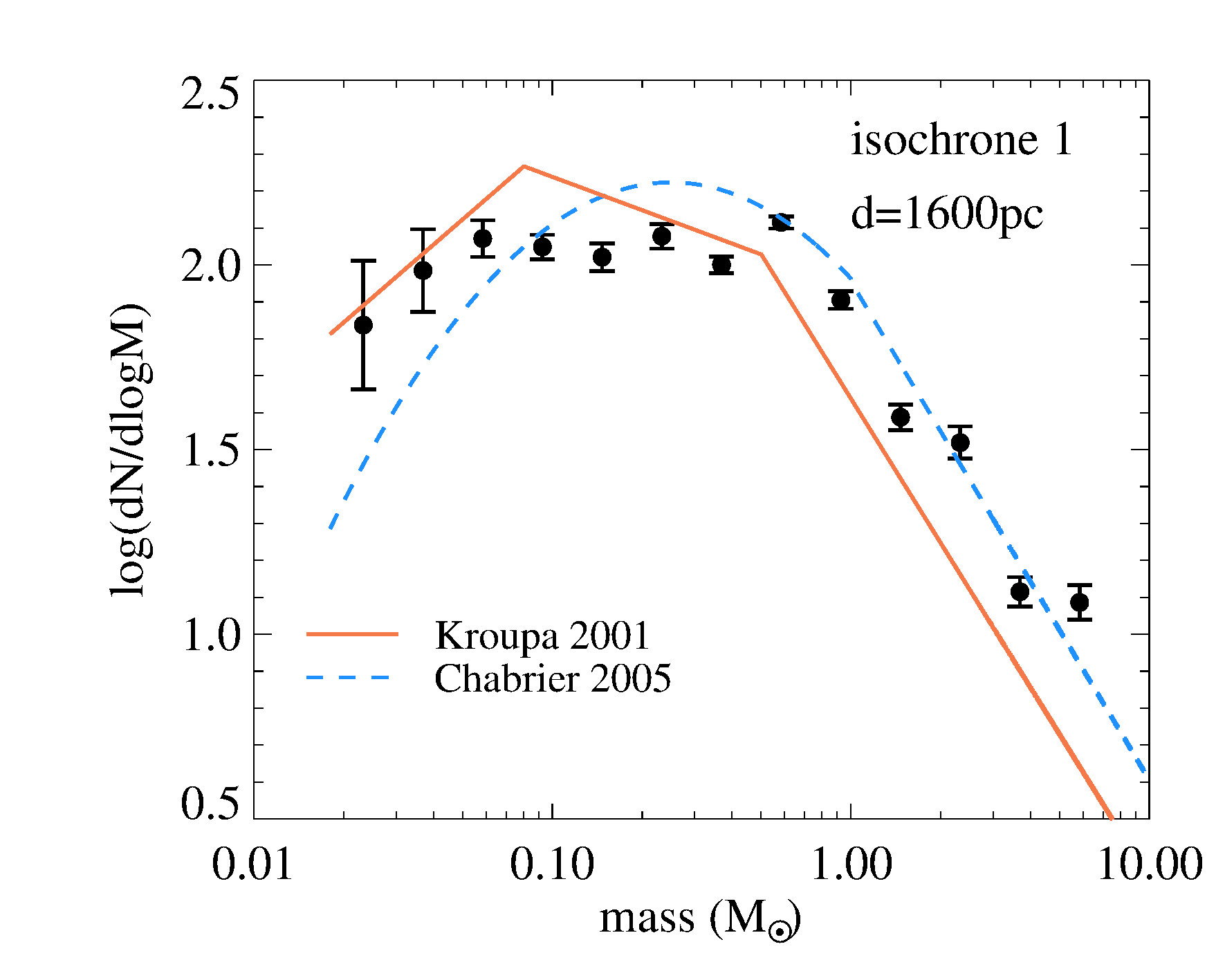}{0.33\textwidth}{}
          \fig{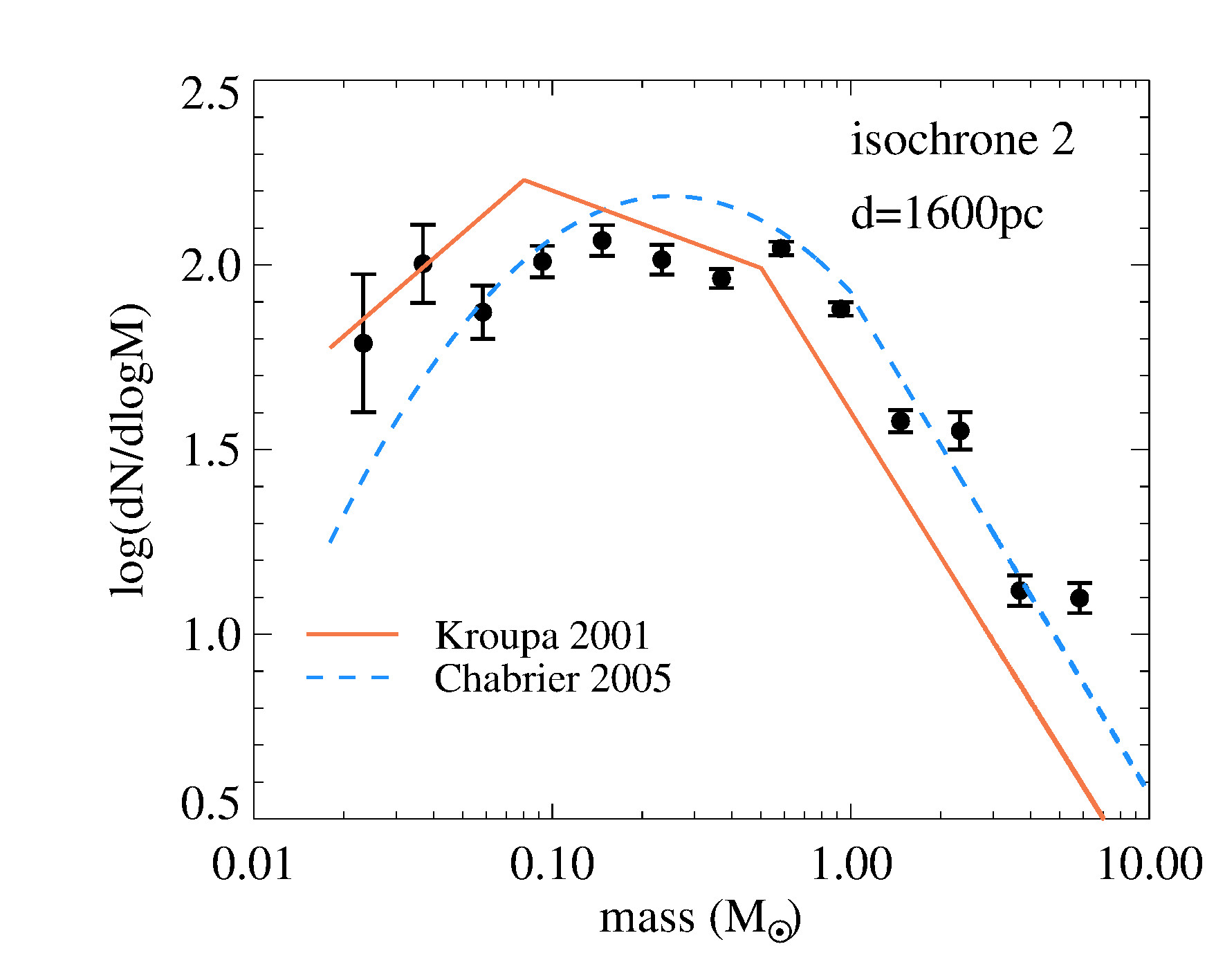}{0.33\textwidth}{}
          \fig{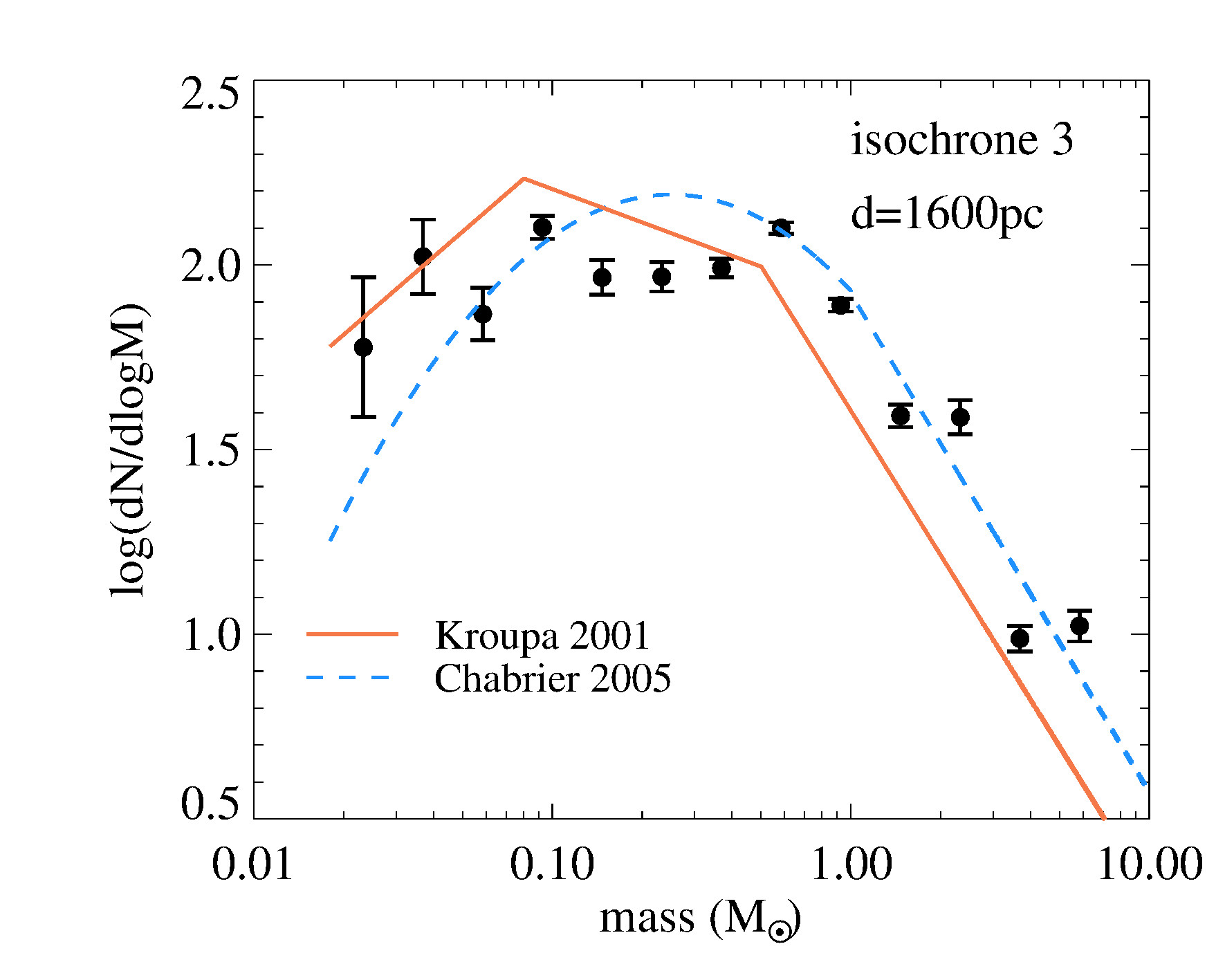}{0.33\textwidth}{}}
\caption{Initial mass function of the central core of NGC\,2244, represented in the equal-size logarithmic bins with size  $\Delta$\,log(m/M$_{\sun}$)=0.2, with the masses derived by comparison to three different isochrones (see Appendix~\ref{append_isochrones} for details). Two different IMF representations are shown, $dN/dm$ (top panels) and $dN/dlog(m)$ (bottom panels). 
Solid grey lines in the top panels show the power law fits with a break at 0.4\,M$_{\sun}$, while the dashed magenta lines show the power law fits in the substellar regimes ($0.02-0.1\,$M$_{\sun}$). The slopes are given in Table~\ref{tab:imf}. The orange solid line is the Kroupa segmented power law mass function \citep{kroupa01}, and the blue dashed line shows the Chabrier mass function \citep{chabrier05}, both normalized to match the total number of objects in the cluster. The distance to the cluster is assumed to be 1600 pc.}
\label{fig_imf_1600}
\end{figure*}

\begin{figure*}
\gridline{\fig{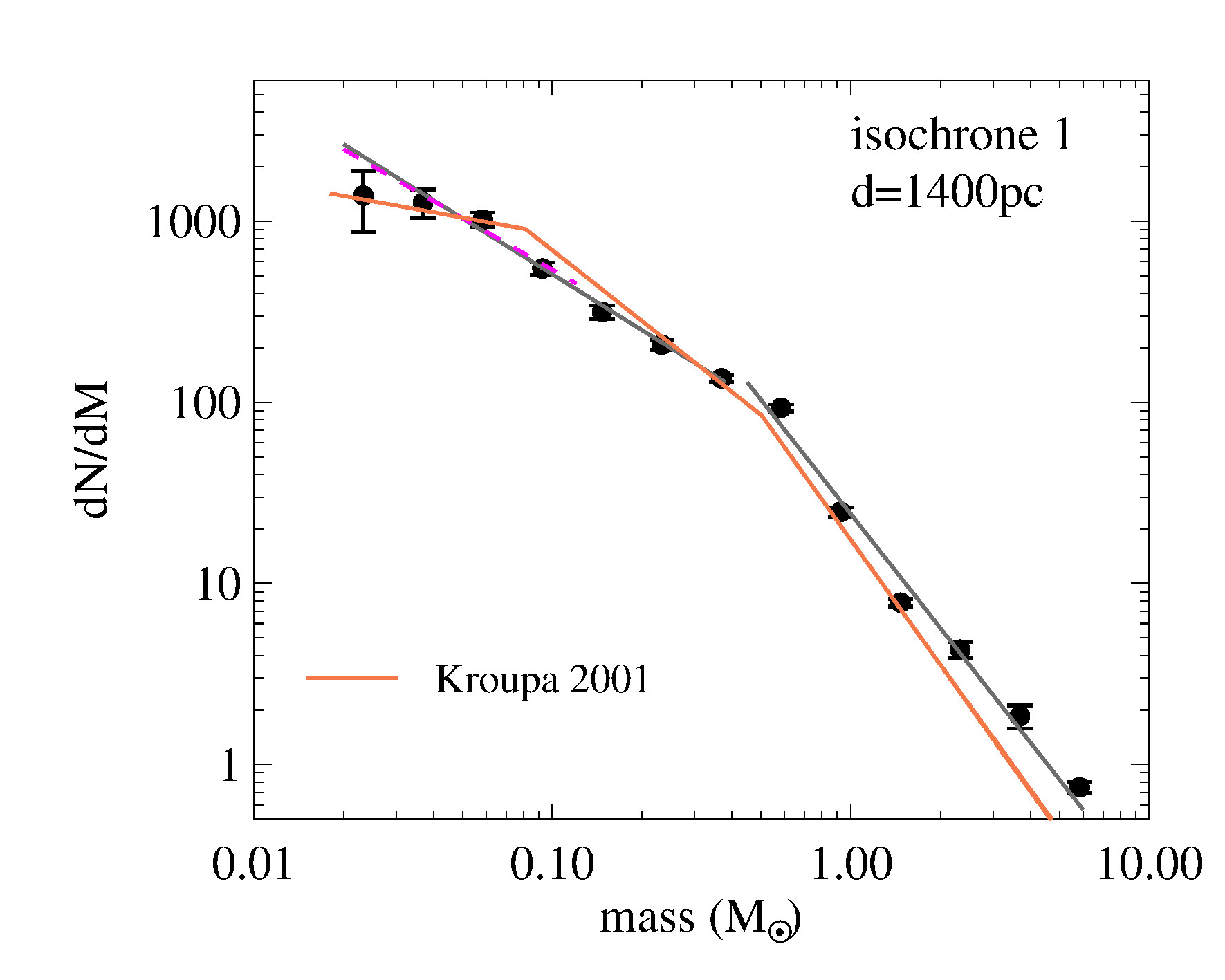}{0.33\textwidth}{}
          \fig{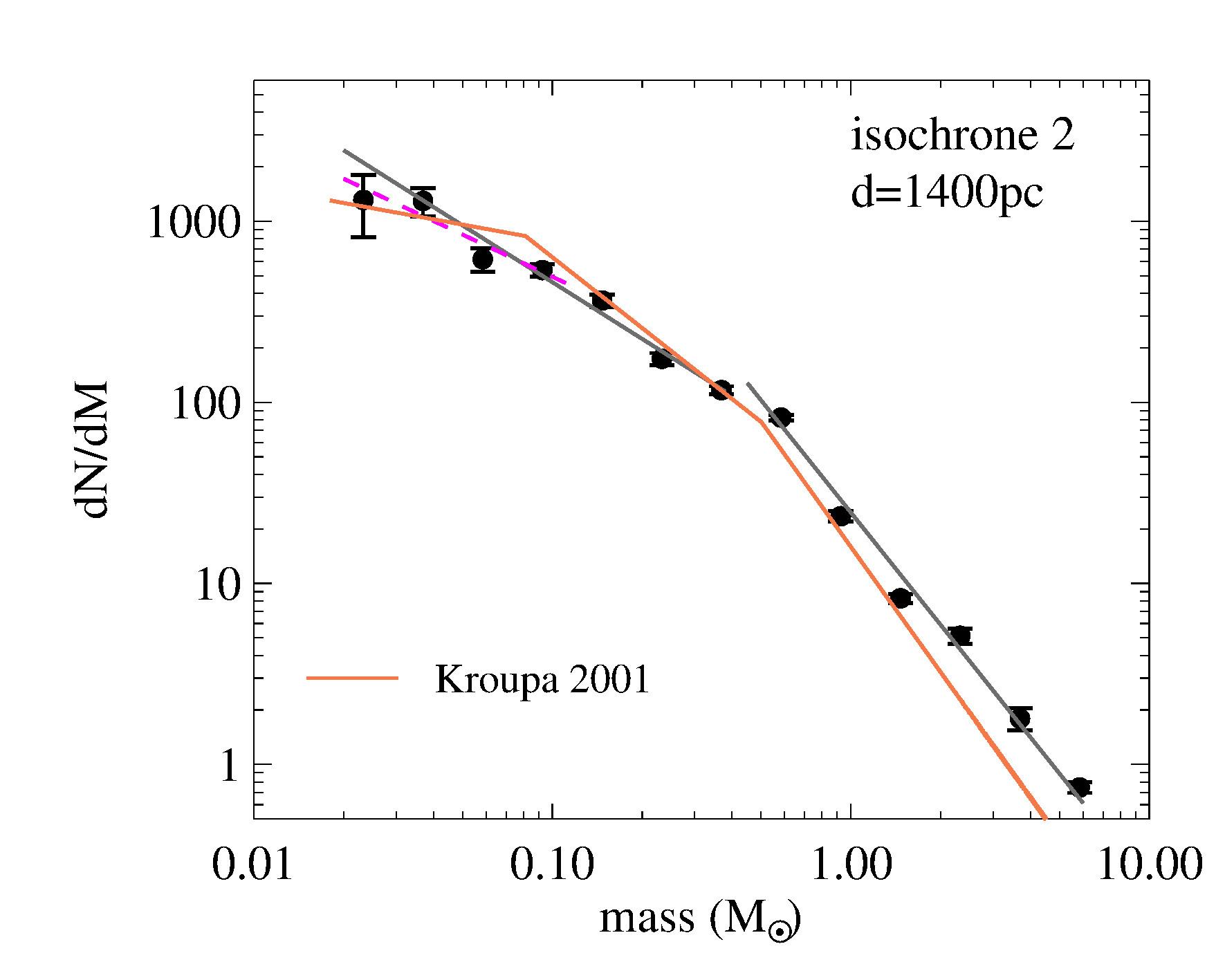}{0.33\textwidth}{}
          \fig{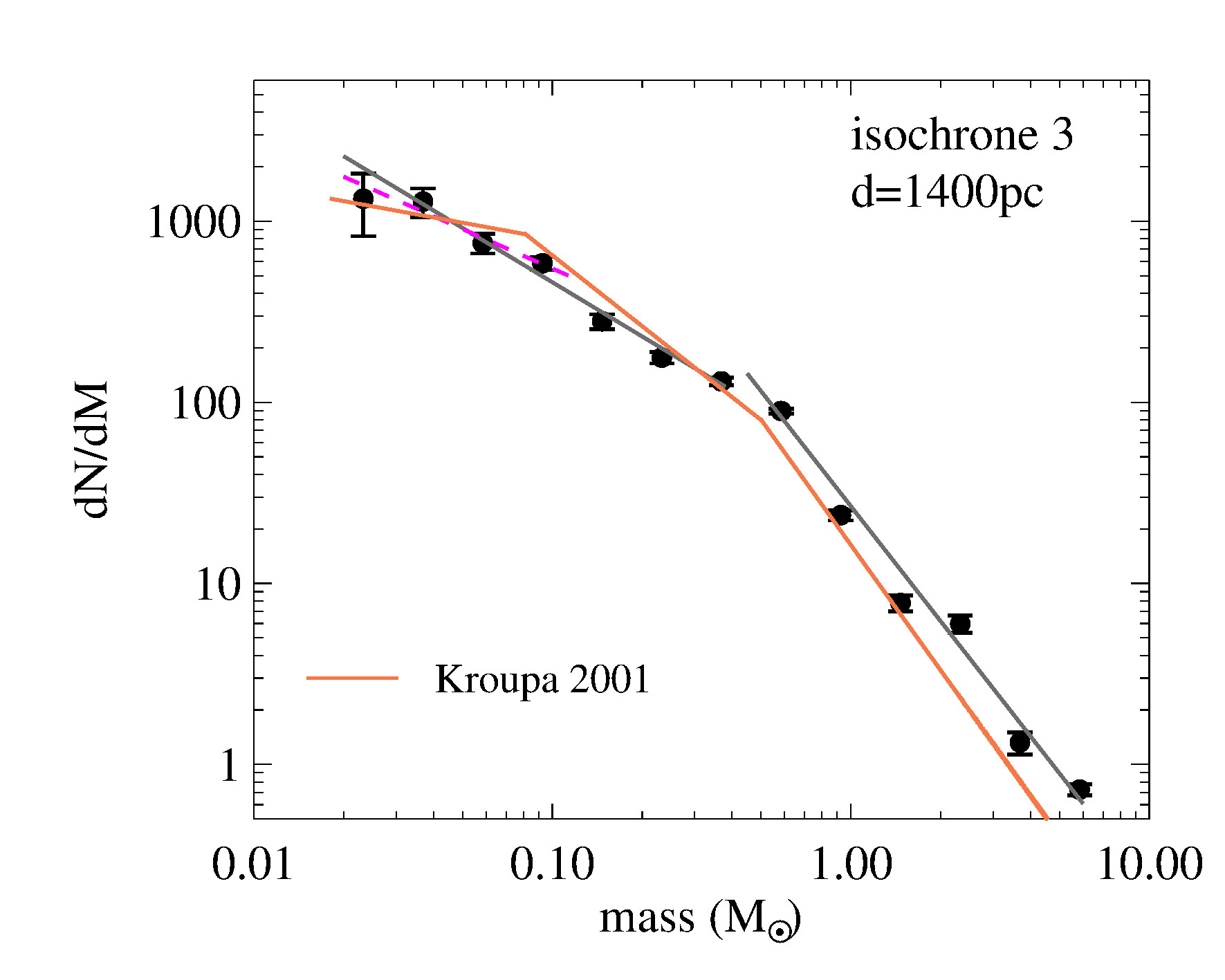}{0.33\textwidth}{}}
\gridline{\fig{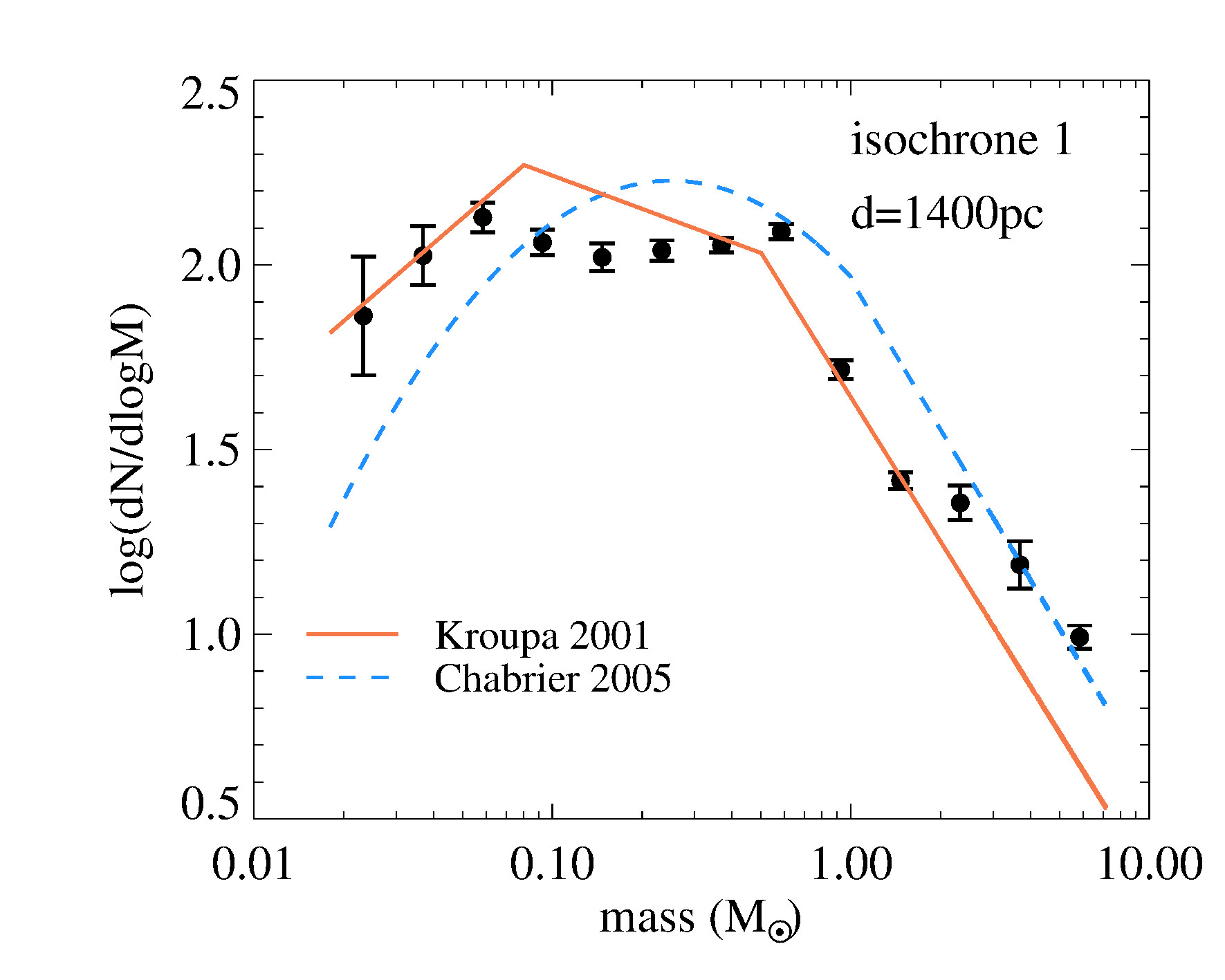}{0.33\textwidth}{}
          \fig{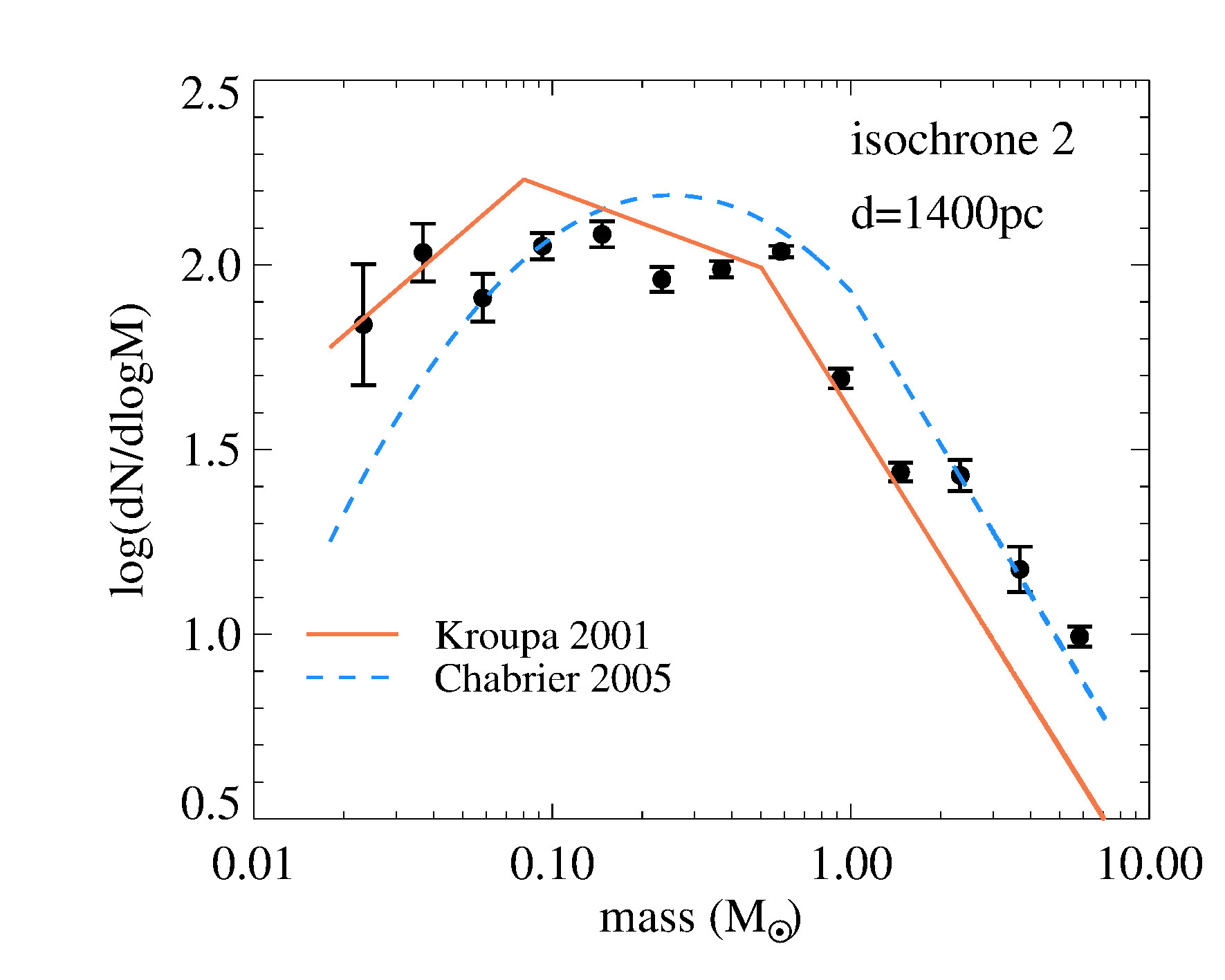}{0.33\textwidth}{}
          \fig{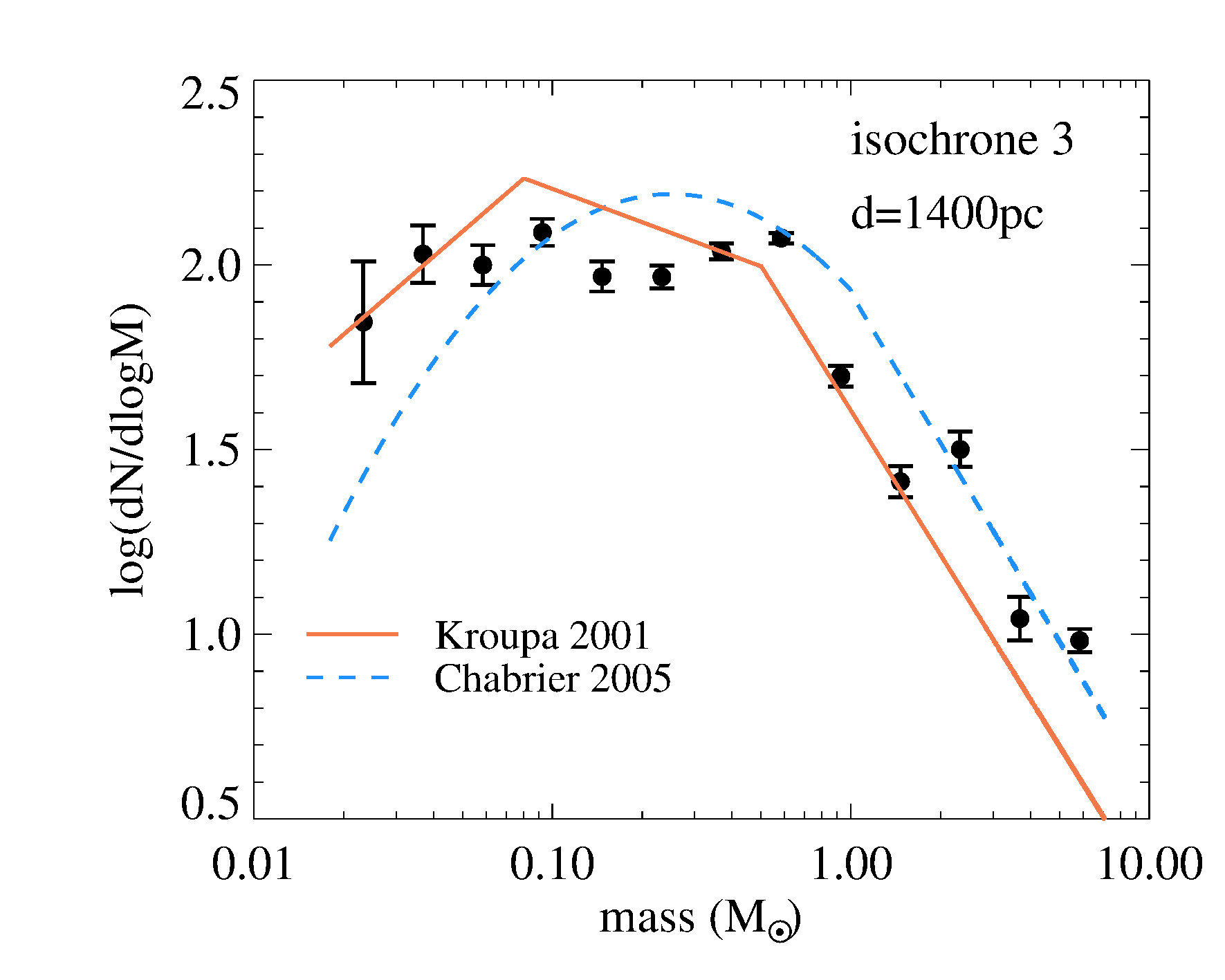}{0.33\textwidth}{}}
\caption{Same as Fig.~\ref{fig_imf_1600}, but for the distance of 1400 pc.}
\label{fig_imf_1400}
\end{figure*}

\begin{deluxetable}{llll}
\tablecaption{The slope of the Initial Mass Function in NGC\,2244 in the power law form and star-to-BD number ratio for the three different sets of isochrones, and the distances of 1600\,pc and 1400\,pc.}
\label{tab:imf}
\tablehead{\multicolumn{4}{c}{\bf{Slope of the IMF}}}
\tablecolumns{4}
\startdata 
\hline
& 0.02 - 0.1\,M$_{\sun}$ &  0.02 - 0.4\,M$_{\sun}$ & 0.4 - 7\,M$_{\sun}$  \\
\hline
\multicolumn{4}{c}{d\,$=1600$\,pc}\\
\hline
	$isochrone$ 1 & $0.86\pm0.19$	& $1.04\pm0.05$ & $2.12\pm0.04$ \\
	$isochrone$ 2 & $0.74\pm0.21$	& $1.01\pm0.06$ & $2.03\pm0.04$ \\
	$isochrone$ 3 & $0.51\pm0.19$  & $1.05\pm0.05$ & $2.18\pm0.03$ \\
	\hline
\multicolumn{4}{c}{d\,$=1400$\,pc}\\
\hline
	$isochrone$ 1 & $0.95\pm0.16$	& $1.02\pm0.04$ & $2.11\pm0.03$ \\
	$isochrone$ 2 & $0.77\pm0.17$	& $1.04\pm0.05$ & $2.06\pm0.03$ \\
	$isochrone$ 3 & $0.73\pm0.17$  & $1.08\pm0.03$ & $2.11\pm0.03$ \\
\hline 
\hline
\multicolumn{4}{c}{\bf{Star-to-BD ratio}}\\
\hline
stars (M$_{\sun}$) & $0.075 - 1$ 	& $0.075 - 1$ 	&  $>0.075$ \\
BDs (M$_{\sun}$)   & $0.03 - 0.075$  & $0.02 - 0.075$  &  $0.02 - 0.075$\\
\hline
\multicolumn{4}{c}{d\,$=1600$\,pc}\\
\hline
	$isochrone$ 1 & $2.72\pm0.41$ & $2.00\pm0.41$ & $2.31\pm0.49$ \\
	$isochrone$ 2 & $3.30\pm0.56$ & $2.27\pm0.52$ & $2.68\pm0.63$\\
	$isochrone$ 3 & $3.21\pm0.54$ & $2.22\pm0.50$ & $2.62\pm0.60$ \\
\hline 
\multicolumn{4}{c}{d\,$=1400$\,pc}\\
\hline
	$isochrone$ 1 & $2.34\pm0.25$ & $1.74\pm0.28$ & $1.95\pm0.32$ \\
	$isochrone$ 2 & $2.95\pm0.39$ & $2.04\pm0.38$ & $2.34\pm0.44$\\
	$isochrone$ 3 & $2.62\pm0.30$ & $1.87\pm0.31$ & $2.15\pm0.36$ \\
\enddata
\end{deluxetable}

In this section we derive the IMF of NGC\,2244, for masses in the range from 0.02\,M$_{\sun}$ to 7\,M$_{\sun}$. 
Stellar masses are binned with logarithmic bin sizes of $\Delta$\,log(m/M$_{\sun}$)=0.2.
As can be appreciated from Fig.~\ref{fig_cmdccd} our lower mass limit is well above the 90\% completeness limit at the average cluster extinction (A$_V$=1.5\,mag), and should also contain the complete dereddened control field population to ensure the lowest-mass bins are not artificially overpopulated.

In Section~\ref{membership} we have derived 324 lists containing statistically cleaned probable members. Each object was assigned a mass distribution using the method described in Section~\ref{sec_masses_nir_excess}.
For each of the 324 member lists, we run a Monte Carlo simulation to derive the values and uncertainties of $dN/dM$ for each bin.
The $dN/dM$ values of the resulting 324 IMF versions are combined (as weighted average) to derive the final cluster IMF.
The Monte Carlo simulation is very similar to the one previously applied in the cluster RCW\,38 \citep{muzic17}. The mass of each star is drawn from the distribution derived in Section~\ref{sec_masses_nir_excess}.
This is performed 100 times, and for each of the 100 realizations we do 100 bootstraps, i.e. random samplings with replacement.
In other words, starting from a sample with N members, in each bootstrap we draw a new sample of N members, allowing some members of the initial sample to be drawn multiple times. This results in 10$^4$ mass distributions, which are used to derive the $dN/dM$ and the corresponding uncertainties.

In the upper panels of Fig.~\ref{fig_imf_1600} and Fig.~\ref{fig_imf_1400} we show the IMF in the power law form ($dN/dM\propto M^{-\alpha}$), derived from three different sets of isochrones and for distances of 1600 pc and 1400 pc, respectively. 
In all cases the IMF is well represented by two power laws, with
a break at $\sim$0.4\,M$_{\sun}$. We fit separate power laws to the mass ranges $0.4 - 7\,$M$_{\sun}$ and $0.02 - 0.4\,$M$_{\sun}$, as well as 
to the one encompassing only the substellar regime ($\sim0.02-0.1\,$M$_{\sun}$).
The least-squares fits are shown as grey lines in Figs.~\ref{fig_imf_1600} and \ref{fig_imf_1400}, and the resulting slopes $\alpha$ are given in Table~\ref{tab:imf}.
For each distance, the $\alpha$ values derived from the three isochrones agree within the uncertainties. The largest variation is seen in the $0.02-0.1\,$M$_{\sun}$ range, where the fit
is more sensitive to single-point variations given the small number of points available for the fit.
We can also observe that the uncertainty in distance of the cluster does not significantly affect our results, which again agree within the 
uncertainties.
For a comparison, we overplot the Kroupa segmented power law IMF, normalized to the total number of objects in the cluster (orange line).

To estimate the effects of the choice of the bin size and location, we repeat the IMF calculation and power law fits (within the same mass limits) for two additional bin sizes, $\Delta$\,log(m/M$_{\sun}$)=0.15 and 0.25. We also repeat the calculation with the same bin size as before ($\Delta$\,log(m/M$_{\sun}$)=0.2), but shift the bin locations by half the bin size. For each run, we combine the obtained slopes from the three isochrones as a weighted average, and show the results in Table~\ref{tab:app_imf} of the Appendix~\ref{append_imf}. The first line in the table is the default one and is shown to facilitate the comparison. There is no major effect on the slope of the IMF in any of the considered mass ranges.

In the lower panels of Figs.~\ref{fig_imf_1600} and \ref{fig_imf_1400} we plot the IMF in the log-normal form ($dN/dlogM$), and overplot the Kroupa and Chabrier \citep{chabrier05} mass functions, 
both normalized to match the total number of objects in the cluster.
Unlike the power law tail at masses $>1\,$M$_{\sun}$, the log-normal part of the Chabrier mass function does not seem to represent the NGC\,2244 mass function very well. Possible implications of this result will be discussed in Section~\ref{sec_discussion}.

For the masses above 0.4\,M$_{\sun}$, $\alpha=2.12\pm0.08$ 
is close to the Salpeter slope ($\alpha=2.35$; \citealt{salpeter55}), and identical to the one derived for X-ray-selected members with masses $>0.5\,$M$_{\sun}$ \citep{wang08}.
Below 0.4\,M$_{\sun}$, a single power law with $\alpha=1.03\pm0.02$ describes well both the low-mass stellar and substellar regimes. We can compare this with the values of other clusters and star forming regions summarized in Table 4 of \citet{muzic17}. The slope of the low-mass IMF in NGC\,2244 is steeper than the one in RCW\,38 (located at 1.7 kpc), where we find $\alpha=0.7\pm0.1$ for the mass range $0.02-0.5\,$M$_{\sun}$. 
The slopes found in nearby star forming regions at distances of up to 400 pc, typically lie in the range 0.5 - 1.0. The slope we derive for NGC\,2244 at masses 0.02-0.4\,M$_{\sun}$ is on the high end of this range, but still in agreement with it, taking into account the statistical uncertainties. Furthermore, there are a number of other 
sources of uncertainty which are more difficult to take into account, such as the errors in age, systematics associated with the evolutionary models, 
extinction laws, and finally the choice of the mass range included in the fit. The latter is demonstrated by the power law slope fits in the substellar regime  
which are somewhat shallower than those encompassing also very-low-mass stars.

\subsection{Star/BD ratio}
\label{sec_ratio}

\begin{figure*}
\gridline{\fig{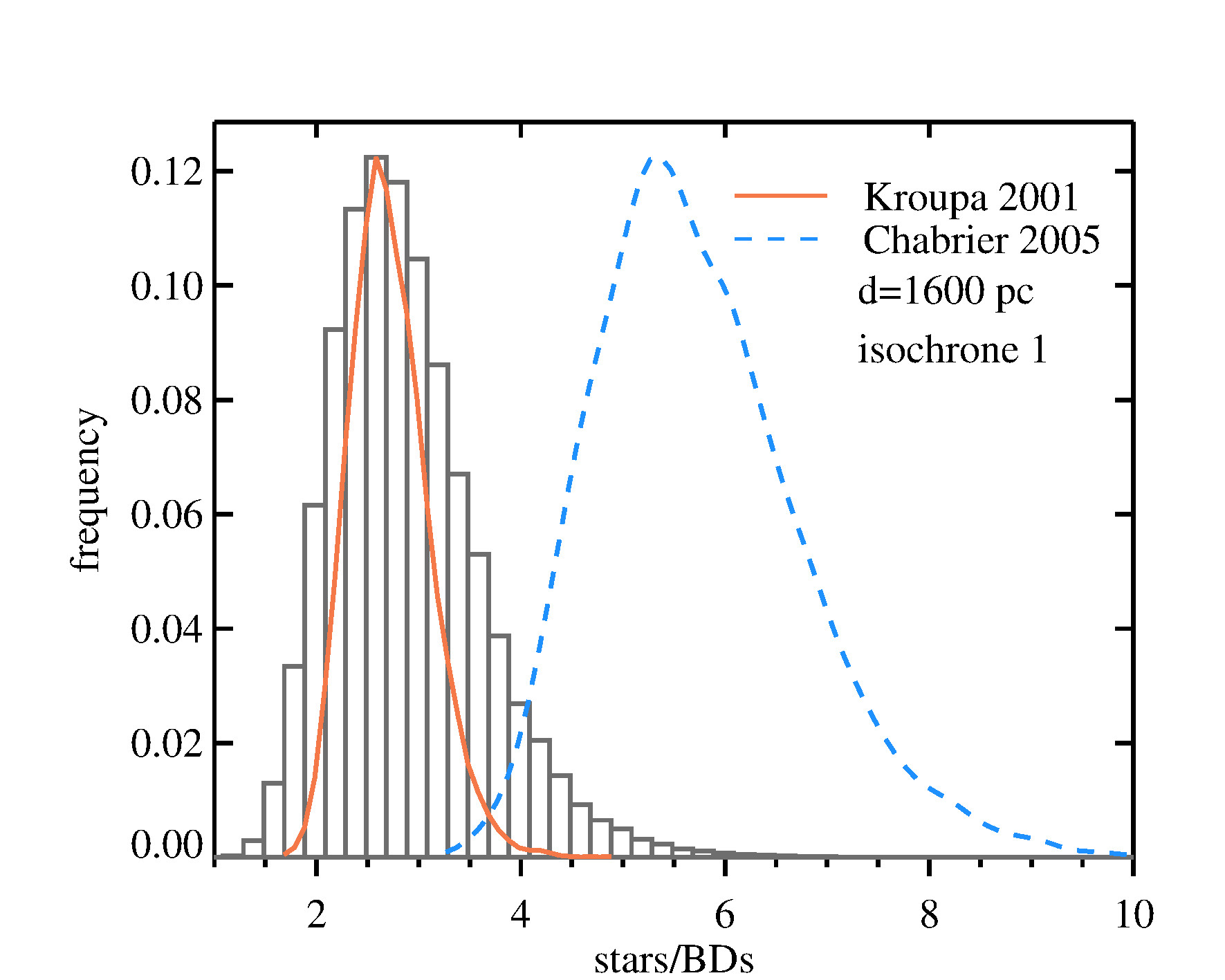}{0.33\textwidth}{}
          \fig{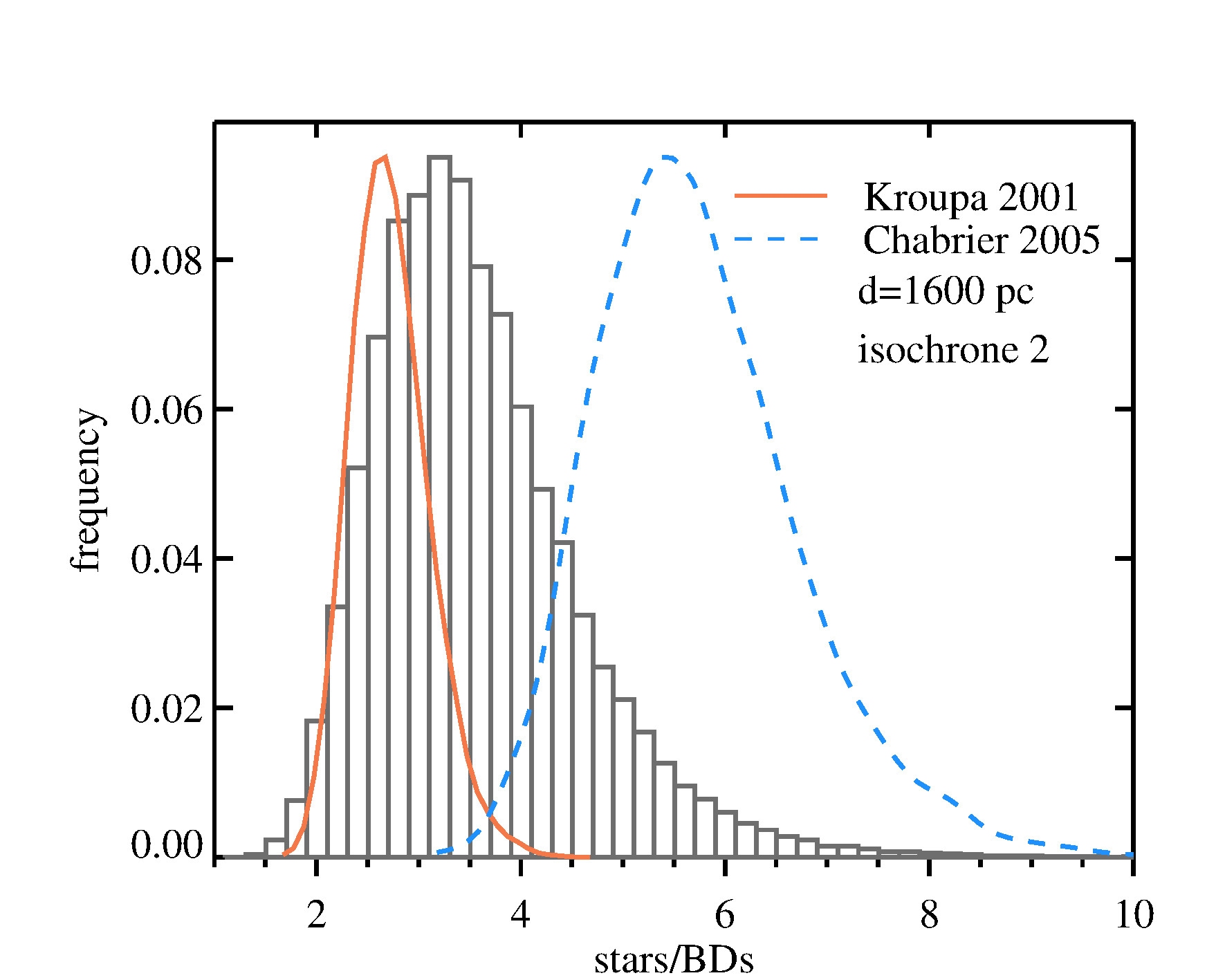}{0.33\textwidth}{}
          \fig{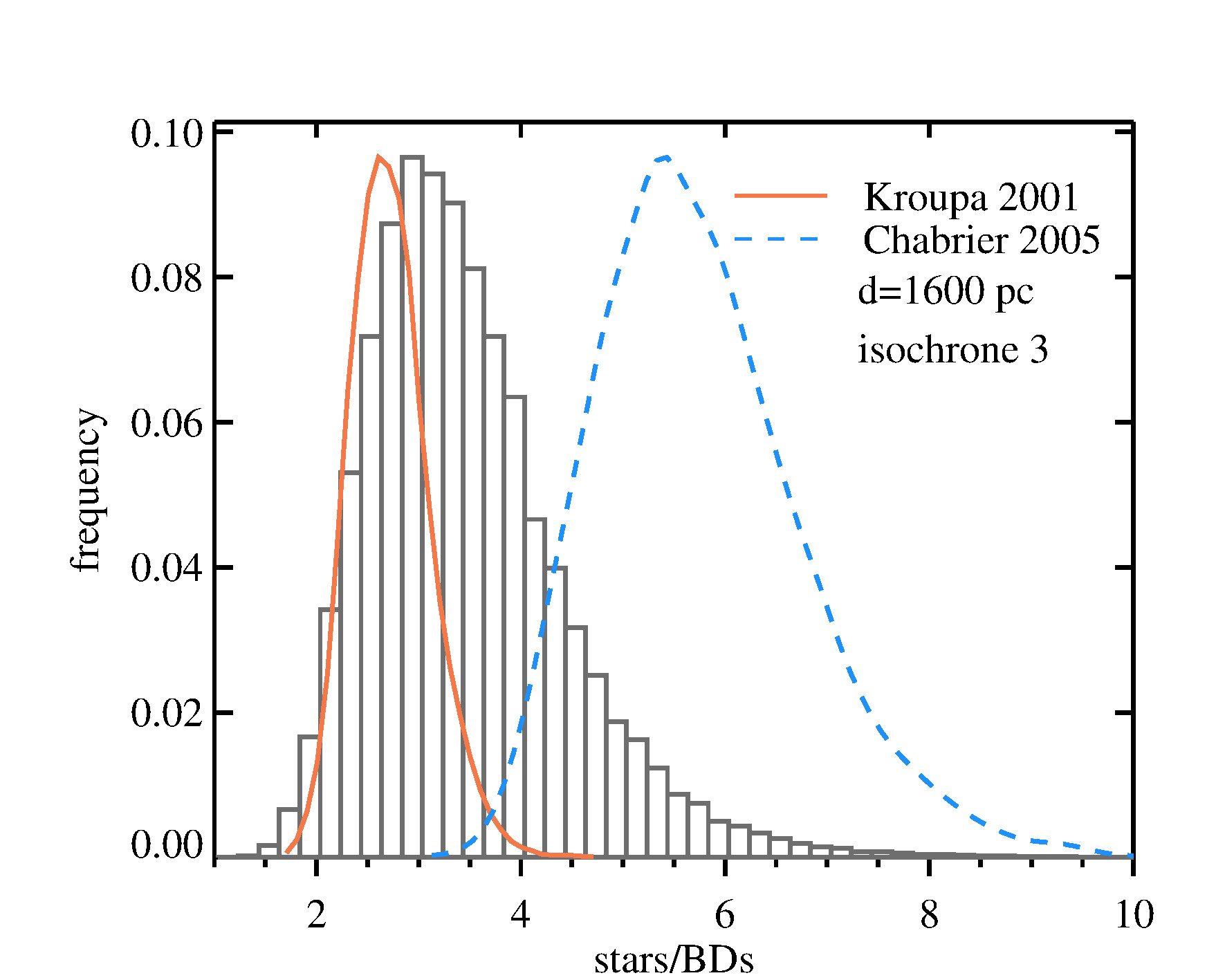}{0.33\textwidth}{}}
\gridline{\fig{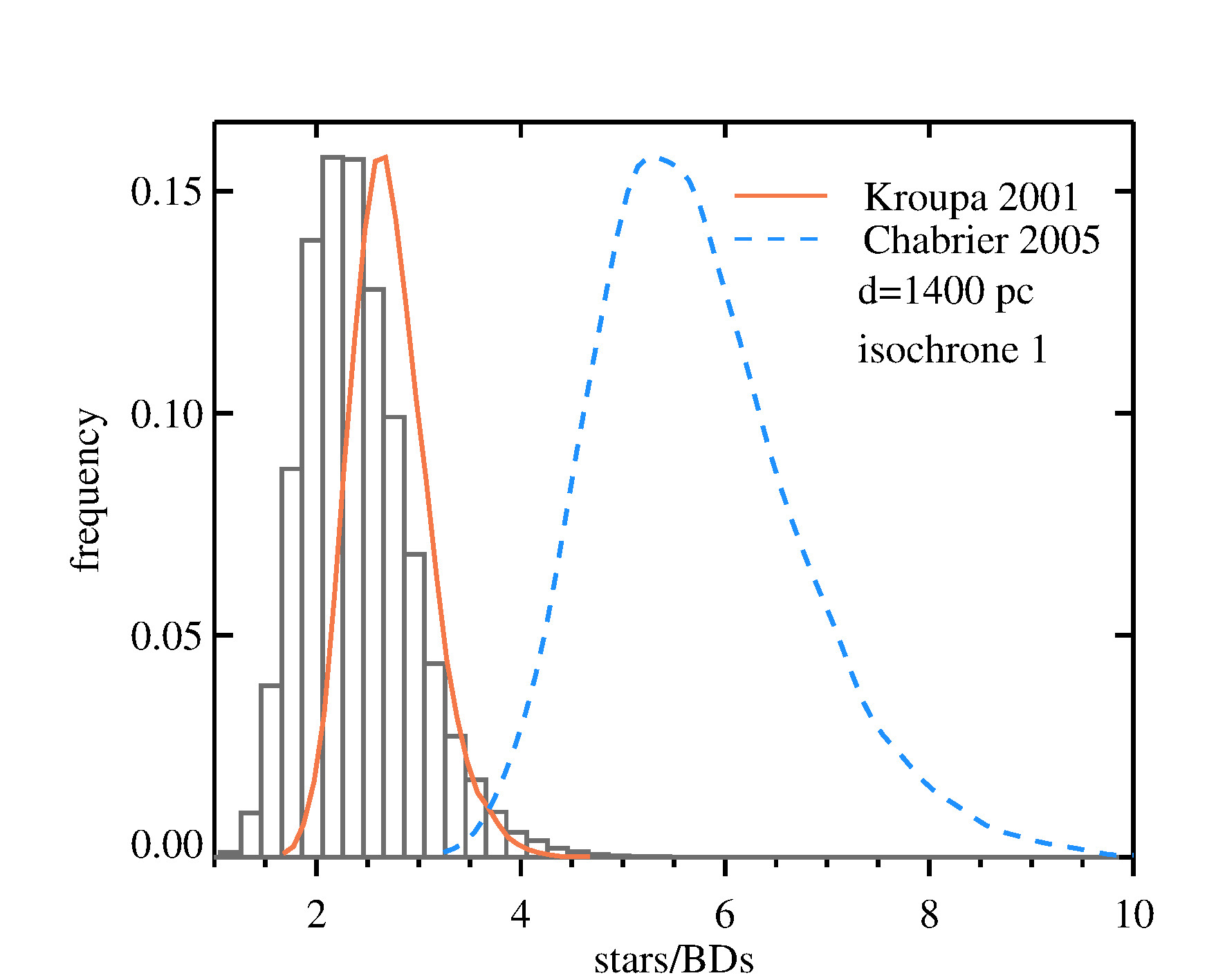}{0.33\textwidth}{}
          \fig{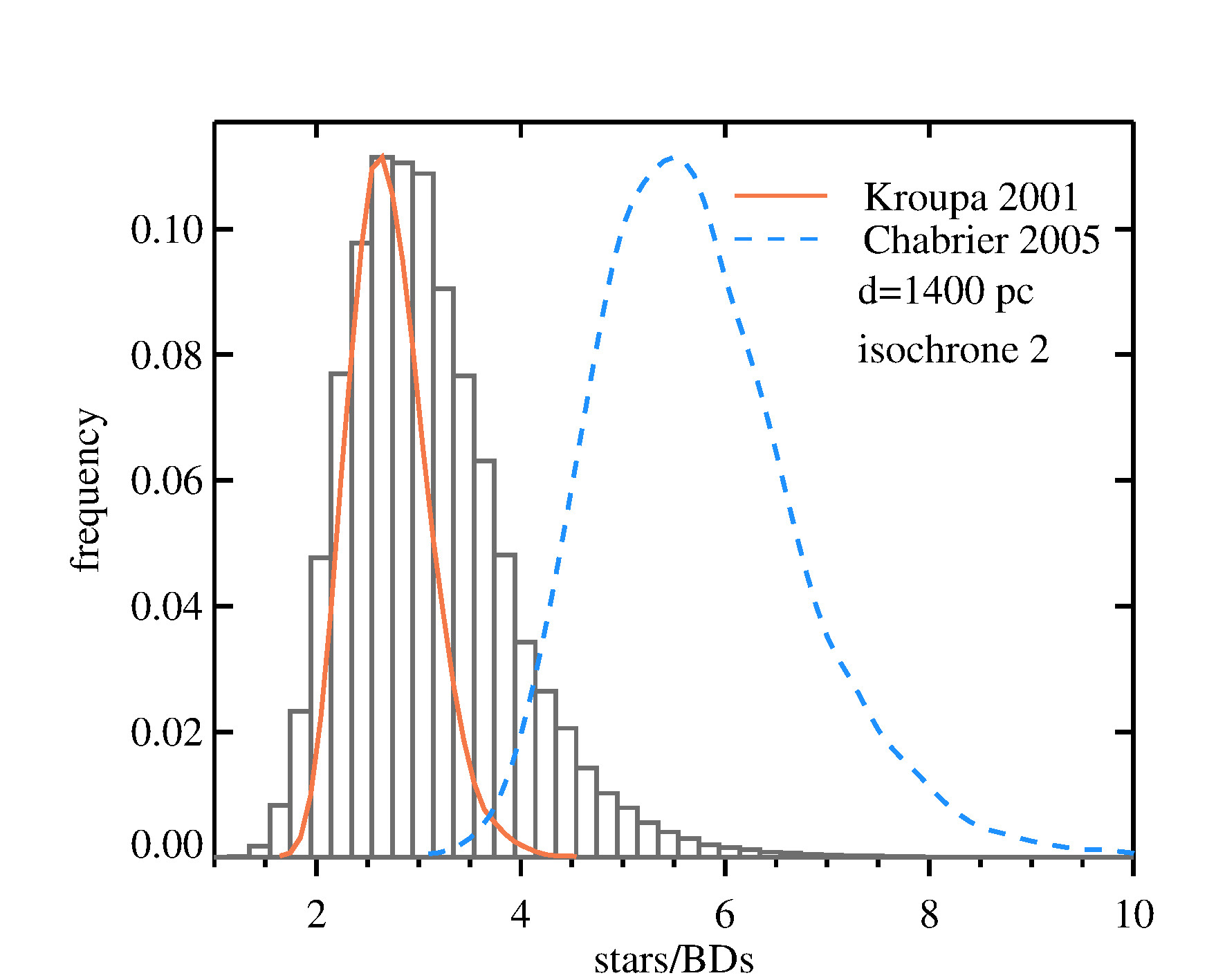}{0.33\textwidth}{}
          \fig{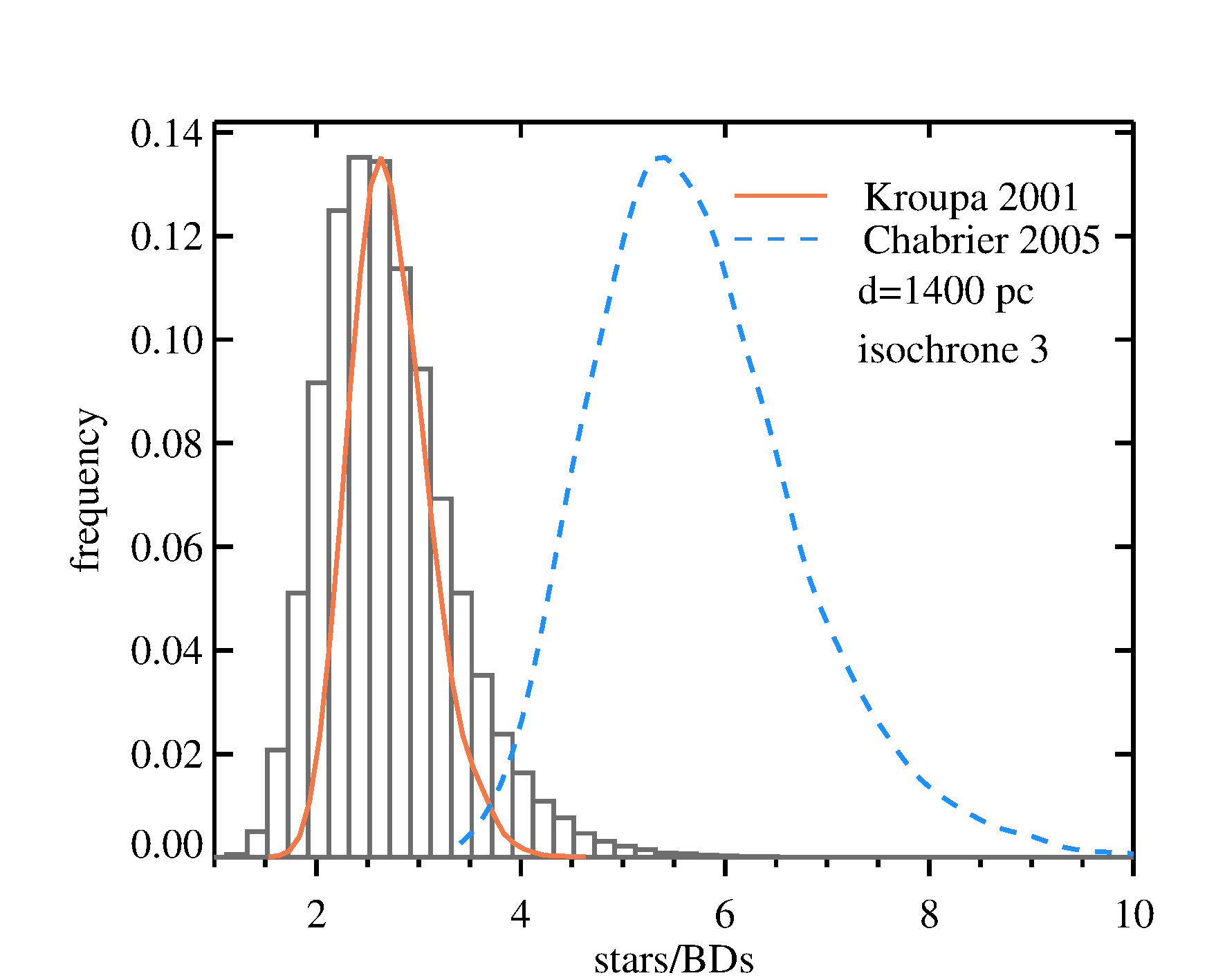}{0.33\textwidth}{}}
\caption{ 
Star to brown dwarf ratio distribution obtained from our data (grey histogram), for the distance of 1600 pc (top panels) and 1400 pc (bottom panels), and for the three different isochrones. 
Blue dashed and orange solid lines show the expected distribution of star-to-BD ratios
for 310 cluster members if their masses were following the Chabrier or Kroupa IMF forms, respectively, and normalized to the peak value of the histogram. }
\label{fig:ratio}
\end{figure*}

\begin{figure}
\centering
\resizebox{0.45\textwidth}{!}{\includegraphics{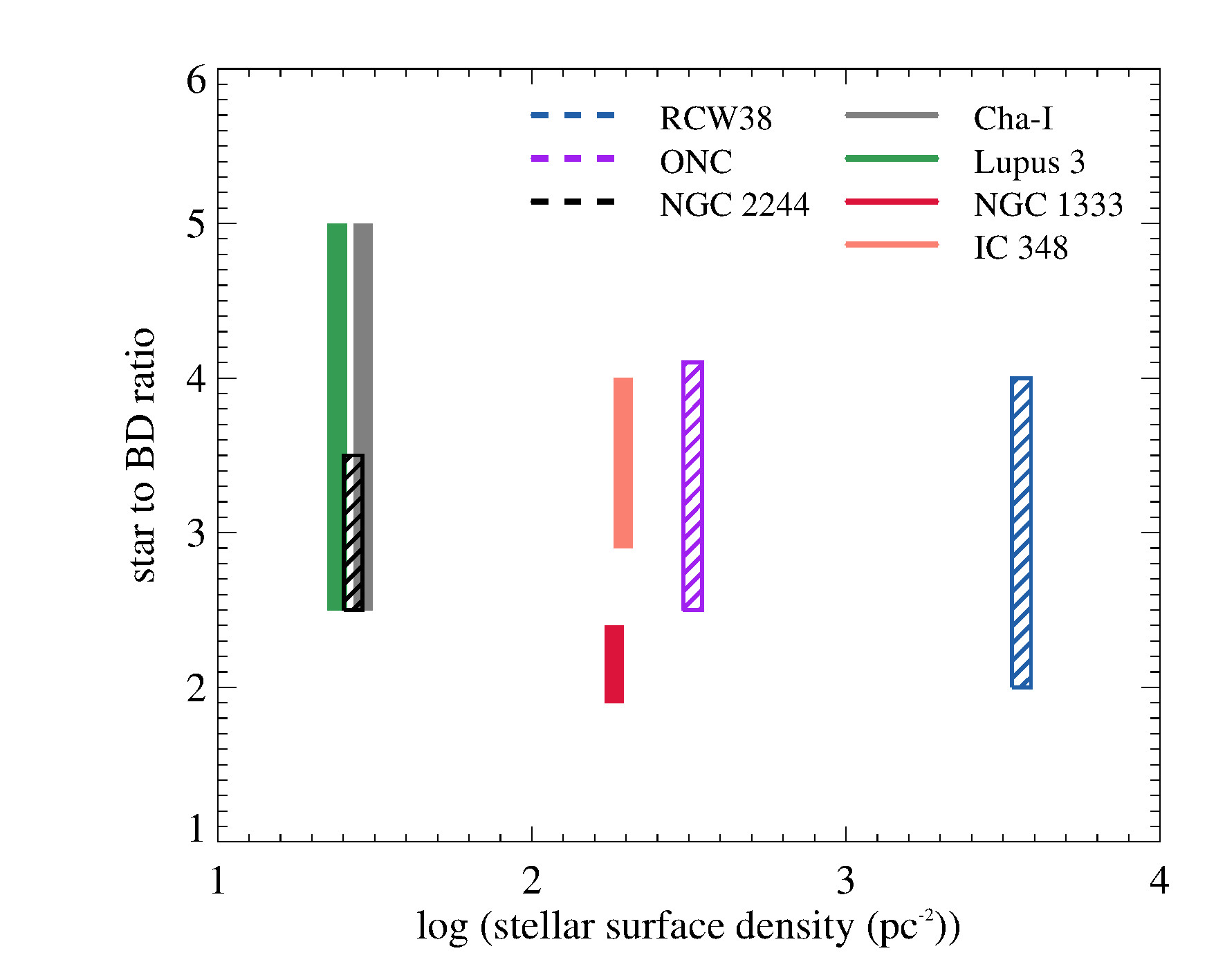}}
\caption{Dependence of the star-to-BD dwarf number ratio on cluster surface density. Different regions are represented by polygons of different colors. The height of each polygon represent either the $\pm1\,\sigma$ range around the mean value, or the range in star-to-BD ratio if given in this form (see text for details). The width of the polygons has no physical meaning. The filled polygons represent the regions with few or no massive stars, while the dashed ones mark the regions with substantial OB star population. 
}
\label{fig_sbd_sigma}
\end{figure}

To estimate the ratio between stars and brown dwarfs, we use the masses 
derived in Section~\ref{sec_masses_nir_excess}, and apply the same method as we did in deriving the IMF, 
generating 10$^4$ mass distributions for each of the 324 member lists. 
The stellar-substellar mass boundary is set to 0.075\,M$_{\sun}$, the value at the solar metallicity. 
We calculate the ratio using two low-mass limits at the BD side, 0.02\,M$_{\sun}$ and 0.03\,M$_{\sun}$. 
On the stellar side, we set an upper limit at 1\,M$_{\sun}$ as commonly found in the literature, but also calculate the ratio by taking into account all stellar candidates. 
The results are given in Table~\ref{tab:imf}, for the three different isochrones used in this work. 
The first column contains the results for the mass ranges 0.03 - 0.075\,M$_{\sun}$ for the BDs, and 0.075 - 1\,M$_{\sun}$ for the stars, which are the limits most commonly
used in the literature and the most suitable for comparison to other works. The star-to-BD number ratio in this mass range for the core of NGC\,2244 is $3 \pm 0.3$.
We find that 
the star-to-BD ratio in NGC\,2244 is slightly larger than that of RCW\,38 ($2.0\pm0.6$; \citealt{muzic17}, however, the values are still in agreement within their uncertainties. 
Furthermore, the star-to-BD ratio in NGC\,1333 was found to be $1.9-2.4$, and in IC\,348 $2.9 - 4$ \citep{scholz13}. For Cha-I and Lupus~3, \citet{muzic15} derive the ratio of $2.5 - 6.0$, although the analysis based
on the completeness levels of the spectroscopic follow-up suggests that these ratios might in reality be on the
lower side of the quoted span.
These numbers are also in agreement with the star-to-BD ratios found in the ONC (3.3$^{+0.8}_{-0.7}$; \citealt{andersen08}) and NGC\,2024 (3.8$^{+2.1}_{-1.5}; $\citealt{andersen08}).

In section~\ref{sec_densities} of the Appendix, we derive stellar surface densities for a number of young clusters and star forming regions, and in Fig.~\ref{fig_sbd_sigma} we show the dependence of star-to-BD ratio on this parameter. Different regions are represented by polygons of different colors. The height of each polygon represent either the $\pm1\,\sigma$ range around the mean value, or the range in star-to-BD ratio if given in this form. The width of the polygons has no physical meaning. The filled polygons represent the regions with few or no massive stars, while the dashed ones mark the regions with substantial OB star population. The star-to-BD ratio does not to seem to depend on neither the stellar density, nor the presence of OB stars.

To compare our results with expectations from the standard mass function forms, we perform a simulation in which we draw 310 stars in the mass range 0.02 - 10\,M$_{\sun}$ (our estimate of the average number of members and mass limits in the data) from the underlying Kroupa or Chabrier mass functions. We then estimate the star-to-BD ratio counting the sources in the 0.03 - 0.075\,M$_{\sun}$, and 0.075 - 1\,M$_{\sun}$ mass ranges. Repeating this process 10$^4$ times, we obtain distributions of star-to-BD number ratio expected from the two standard IMFs. 
The results are shown in Fig.~\ref{fig:ratio}, where the grey histograms shows the distribution obtained for NGC\,2244 from the 324 member lists (total $324\times10^4$ data points). 
The dashed blue and solid orange lines distributions expected from the underlying Chabrier, and the Kroupa IMFs, normalized to the peak value of the histogram. The six panels of Fig.~\ref{fig:ratio} show the star-to-BD ratios derived for the three isochrones, and two distances (1600\,pc and 1400\,pc). 
As expected from the IMF, the Kroupa mass function provides a better representation of the NGC\,2244 star-to-BD ratio than the Chabrier one, which predicts higher values of the ratio than what we find in NGC\,2244.

The low-mass end IMF slope in RCW\,38 is shallower than that of NGC\,2244, while the star-to-BD ratios we derived for RCW\,38 is lower than the one derived for NGC\,2244. This is the opposite of what is expected, since a shallower IMF slope should in principle result in less BDs with respect to stars (i.e. higher star-to-BD number ratio). The reason for this is the way we treated the incomplete mass bins in the RCW\,38 survey. 
The data in the lowest mass bins were corrected for incompleteness in the IMF calculation, according to the average $K$-band magnitude of the sources belonging to it. However, when calculating the star-to-BD ratio, all BDs are counted in a single bin and therefore this approach could not be applied. In this case we added ``missing" objects for each source if its $K$ magnitude was in the incomplete magnitude range. The mass function we derived for RCW\,38 is consistent with a slightly higher star-to-BD ratio ($\sim 4$). The value is still within the range expected from other star forming regions, but we note that the way the incompleteness is treated is clearly another 
important source of uncertainty. In this work, however, we do not encounter this issue since the analysis was restricted to the levels where our photometry is more than $90$\%complete.

\section{Discussion}
\label{sec_discussion}

The main goal of this work is to study the low-mass part of the IMF, and compare it with the well studied
mass distributions in nearby star forming regions, and with the results of our recent study in another massive young cluster RCW\,38. According to different BD formation theories, increased gas or stellar density, as well as 
the presence of massive OB stars are the factors that could lead to an increase in BD frequencies. 
To test these predictions, the first cluster we selected is the densest stellar system within 4 kpc from us (more than an order of magnitude denser than the nearby star forming regions) and at the same time rich in massive stars (RCW\,38; \citealt{muzic17}). The second cluster we studied is NGC\,2244, which exhibits low stellar densities similar to
nearby star forming regions (e.g. Chamaeleon), but hosts a significant population of massive stars. 
When combined, these observations provide meaningful constraints on brown dwarf formation models, as outlined in the following. 

The main result of the two studies is that the slopes of the IMF in the power law form, as well as the star-to-BD ratios, agree between each other, and also agree with the
typical values derived for the nearby star forming regions. If there are any variations in the low-mass mass function, they must be
more subtle than the error bars allow us to discern. 
What is the level of variations that we expect from theory, and how does that compare to our results?
In the scenario by \citet{bonnell08}, BDs are preferentially formed in regions with high stellar density. An increase of object density by an order of magnitude would result in an increase in the BD fraction by a factor of about two. There is, however, no evidence for a change at this level from the observational data.
The stellar densities in RCW\,38 and the ONC are $\sim\,$25 and $\sim\,$10 times higher than those in NGC\,2244, respectively, but the measured star-to-BD ratios are similar (2-4 for RCW\,38, $3\pm0.3$ for NGC 2244, 3.3$^{+0.8}_{-0.7}$ for the ONC; see Fig.~\ref{fig_sbd_sigma}). This scenario therefore seems to be ruled out by the available observations.

The predicted effect is more subtle in the hydrodynamical simulations of molecular cloud collapse by \citet{jones18}, 
where BDs are formed after ejection from multiple systems or disks. Here an increase of gas density by a factor of 100 leads to BD frequencies larger by $\sim 35$\% (assuming the mass bins 0.03 - 0.075\,M$_{\sun}$ for the BDs, and 0.075 - 1\,M$_{\sun}$). Gas surface density in the Milky Way's clouds is linearly related to their star formation rate \citep{heiderman10}, therefore we expect a similar increase in BD frequencies with the stellar density. 
Even if the predicted effect were present in our data, the errors involved in the calculation of the star-to-BD ratio would mask it. 

Finally, the turbulent fragmentation scenario also predicts an increase in BD production with gas density \citep{padoan02, padoan04, hennebelle09}. However, it is not trivial to quantify this increase, because the outcome of the simulations depends also on other factors (Mach number, scale of the turbulence). Furthermore, the simulations of turbulent fragmentation result in core masses, rather than the stellar masses, and it is unclear how exactly the core mass function (CMF) corresponds to the IMF. We can nevertheless try to make an estimate, by taking two CMFs calculated for a constant Mach number and different gas densities (Fig. 1 of \citealt{padoan02}), and assuming that there exists a direct mapping from the CMF to the IMF, with a shift by a factor of 3 in mass \citep{alves07, offner14}. With this assumption, we find that the increase in gas density by a factor of 25 results in a decrease in star-to-BD ratio by a factor of $\sim 5$. This is clearly not supported by our data, but it has to be stressed that this prediction might vary significantly if the initial conditions in the simulations are changed, or the CMF/IMF mapping takes a different form.

Apart from the density, a factor that could play a role in BD formation is a presence of OB stars. 
If a massive core that is significantly denser than its surroundings but
not as yet gravitationally unstable is overrun by an HII region, its outer material can be photoionized, while at the same time a shock front compresses the inner regions of the core. The final mass of the object depends on the competition of these two processes \citep{hester96,whitworth04,whitworth18}. Although the photoevaporation process is considered inefficient (a massive core is necessary to form a single BD) and cannot be a dominant process in BD formation, it could still influence the BD production in centers of massive clusters rich in OB stars. The two clusters we studied, RCW\,38 and NGC\,2244, are both rich in OB stars, with stellar densities differing by a factor of $\sim 25$. 
In RCW\,38 we found an IMF in agreement with low-mass star forming regions, but this still left an option that the two factors (presence of OB stars and high stellar densities) might somehow play an opposite role and cancel out the potential differences. In NGC\,2244, we test low stellar densities coupled with a presence of OB stars, but still we see no clear change in the IMF. We conclude that the presence of OB stars is unlikely to play any significant role in the formation of BDs. 

Another interesting observation we can make is that the IMF does not seem to be well described by the log-normal function (Figs. ~\ref{fig_imf_1600} and \ref{fig_imf_1400}), being much flatter below 1\,M$_{\sun}$ than at higher masses. A comparison of the Chabrier (empirical) mass function with simulations of turbulent fragmentation in molecular clouds with a variety of initial conditions is shown in Figs. 8 and 9 of 
\citet{hennebelle09}. A log-normal mass distribution is a natural outcome of the turbulent fragmentation scenario, but basically in all cases the simulation underpredicts the number of BDs, i.e. the predicted IMF is steeper in the very-low mass regime than the Chabrier one. On the other hand, the IMF we measure in NGC\,2244 is even shallower than the Chabrier mass function in this mass regime, which cannot be reproduced by the current simulations of turbulent fragmentation. Qualitatively, our IMF below 1\,M$_{\sun}$ looks more similar to that resulting from the simulations of gravitational collapse of molecular clouds by \citet{jones18} at low gas densities, although the same simulation predicts more intermediate-mass stars than we see.
 
Finally, we should mention that mass segregation has been reported for NGC\,2244 at very large radii ($r>14\arcmin$; \citealt{chen07}), manifesting itself through an excess of bright stars located inside this radius. At radii smaller than $12\arcmin$, however, no difference is seen between the radial distributions of high and low mass stars \citep{wang08}. Thus, mass segregation should not have an impact on our results.

\section{Summary and Conclusions}
\label{summary}

In this work we present new, deep near-infrared observations of the central $\sim 2.5\,$pc$^2$
of the young (2 Myr) cluster NGC\,2244, located at the heart of the Rosette Nebula.
The data are complete down to $\sim\,0.02$\,M$_{\sun}$, allowing, for the first time, an analysis of the 
candidate substellar population of this cluster.
The distance to NGC\,2244 was derived using $Gaia$ DR2 parallaxes, and estimated to be 1.59 kpc, with errors of 1\% (statistical) and 11\% (systematic).

For the stellar portion of the cluster, we queried the $Gaia$ DR2 and Pan-STARRS1 data bases and selected the candidate members based on color, proper motion and parallaxes, where available. For the object fainter than J$\sim$18 ($\sim 0.1$\,M$_{\sun}$), we only have the NIR photometry available.
Therefore, a statistical membership determination was performed through a comparison of the cluster CMD with that of 
a nearby control field. We estimate a field contamination of $61\pm2$\%, and a cluster population in the observed area of $\sim310$ members. 

According to different BD formation theories, increased gas or stellar density, as well as the presence of massive OB stars are the factors that could lead to an increase in BD frequencies. 
To test these predictions, we aim at comparing the IMF and star-to-BD number ratio in clusters with different environmental characteristics: the nearby star forming regions (e.g. Chamaeleon, NGC 1333), and the two massive clusters RCW\,38, and NGC\,2244. The first of the two massive clusters is characterized by high stellar densities (more than  an order of magnitude denser than the nearby star forming regions), and is rich in massive OB stars. NGC\,2244, on the other hand, exhibits low stellar densities similar to, e.g., Chamaeleon, but at the same time hosts a significant population of massive stars. 

We find that the IMF in NGC\,2244 can be well represented by a broken power law ($dN/dM \propto M^{-\alpha}$), with a break mass around 0.4\,M$_{\sun}$. A log-normal functional form of the IMF \citep{chabrier05} does not provide a good representation of the observed data. On the high-mass side ($0.4 - 7\,$M$_{\sun}$) we obtain $\alpha=2.12\pm0.08$, which is close to the Salpeter's slope. In the low-mass range ($0.02 - 0.4\,$M$_{\sun}$), we get $\alpha=1.03\pm0.02$, which is on the high side of the range of values values obtained in nearby star forming regions and RCW\,38 ($\alpha=0.5-1.0$), but still in agreement within the uncertainties.
Our results reveal no clear evidence for variations in the formation efficiency of brown dwarfs and very-low mass stars due to the lack or presence of OB stars or a change in stellar densities. If the gas or stellar densities have an influence on BD formation, this must be on a much more subtle level than the observational uncertainties currently permit us to measure.
This rules out the formation of significant numbers of BDs  via photoerosion of cores near OB stars \citep{whitworth04} and via gravitational collapse in infalling filaments \citep{bonnell08}.

In the future, a spectroscopic follow-up of the substellar candidates should be performed to confirm our results and to identify individual low-mass stellar and substellar members.

\vspace{0.5cm} 

KM acknowledges funding by the Science and Technology Foundation of Portugal (FCT), grants No. IF/00194/2015 and PTDC/FIS-AST/28731/2017.
Part of the research leading to these results has received funding from the European Research Council 
under the European Union's Seventh Framework program (FP7/2007-2013) / ERC grant agreement No. [614922]. RJ acknowledges support from NSERC grants. AS's work is supported by the STFC grant no. ST/R000824/1. L. C. acknowledges support from CONICYT-FONDECYT grant No 1171246. KPR acknowledges CONICYT PAI Concurso Nacional de Inserci\'on en la Academia, Convocatoria 2016 Folio PAI79160052.
This research has made use of the Spanish Virtual Observatory (http://svo.cab.inta-csic.es) supported from the Spanish MINECO/FEDER through grant AyA2014-55216.




\bibliographystyle{aasjournal}
\bibliography{ngc2244_v4} 



\appendix

\section{Isochrones}
\label{append_isochrones}

The basic products of the stellar models are the bolometric luminosity and effective temperature. To allow a comparison with observational data, these quantities have to be converted into magnitudes and colors, by applying bolometric corrections (BC) and $T_{\mathrm{eff}}$-color relations.  
Various groups provide stellar and substellar theoretical isochrones, employing the BCs and colors derived from the theoretical spectra, which are products of atmosphere models. A clear advantage of this approach is the availability of theoretical spectra for a wide range of $T_{\mathrm{eff}}$, metallicity and gravity,
which is substantially more difficult (or even non-feasible) to obtain through observations. 
Several authors, however, reported discrepancies between the model-predicted and the observed colors of pre-main-sequence (PMS) stars, stressing the need for using more realistic (empirical) BCs and color relations (e.g. \citealt{scandariato12, bell14, herczeg15}). Here we decide to use both approaches, using the color and magnitudes supplied by the modelers, and those derived from model-supplied luminosities by applying empirical BCs and colors.

The models we use are the PARSEC stellar models \citep{bressan12,marigo17} extending from 350\,M$_{\sun}$ down to $\sim$0.09\,M$_{\sun}$, and the Lyon models, consisting of BT-Settl \citep{baraffe15} in the range 1.4 -- 0.01\,M$_{\sun}$ and AMES-Dusty models \citep{allard01, chabrier00} for masses below 0.01\,M$_{\sun}$. In both cases we assume an age of 2 Myr and the solar metallicity.
Using the observed colors of member stars in young clusters with well-established age, distance and reddening, \citet{bell14} created a set of semi-empirical PMS isochrones, extending from 9 M$_{\sun}$ down to $\sim$0.09\,M$_{\sun}$ at 2 Myr. 
To perform the exercise ourselves, we use the BCs and colors appropriate for young stars from \citet{pecaut13} to convert the model luminosities to observables. They are available only for the stars with spectral types F0 - M5 ($T_{\mathrm{eff}}\approx\,$7300 - 2900\,K). Due to the lack of BCs suitable for young brown dwarfs, for the objects below the substellar limit we use the BC - $T_{\mathrm{eff}}$, and color - $T_{\mathrm{eff}}$ relations from \citet{golimowski04,stephens09}, and colors from \citet{hewett06}. Although these relations were derived for field objects, \citet{penaramirez12,penaramirez16} have shown that the isochrone obtained in this way describes the low-mass sequences of $\sigma\,$Ori (3 Myrs) and Upper-Sco (5-10 Myrs) very well. 
Using these BCs and colors, we modify the Lyon models in the range 1.4 - 0.001\,M$_{\sun}$, and the PARSEC models in the range 3 - 0.09\,M$_{\sun}$, limited by the availability of the PMS BCs and the model range. 
We note that the transition from the BT-Settl+old dwarf correction to the BT-Settl+young dwarf correction at 2900 K is not smooth, but the transition to the \citet{bell14} semi-empirical isochrone is.

We create the following sets of isochrones (Fig.~\ref{fig_iso}): 
\begin{itemize}
\item $isochrone$\,1: 
 \begin{itemize}
   \item m$\geq$\,1~M$_{\sun}$: original PARSEC 
   \item m$<$1\,M$_{\sun}$: original Lyon
  \end{itemize}
\item  $isochrone$\,2:
 \begin{itemize}
   \item m$\geq$\,3~M$_{\sun}$: original PARSEC 
   \item 0.07\,M$_{\sun}<$m$<3\,$M$_{\sun}$: combination of PARSEC and Lyon models modified by \citet{pecaut13} BCs (we average the two modified isochrones where they overlap)
   \item m$<$0.07~M$_{\sun}$: ($T_{\mathrm{eff}}\approx\,$2900\,K) the average between the original Lyon and the one modified by using the old dwarf colors and BCs.
  \end{itemize}
\item   $isochrone$\,3:
 \begin{itemize}
 \item m$\geq$\,9\,M$_{\sun}$: original PARSEC
 \item 0.09\,M$_{\sun}<$m$<$\,9M$_{\sun}$: \citet{bell14} isochrone
 \item  m$<$0.07~M$_{\sun}$: Lyon+old dwarfs BCs correction.
 \end{itemize}
\end{itemize}

\begin{figure*}
\centering
\resizebox{17cm}{!}{\includegraphics{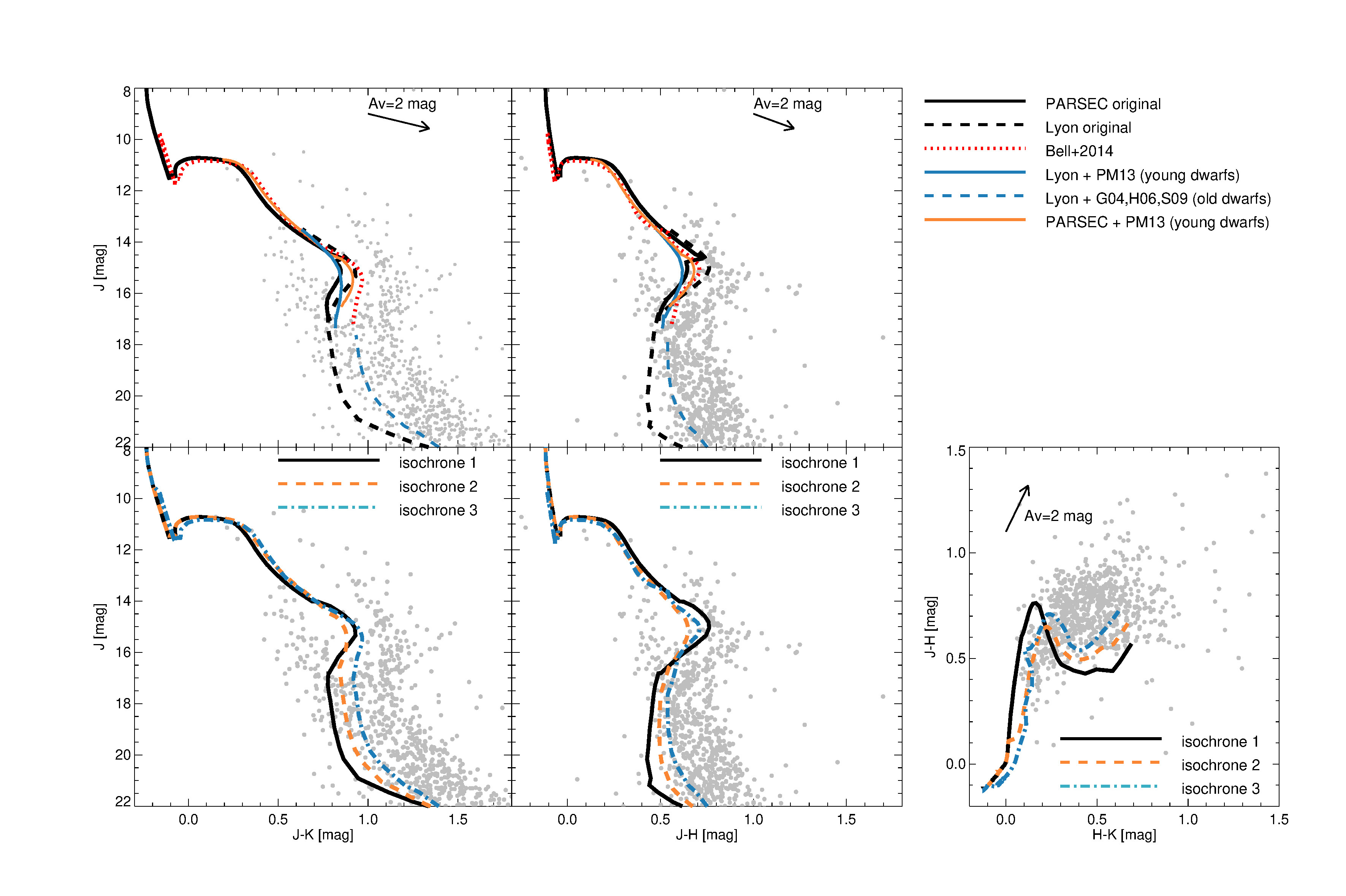}}
\caption{ 
{\bf Top panels:} Comparison of isochrones from different sources (see Appendix~\ref{append_isochrones} for detailed explanation).
{\bf Bottom panels:} Three final isochrones used to derive masses in this work. All the isochrones were shifted to the distance of 1600\,pc. Grey dots show our photometry towards NGC\,2244.}
\label{fig_iso}
\end{figure*}

\section{The effect of the bin size and location on the IMF}
\label{append_imf}

In Table~\ref{tab:app_imf} we show the results for the IMF slopes by varying the bin size and location. The first line in the table is the default one and is shown to facilitate the comparison.

\begin{deluxetable*}{cclll}
\tablecaption{The slope of the Initial Mass Function in NGC\,2244 in the power law form, for the distances of 1600\,pc and 1400\,pc, and for different bin sizes and bin locations.}
\label{tab:app_imf}
\tablehead{
\colhead{$\Delta$\,log(m/M$_{\sun}$)} &
\colhead{m$_{min}$(M$_{\sun}$)} &
\colhead{0.02 - 0.1\,M$_{\sun}$ } &
\colhead{0.02 - 0.4\,M$_{\sun}$} &
\colhead{0.4 - 7\,M$_{\sun}$}}
\tablecolumns{5}
\startdata 
\multicolumn{5}{c}{d\,$=1600$\,pc}\\
\hline
0.2 & 0.02 & $0.70\pm0.18$	& $1.03\pm0.02$ & $2.12\pm0.08$ \\
0.2 & 0.01 & $0.74\pm0.27$	& $1.01\pm0.03$ & $2.03\pm0.07$ \\
0.15 & 0.02 & $0.70\pm0.27$  & $0.99\pm0.02$ & $2.15\pm0.13$ \\
0.25 & 0.02 & $0.92\pm0.34$  & $0.97\pm0.04$ & $2.17\pm0.04$ \\
\hline
\multicolumn{5}{c}{d\,$=1400$\,pc}\\
\hline
0.2 & 0.02 & $0.82\pm0.12$	& $1.05\pm0.03$ & $2.09\pm0.03$ \\
0.2 & 0.01 & $0.61\pm0.13$	& $1.07\pm0.02$ & $2.13\pm0.10$ \\
0.15 & 0.02 & $0.72\pm0.12$  & $0.94\pm0.05$ & $1.97\pm0.02$ \\
0.25 & 0.02 & $0.69\pm0.23$  & $0.99\pm0.02$ & $2.16\pm0.07$ \\
\hline 
\enddata
\end{deluxetable*}

\section{Stellar surface densities}
\label{sec_densities}

\begin{figure*}
\centering
\resizebox{17.5cm}{!}{\includegraphics{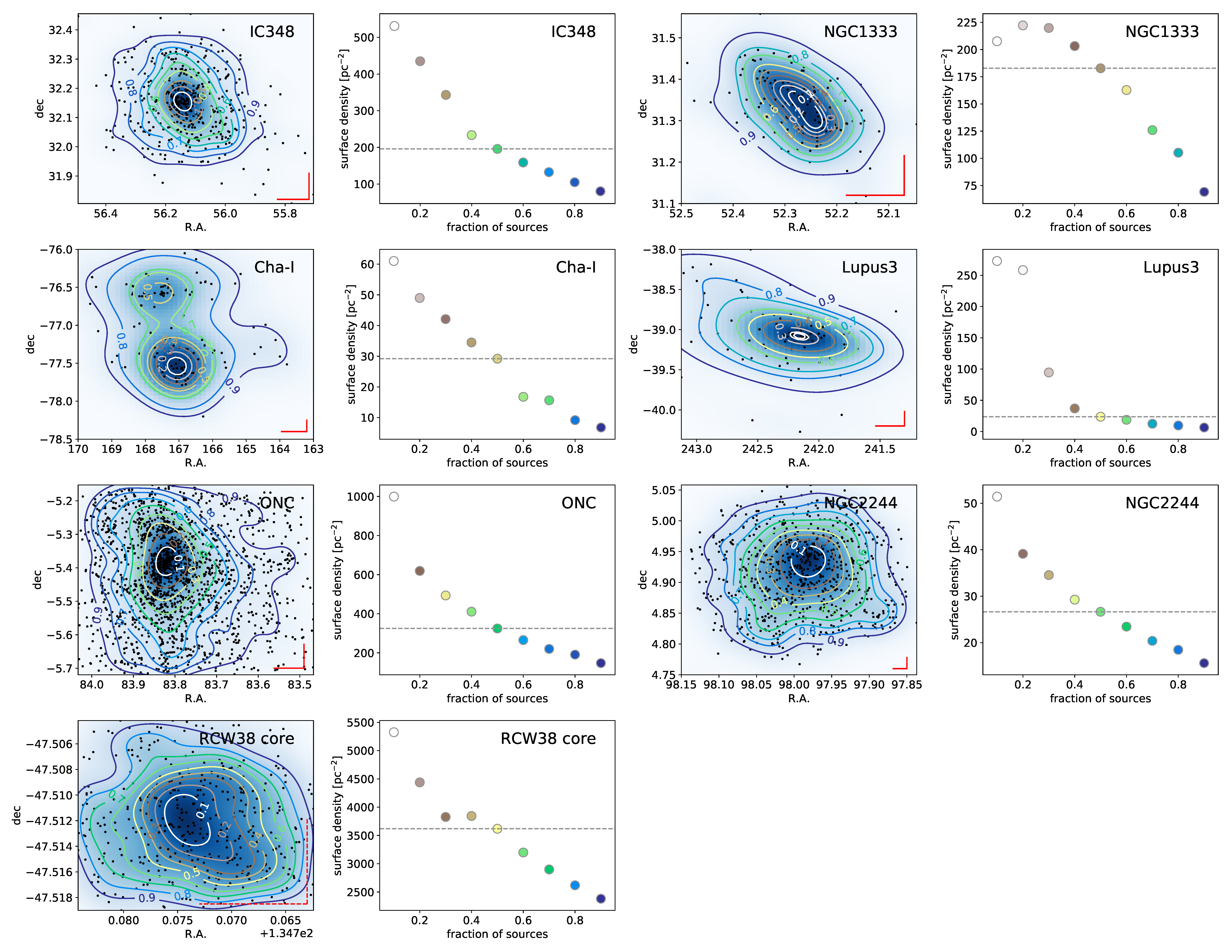}}
\caption{Spatial distribution of members in various clusters. The color maps are two-dimensional kernel density estimations (KDEs) of the member distributions, and the contours represent the levels containing different percentages of the sources ($10\%$ to $90\%$, in steps of $10\%$). The stellar surface densities for the area contained inside each contour are shown to the right of the KDE plots for each cluster. The surface densities associated with the 50\% contour are marked by the horizontal dashed lines. The red bars shown in the lower right corner represent the 0.5\,pc scale at the distance of each cluster, except for RCW\,38 where the scale shown is 0.2\,pc (dashed red lines).}
\label{fig_dens_kde}
\end{figure*}

To estimate the stellar densities, we plot the 2-dimensional kernel density estimations (KDE), using Gaussian kernels (Python Module \textit{scipy.stats.gaussian\_kde}). We then
identify the density contours that contain a certain percentage of cluster members (between $10\%$ and $90\%$, with a step of $10\%$; see Fig.~\ref{fig_dens_kde}). We calculate the stellar surface density ($\Sigma$) for the area inside each contour, shown to the right of the KDE distributions for each cluster. We chose the $\Sigma$ associated with the 50$\%$ contour level as a parameter to be used in comparisons of star-to-BD ratio with respect to cluster density. This parametrization is preferred to the one we used in \citet{muzic17}, where we use a fixed circular radius of 0.2\,pc, because the regions considered here show a diversity of stellar distributions (centrally concentrated, clumpy) and spatial scales. The red bars shown in the lower right corner represent the 0.5\,pc scale at the distance of each cluster, except for RCW\,38 where the scale shown is 0.2\,pc (dashed red lines).

As in \citet{muzic17}, we take into account only stars more massive than 0.1\,M$_{\sun}$, to avoid errors due to incompleteness.
For Cha-I, we consider the masses calculated in \citet{muzic15}, which are based on the census combined from several works \citep{luhman04a,luhman07,luhman&muench08, daemgen13, muzic15}.
For NGC\,1333 and IC\,348 we take the census from \citet{luhman16}, and exclude all the objects with the spectral type M7 or later. An M7 object should have the Teff$\sim$2900-3000\,K \citep{muzic14}, 
equivalent to 0.1\,M$_{\sun}$ at 1-3 Myr, according to the BT-Settl models. 
The NIR sources without spectral classification brighter than $Ks$=12.2 in NGC\,1333 and $Ks$=13.0 in IC\,348 are kept, since they
are expected to be more massive than 0.1\,M$_{\sun}$ already at A$_V=0$ (the difference in cut in $Ks$ stems from the differences in distance and age of the two clusters).  
For the ONC, we take the masses from \citet{dario12}, using the models of \citet{baraffe98}.
For Lupus~3 we take the census of M-type spectroscopically confirmed members from \citet{muzic14}, complemented with the earlier type members summarized in \citet{comeron08}. 

For NGC\,2244, we use the members identified in \citet{meng17} because their analysis contains member candidates over a significantly wider field and they are more representative of a cluster as a whole than the small region presented in this work. We note that the faintest member candidates from \citet{meng17} are only slightly brighter than the 0.1\,M$_{\sun}$ limit, but the effect on the surface density is negligible. If we take instead the candidate members from our $Gaia/PS1$ selection, we obtain $\Sigma\sim$60\,pc$^{-2}$ for a spherical area with the radius of 0.2\,pc. This core density is comparable to the peak of densities shown in  Fig.~\ref{fig_dens_kde} for NGC\,2244.
We note that the core density of NGC\,2244 may be slightly higher, since the central star HD46150 and a few bright stars in its vicinity (r$\leq7''$) appear saturated both in our and UKIDSS data, and therefore are not contained in our catalog. These five stars have been detected in the MYStIX X-ray data \citep{kuhn15}, and by adding them to the star counts we obtain $\Sigma\sim$95\,pc$^{-2}$. This value is in agreement with the surface densities reported by the MYStIX survey. 

For RCW\,38, we use the sources identified in \citet{muzic17}. This work only covered the central part of the cluster, but there is no other candidate member list over a wider field that would be comparable in depth and not biased toward disk-bearing objects. Therefore, the $\Sigma$ quoted here is probably somewhat overestimated, however, even by taking the lowest contour shown shown in Fig.~\ref{fig_dens_kde}, we obtain $\Sigma$ that is several times higher than in e.g. the ONC, which is the second densest cluster considered here. The surface density in the central circle with the radius of 0.2\,pc is $\Sigma\sim$2500\,pc$^{-2}$ \citep{muzic17}. 

Finally, we note that the numbers derived here depend on how each cluster area is defined (extent of the survey), and on member sample used. For example, if for the two clusters in Perseus (IC\,348 and NGC\,1333) we use the census from \citet{young15} corrected by the disk fractions \citep{luhman16}, we obtain $\sim35\%$ higher surface density for NGC\,1333 than for IC\,348.

\end{document}